%% file: main.tex
\documentclass[a4paper,11pt]{article}
\pdfoutput=1 

\usepackage[T1]{fontenc} 

\usepackage{multirow}
\usepackage{graphicx}
\graphicspath{{Figures/}}
\usepackage{amsmath}
\usepackage{amssymb}
\usepackage{array}
\usepackage{bm}
\usepackage{color,colortbl}
\usepackage[dvipsnames]{xcolor}
\usepackage[tikz]{bclogo}
\usepackage{microtype}
\usepackage{paralist}
\usepackage[section]{placeins} 
\usepackage{slashed}
\usepackage[
colorlinks=true
,urlcolor=blue
,anchorcolor=blue
,citecolor=blue
,filecolor=blue
,linkcolor=blue
,linktocpage=true
]{hyperref}

\usepackage{jcappub} 

\usepackage{subfig}

\usepackage{float}
\usepackage{tcolorbox}
\usepackage{framed}
\usepackage[version=4]{mhchem}
\usepackage{tabu}

\usepackage{tabularx}
\usepackage{arydshln}
\newcommand{\overbar}{\overline}
\title{An Extended Analysis of Heavy Neutral Leptons during Big Bang Nucleosynthesis}

\author[a]{Nashwan Sabti,}
\author[b]{Andrii Magalich,}
\author[c]{and Anastasiia Filimonova}

\affiliation[a]{King's College London, Department of Physics, Strand, London WC2R 2LS, UK}
\affiliation[b]{Lorentz Institute, Leiden University, Niels Bohrweg 2, Leiden, NL-2333 CA, The Netherlands}
\affiliation[c]{Institute for Theoretical Physics, Heidelberg University, D-69120 Heidelberg, Germany}

\emailAdd{nashwan.sabti@kcl.ac.uk}
\emailAdd{magalich@lorentz.leidenuniv.nl}
\emailAdd{filimonova@thphys.uni-heidelberg.de}

\abstract{Heavy Neutral Leptons (HNLs) are strongly motivated by theory due to their capability of simultaneously explaining the observed phenomena of dark matter, neutrino oscillations and the baryon asymmetry of the Universe. The existence of such particles would affect the expansion history of the Universe and the synthesis of primordial abundances of light elements. In this work we review, revise and extend the phenomenology of HNLs during the Big Bang Nucleosynthesis (BBN) epoch  for masses up to 1 GeV. This is of great importance, as BBN is able to provide complementary bounds to those from upcoming and proposed laboratory experiments. To this end we have developed a high-precision Boltzmann code that simulates BBN in the presence of HNLs and takes into account all relevant HNL decay channels, as well as subsequent interactions of decay products (thermalization and decay showers), dilution due to QCD phase transition, active neutrino oscillations and corrections to the weak reaction rates. We present robust bounds on the lifetime and mixing angles of HNLs for masses $3\,\mathrm{MeV}\leq m_N \leq 1\,\mathrm{GeV}$ and show that BBN is able to constrain HNL lifetimes down to $0.03 - 0.05$ s, depending on the mixing pattern. Moreover, combining our results with current experimental searches, we can exclude HNLs that mix purely with electron neutrinos up to ${\sim}$450 MeV and those that mix purely with muon neutrinos up to ${\sim}$360 MeV, in both cases for lifetimes up to at least a few tens of seconds. Finally, we compare the BBN constraints with those obtained from Cosmic Microwave Background observations and explore how our results will be improved by a number of upcoming and proposed laboratory experiments.
}

\begin{document}
\hspace{13 cm}KCL-2020-09\\
\maketitle
\flushbottom

\section{Introduction}
There is no doubt today that the Standard Model (SM) is deficient in explaining a number of observed phenomena in particle physics, cosmology and astrophysics. Examples of these phenomena are neutrino oscillations, dark matter and the baryon asymmetry of the Universe. Major efforts are being made from both a theoretical and an experimental perspective to address these issues. From a theoretical point of view, new particles beyond the SM (BSM) are proposed to accommodate for one or more of these issues. In particular, the existence of neutrino oscillations \emph{requires} the addition of new states to the SM (see e.g. \cite{Gelmini:1994az} for a review). Similarly, no particle in the SM can comprise all observed dark matter.
\vspace{2mm}

Among the suggested BSM candidates that have gained increasing interest in recent years, are Heavy Neutral Leptons (HNLs, also known as sterile neutrinos). HNLs are the right-handed companions to the SM neutrinos (active neutrinos) that can play the role of dark matter~\cite{Boyarsky:2018tvu,Gelmini:2009xd}, give masses to SM neutrinos~\cite{Ma:1998dn,Mohapatra:2006gs} or generate a baryon-asymmetry~\cite{Akhmedov:1998qx, Canetti:2012kh, Cline:2020mdt}. Moreover, it is possible to accommodate for these phenomena simultaneously by considering an extension of the SM with three HNLs -- the Neutrino Minimal Standard Model ($\nu$MSM)~\cite{Asaka:2005an,Asaka:2005pn,Shaposhnikov:2008pf,Canetti:2010aw,Ghiglieri:2019kbw}.
\vspace{2mm}

In this work we consider Heavy Neutral Leptons that are singlets with respect to the SM gauge group and couple to the SM via the neutrino portal as~\cite{Mohapatra:1979ia}:
\begin{align}
  \label{eq:neutrino_portal}
  \mathcal{L}_\text{neutrino portal} = F_{\alpha}(\overline{L}_\alpha\widetilde{H})N  + \mathrm{h.c.}\, ,
\end{align}
with $\alpha = \{e, \mu, \tau\}$, $F_{\alpha}$ dimensionless Yukawa couplings, $L_\alpha$ the SM lepton doublet, $\widetilde{H} = i\sigma_{2}H^{*}$ the Higgs doublet in the conjugated representation and $N$ the HNLs. The true singlet nature of HNLs allows for the inclusion of an HNL Majorana mass term that obeys the gauge invariance of the SM. As a consequence, after the SM Higgs acquires a non-zero vacuum expectation value $v = 246\,\mathrm{GeV}$ due to electroweak symmetry breaking, the HNL type-I seesaw Lagrangian reads:
\begin{align}
  \label{eq:seesaw_Lagrangian}
  \mathcal{L}_\text{HNL} = i\overline{N}\slashed\partial N - \left(
  M^{\mathrm{D}}_{\alpha}\overline{\nu}_\alpha N - \frac{1}{2} \overline{N^c}M_{N}N + \mathrm{h.c.}\right)\, ,
\end{align}
where $M^\mathrm{D}_\alpha \equiv F_\alpha v/\sqrt{2}$ is the Dirac mass matrix, $M_N$ is the HNL Majorana mass matrix and it is assumed that $M_N \gg M_\mathrm{D}$. The Dirac mass term leads to a mixing between the HNL and active neutrino sectors, which is characterized by small mixing angles and large splitting in masses. Therefore, neutrino gauge states contain a small contribution from HNLs. 
Consequently, this induces neutrino-like interactions for HNLs,
\begin{align}
  \label{eq:HNL_effective_L}
  \mathcal{L}_\text{int} = \frac{g}{2\sqrt 2} W^+_\mu \overline{N^c} \sum_\alpha \theta_{\alpha}^* \gamma^\mu (1-\gamma_5) \ell^-_\alpha + \frac{g}{4 \cos\theta_W}Z_\mu \overline{N^c} \sum_\alpha \theta_{\alpha}^* \gamma^\mu (1-\gamma_5) \nu_\alpha + \text{h.c.}\, ,
\end{align}
\vspace{-1mm}

which are strongly suppressed by the mixing angles
\begin{align}
  \theta_\alpha = |M^\mathrm{D}_{\alpha}| M^{-1}_{N}\,.
\end{align}

Here we are interested in minimal type-I seesaw extensions with HNL masses below electroweak scale, i.e., where direct searches of such HNLs are possible at present and future colliders. One way to realize low-scale seesaw scenarios is by imposing a `lepton number'-like symmetry that gets slightly broken in order to accommodate for neutrino masses~\cite{Shaposhnikov:2006nn, Sierra:2012yy}. Frequently studied models that fall into this category are the Neutrino Minimal Standard Model~\cite{Shaposhnikov:2006nn}, linear seesaw~\cite{Akhmedov:1995ip,Akhmedov:1995vm}, inverse seesaw~\cite{Mohapatra:1986bd} and minimal flavour violation models~\cite{Gavela:2009cd, Cirigliano:2005ck, Alonso:2011jd, Dinh:2017smk}. In the case of two Majorana HNLs with nearly degenerate masses and small `lepton symmetry'-breaking terms, the Majorana pairs can be conveniently combined to form a pseudo-Dirac HNL~\cite{Petcov:1982ya}, which is the type of particle we will consider from this point onward.
\vspace{2mm}

HNLs can provide solutions to the observed phenomena mentioned before for a wide range of masses and couplings. Additionally, they can also serve as a feasible explanation of the reported excess in baseline experiments \cite{Masip:2012ke,Fischer:2019fbw} and possibly alleviate the tension in measurements of the Hubble expansion rate \cite{HNLs_Neff}. Their parameter space can be constrained by seesaw considerations \cite{Asaka:2011pb}, experimental searches (see \cite{Deppisch_2015} for a review) and cosmological and astrophysical observations \cite{Dolgov_1988IJMPA3267D, Fuller:2009zz, Fuller:2011qy, Raffelt:2011nc,Albertus:2015xra,Mastrototaro:2019vug, Dolgov:2000jw, Ruchayskiy:2012si} (see also \cite{Drewes:2015iva,Chrzaszcz:2019inj} for combined analyses). Strong experimental constraints on HNLs in the MeV $-$ GeV mass range were previously obtained in studies of pion and kaon decays, and neutrino-nucleus scatterings \cite{Bernardi:1987ek,Bergsma:1985is,Vaitaitis:1999wq,Orloff:2002de,Shaykhiev2011,Artamonov:2014urb,Aguilar-Arevalo:2017vlf}. Further reach in mass was possible using decays of heavier particles at $e^+e^-$ colliders \cite{Abreu:1997uq,Liventsev:2013zz}. Complementary searches of active neutrino mixing patterns are summarized in~\cite{Esteban:2018azc} and can be used to put constraints on right-handed neutrinos (see e.g.~\cite{Chrzaszcz:2019inj}). More recently, a comprehensive experimental program has been developed to search for HNLs at collider, collider-based and neutrino experiments \cite{deGouvea:2015euy,Aad:2015xaa, Antusch:2015mia, Aad:2019kiz,Sirunyan:2018mtv,Sirunyan:2018xiv,Verbeke:2317083,Bondarenko:2019tss,Shuve:2016muy,Aaij:2016xmb,Antusch:2017hhu,beacham2019physics,Lurkin:2017tmu,Drewes:2019vjy, Abratenko:2019kez, Ballett:2016opr, Abe:2019kgx, Aad:2020srt, NA62:2020mcv}, together with a vast effort in the theoretical community in reinterpreting and combining constraints from various experimental sources \cite{Atre:2009rg, Ruchayskiy:2011aa,Antusch:2014woa,Fernandez-Martinez:2015hxa,Abada:2015trh,Abada:2016awd,Drewes:2016jae,Fernandez-Martinez:2016lgt, Das:2017zjc,Cvetic:2018elt,Bryman:2019ssi,Bryman:2019bjg,Chun:2019nwi, Zamora-Saa:2019naq, Bolton:2019pcu}. In addition, a wide range of experiments has been proposed to further extend the parameter space that has already been probed~\cite{Blondel:2014bra,Gligorov:2017nwh,Curtin:2018mvb,antusch2018probing,SHiP:2018xqw,Kling:2018wct,Bondarenko:2019yob,Ariga:2019ufm,Ballett:2019bgd,beacham2019physics, Dercks:2018wum, Tastet:2019nqj, Hirsch:2020klk, Gorbunov:2020rjx}.
\vspace{2mm}

From a cosmological perspective, both the Cosmic Microwave Background (CMB) and Big Bang Nucleosynthesis (BBN) can provide bounds that are complementary to those from laboratory experiments. This makes them relevant when defining goals for proposed accelerator experiments. For example, a change in the effective number of relativistic species due to additional long-lived particles and their decay products can be probed by CMB measurements (see e.g.~\cite{Aghanim:2018eyx, Escudero:2018mvt, Escudero:2020dfa}).
Similarly, BBN provides a window into the very early stages of the Universe, when several cosmologically relevant reactions made the delicate transition from the equilibrium to the non-equilibrium regime. Given the outstanding agreement between BBN predictions in the SM and measurements of primordial abundances, this makes BBN a powerful probe that has been widely used to constrain new physics \cite{Khlopov, Sarkar:1995dd, Kawasaki:1999na, Iocco:2008va,Jedamzik:2009uy,Pospelov:2010hj,Hufnagel:2017dgo, Kawasaki:2017bqm, Forestell:2018txr}.
\vspace{2mm}
\newpage

\vspace*{-0.75cm}
\enlargethispage{1.5cm}
The impact of short-lived HNLs on BBN has been studied in detail for low masses and bounds have been derived for masses below the pion mass in \cite{Dolgov:2000pj,Dolgov:2003sg,Ruchayskiy:2012si} and up to ${\sim}200\, \mathrm{MeV}$ in~\cite{Dolgov:2000jw}. In general, it has been shown that BBN can be used to constrain models in which the particle's lifetime exceeds ${\sim}0.1\,\mathrm{s}$. Currently, the bounds in the literature are extrapolated to higher masses and do not take into account various effects that become relevant in this domain, such as new decay modes, subsequent interactions of decay products, dilution due to QCD phase transition and a number of corrections to the weak reaction rates.
\vspace{2mm}

In this work we review and revise the impact of short-lived, thermally decoupled HNLs on BBN, update the currently existing bounds on HNL lifetimes and mixing angles, and extend the analysis for masses up to 1 GeV. The goal here is to establish a robust estimate of the BBN bound in this mass range. To this end, we have developed a Boltzmann code that accurately accounts for the impact of sub-GeV scale HNLs on the cosmological evolution during the BBN epoch. We include all relevant effects as mentioned above, use up-to-date measurements of the primordial element abundances and marginalize over the baryon density to get pure BBN constraints. Additionally, we make a direct comparison with the previous works \cite{Ruchayskiy:2012si,Dolgov:2000jw} and the more recent analysis presented in~\cite{MV_paper}. We also comment on the models for which our bounds apply and on future prospects.
\vspace{2mm}

This paper is organized as follows: In Section \ref{sec:new-physics-and-BBN}, we discuss the relevant interactions of HNLs with SM particles, the impact of such particles on BBN and the parameter space we expect to probe with BBN. In Section \ref{sec:methodology}, we describe the methodology used to model HNLs during BBN and the statistical procedure deployed to set bounds on their lifetime. In Section \ref{sec:results}, the results of this work are presented. In Section \ref{sec:discussion}, we discuss the robustness and applicability of the obtained bounds, make comparisons with previous literature and bounds from the CMB, and comment on future cosmological constraints. Finally, we present our conclusions in Section \ref{sec:conclusion}. A complete set of bounds and further technical details are included in the Appendices \ref{app:HNL_bounds_theta} --
\ref{app:numerical-methods}.

\vspace{-0.1cm}
\section{HNLs During the BBN Epoch}
\label{sec:new-physics-and-BBN}
\vspace{-0.1cm}
\subsection{Interactions with the SM Plasma}
\label{subsec:HNL_interactions_plasma}
In the expanding Universe, the effectiveness of interactions is controlled by the ratio of the interaction rate and the Hubble expansion rate: $\Gamma / H$. Interactions with $\Gamma \gg H$ are efficient in maintaining particles in thermodynamic equilibrium, while interactions with $\Gamma \ll H$ effectively do not happen, leaving a particle out of equilibrium (decoupled). In the special case when $\Gamma \sim H$, the system enters an intricate non-equilibrium regime. Primordial nucleosynthesis is characterized by a near coincidence of the decoupling temperature of weak reactions and the temperature at which nucleosynthesis begins. This makes BBN a sensitive probe of new physics at temperatures $T \sim1\,\mathrm{MeV}$.
\vspace{2mm}

In this work we consider Heavy Neutral Leptons with masses up 1 GeV. These particles interact with the SM plasma in a similar way as active neutrinos, but with an additional suppression due to the mixing angle. The typical HNL interaction rate $\Gamma_N$ in the relativistic limit can then be obtained from the active neutrino interaction rate $\Gamma_\nu$ by $\Gamma_N = |\theta|^2\Gamma_\nu$. This naturally implies that HNLs decouple like active neutrinos, but at a higher temperature. 
In general, the mixing angle depends on the state of the medium, making it possible for HNLs to initially be created out-of-equilibrium, but later thermalise and decouple~\cite{Asaka:2006nq, Shaposhnikov:2008pf, Ghiglieri:2016xye}. A refined calculation that takes into account all leading-order reactions in the equilibration rate, shows that HNLs with masses up to 1 GeV and lifetimes of interest ($\tau_N\lesssim 0.1\,\mathrm{s}$) enter equilibrium at temperatures $T \gtrsim 5$ GeV \cite{Ghiglieri:2016xye}. We adopt this result and assume that all HNLs considered here have been in equilibrium at some point

\newpage
\begin{figure}[t!]
    \vspace{-4.5mm}
    \centering
    \includegraphics[trim={0 0 1cm 0}, clip, width=0.7\textwidth]{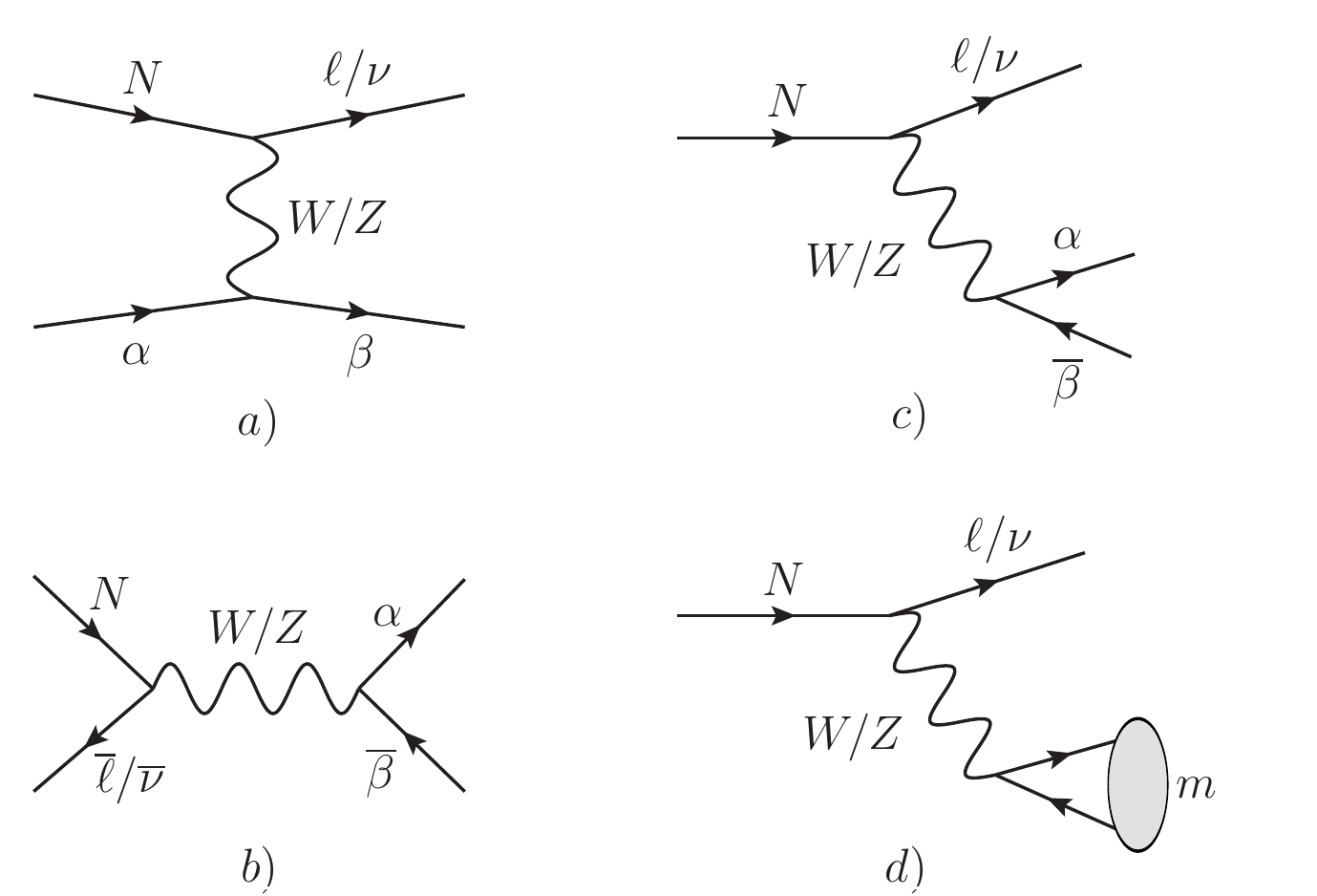}
    \caption{Tree-level diagrams of HNL interactions mediated by charged and neutral currents: \emph{a)} scattering with, \emph{b)} annihilation into and \emph{c)} decay into leptons/quarks, and \emph{d)} decay into mesons. The particles $\alpha$ and $\beta$ together represent a lepton or quark pair. $\alpha = \beta$ for neutral currents and $\alpha \neq \beta$ for charged currents.}
    \label{fig:HNL_diagrams}
\end{figure}

\begin{figure}[t!]
\vspace*{-0.4cm}
\begin{center}
\includegraphics[width=\textwidth]{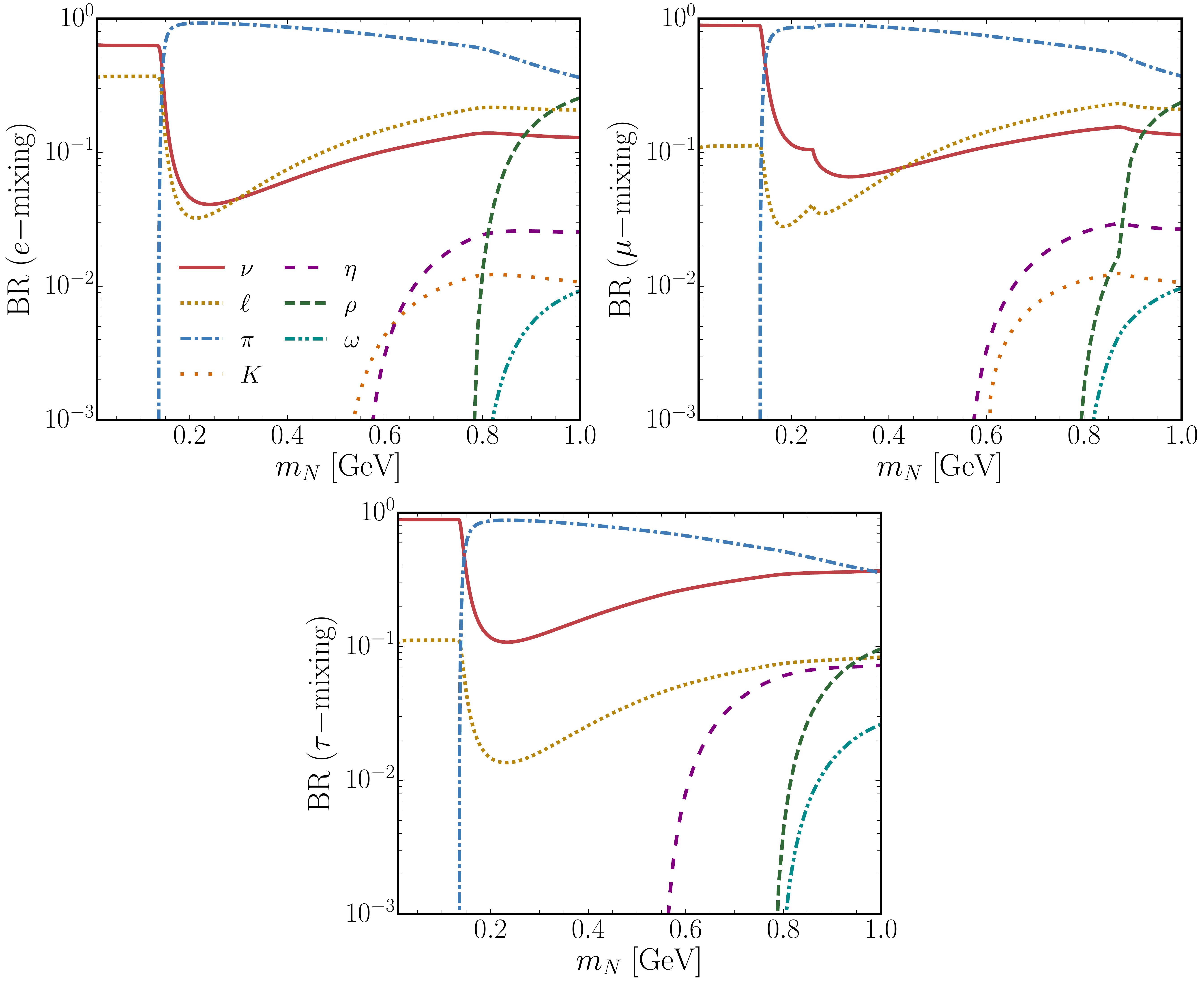}
\end{center}
\caption{Branching ratios of HNL decays below QCD-scale involving neutrinos $\nu$ only, charged leptons $\ell$, pions $\pi$, kaons $K$, $\omega$-, $\rho$- and $\eta$-mesons~\cite{Bondarenko:2018ptm}. \emph{Top left}: mixing with electron neutrinos only. \emph{Top right}: mixing with muon neutrinos only. \emph{Bottom}: mixing with tau neutrinos only. Note that the branching ratios of decays into kaons and $\omega$-mesons hardly exceed the percent level for the $e$- and $\mu$-mixing cases and are hence neglected in our analysis.}
\label{fig:HNL_BR}
\end{figure}

All interactions of HNLs with SM particles are mediated by charged and neutral currents. Since the interaction rates are suppressed by the mixing angle $\theta \ll 1$, only tree-level Fermi-type interactions are considered in this work (see Figure \ref{fig:HNL_diagrams}). At all temperatures, there are scattering, annihilation and decay reactions involving HNLs, active neutrinos and charged leptons. However, interactions with the hadronic matter need to be considered differently, depending on whether HNLs decouple above or below the QCD transition temperature $\Lambda_\mathrm{QCD}$. For temperatures $T > \Lambda_\mathrm{QCD}$, HNLs scatter with, annihilate and decay into free quarks, while for $T < \Lambda_\mathrm{QCD}$ the quarks are replaced by mesons.
The number of HNL decay modes grows with the mass of the particle and multi-meson final states become important for masses larger than 1 GeV \cite{Bondarenko:2018ptm}, while for smaller masses it is enough to consider only single-meson channels as the hadronic decay modes. In this work we will only consider decay channels that have a branching ratio larger than ${\sim}1\%$ (see Figure~\ref{fig:HNL_BR}).
\vspace{2mm}

The decay products of HNLs in their turn will interact with the SM plasma. The unstable products can be divided into particles that live long enough to interact with the plasma prior to decaying (muons, charged pions, kaons) and those that effectively decay immediately (neutral pions, $\eta$-mesons, neutral/charged $\rho$-mesons and $\omega$-mesons). Subsequent decays of heavy unstable HNL decay products can lead to a shower of SM particles, each affecting the process of nucleosynthesis in a certain way.
An overview of all relevant SM interactions and interactions involving HNLs, together with their corresponding matrix elements, is given in Appendix \ref{app:MatrixElements}.

\subsection{Influence of HNLs on BBN}
The two main observables of BBN are the primordial helium, $Y_\mathrm{P}$, and deuterium, $D/H|_\mathrm{P}$, abundances \cite{Cyburt:2015mya, Fields:2019pfx}. The primordial helium abundance is mainly sensitive to the neutron-to-proton ratio $n_n/n_p$ at the start of nucleosynthesis via the relation $Y_\mathrm{P} = \frac{2n_n/n_p }{n_n/n_p + 1}$ and depends only weakly on the baryon-to-photon ratio \cite{Pitrou:2018cgg, Fields:2019pfx}. In Standard Model BBN (SBBN), the neutron-to-proton ratio at the start of nucleosynthesis is determined by the decoupling of the reactions
\enlargethispage{0.3cm}
\begin{align}
    \label{eq:np_reactions}
    n + \nu_e &\leftrightarrow p + e^-\nonumber\\
    n + e^+ &\leftrightarrow p + \overline{\nu_e}\\
    n &\leftrightarrow p + e^- + \overline{\nu_e}\nonumber
\end{align}
and the expansion history between this decoupling and start of nucleosynthesis (see Appendix~\ref{app:npratio}). On the other hand, the primordial deuterium abundance is primarily sensitive to the baryon-to-photon ratio due to the deuterium photo-dissociation threshold (deuterium bottleneck) \cite{Pitrou:2018cgg, Fields:2019pfx}. Note that an increase in the baryon-to-photon ratio leads to a decrease in $D/H|_\mathrm{P}$. To put it concisely, the decay of a new particle into the neutrino sector of the plasma can affect the primordial abundances differently than if the decay happens into the electromagnetic sector~\cite{Sabti:2019mhn}. 
\newpage

\vspace*{-0.5cm}
The SM plasma is in thermal equilibrium before neutrino decoupling, which means that any signatures of HNLs are washed out if they decay before this event. Naturally, this sets a \emph{lower} (\emph{upper}) limit on the lifetime (mixing angles) of HNLs, below (above) which BBN cannot be a constraining probe. However, if their decays happen during or after neutrino decoupling, they will affect BBN by

\begin{itemize}
    \item their contribution to the cosmological expansion rate.\\
    The additional energy density of HNLs in the system leads to an earlier decoupling of the neutron-proton conversion reactions, but also to a shorter timescale during which neutrons can decay. The neutron-to-proton ratio at the start of BBN, and consequently $Y_\mathrm{P}$, will be larger. Similarly, a larger neutron fraction leads to a smaller value of $D/H|_\mathrm{P}$.
    
    \item the interactions of their decay products with the plasma.\\
    HNLs inject entropy into different sectors of the SM plasma, with particles that may have different energies than those found in the plasma.
    \begin{itemize}
        \item Decay into $\nu_e$ leads to two effects: \emph{i)} increase of the expansion rate, which leads to larger $Y_\mathrm{P}$ and smaller $D/H|_\mathrm{P}$, and \emph{ii)} additional interactions that preserve the equilibrium between neutrons and protons longer. The latter effect would make the neutron-to-proton ratio smaller if decay happens sufficiently early and neutrinos have low energies. If this is not the case, neutrino spectral distortions will dominate the neutron-proton conversion rates and cause higher values of $Y_\mathrm{P}$.

        \item Decay into $\nu_{\mu/\tau}$ will increase the expansion rate similarly to $\nu_e$. Moreover, neutrino-neutrino interactions and neutrino oscillations 
        will transfer part of the injected energy to electron neutrinos and affect BBN like described above.
        
        \item Decay into $e^\pm$ and $\gamma$ will inject more energy into the electromagnetic part of the plasma and heat it up. This will lead to an increased expansion rate and effectively dilute decoupled species. The dilution can be so severe, that the Hubble rate may actually become \emph{lower} compared to SBBN (see Section \ref{subsec:spectral_dilution}). The result is a decrease in $D/H|_\mathrm{P}$ (for fixed final baryon-to-photon ratio) and a slight increase in $Y_\mathrm{P}$ due to neutrino spectral distortions.
        
        \item Decay into pions and kaons will lead to these particles participating in neutron-proton conversion reactions prior to their subsequent decay. This will increase the neutron-to-proton ratio and thus $Y_\mathrm{P}$~\cite{Boyarsky:2020dzc}. Additionally, if HNLs are long-lived and these mesons are present around the time the nuclear reaction network commences, then they would be able to destroy light nuclei and interfere with the aforementioned effect~\cite{Kawasaki:2004qu, Pospelov:2010cw}.
    \end{itemize}
\end{itemize}

\begin{figure}[t!]
\vspace{-0.6cm}
\begin{center}
\includegraphics[width=\textwidth]{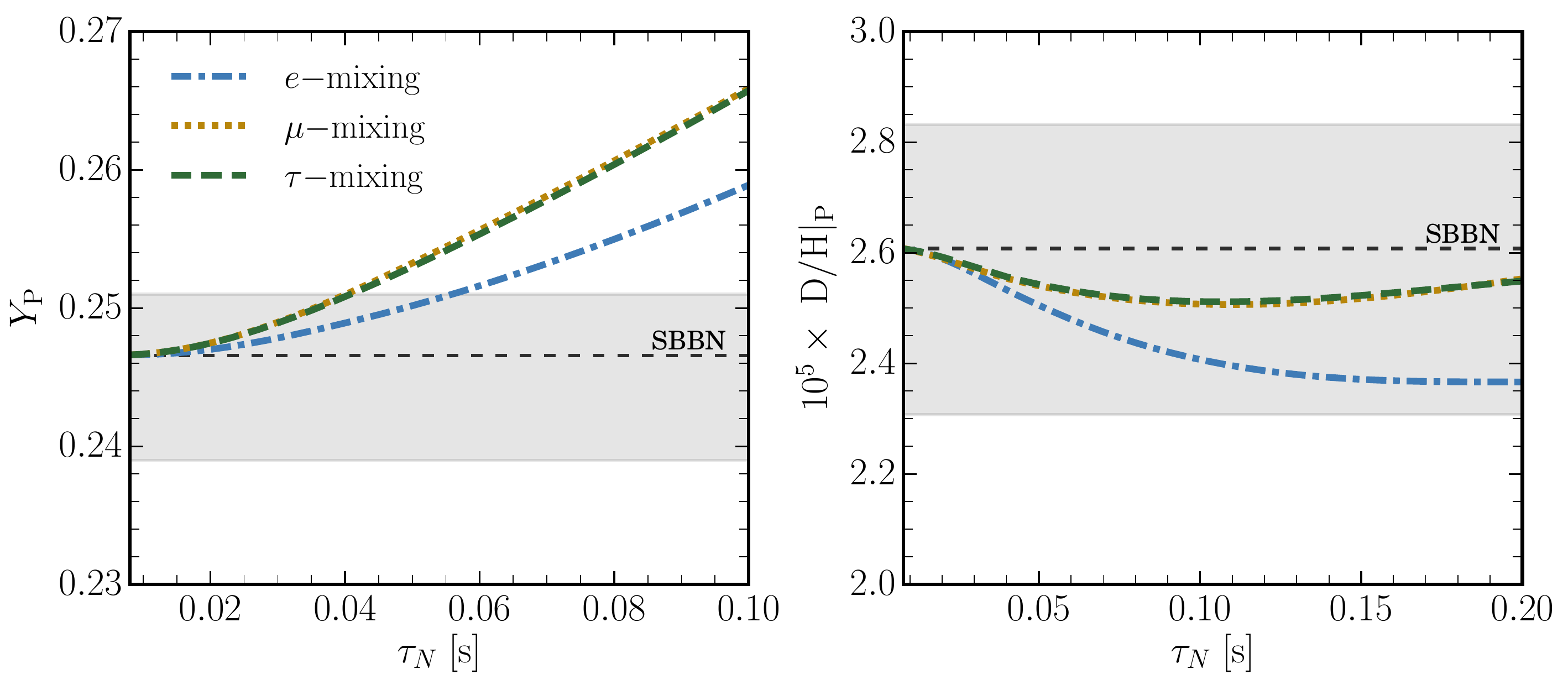}
\end{center}
\vspace{-0.2cm}
\caption{Impact of a $200\,\mathrm{MeV}$ HNL on the primordial helium (\emph{left}) and deuterium (\emph{right}) abundances as a function of its lifetime. The blue, yellow and green lines correspond to the purely electron, muon and tau neutrino mixing case respectively and the horizontal dashed lines depict the abundances in SBBN. The grey bands represent the mean $\pm2\sigma$ errors used in our BBN analysis (see Eqs.~\eqref{eq:Yp} -- \eqref{eq:sigmaDH}). These predictions are computed with neutron lifetime $\tau_n = 880.2$ s and baryon density $\eta = 6.09\times10^{-10}$.}
\label{fig:abundances}
\end{figure}

In Figure~\ref{fig:abundances} we show the change in the primordial helium (left panel) and deuterium (right panel) abundances with respect to the lifetime of a 200 MeV HNL that mixes with either electron, muon or tau neutrinos. The increase in $Y_\mathrm{P}$ is due to neutrino spectral distortions, while the decrease in $D/H|_\mathrm{P}$ is mainly caused by the dilution effect (see also Section~\ref{subsec:spectral_dilution}). Note that in the muon and tau neutrino-mixing cases the increase in the helium abundance is the strongest, since the branching ratio into neutrinos is the largest (see Figure~\ref{fig:HNL_BR}). On the other hand, in the electron neutrino-mixing case the deuterium abundance decreases more rapidly, since electron neutrinos from HNL decays are more efficient than muon and tau neutrinos in transferring energy from the neutrino sector to the electromagnetic sector of the plasma, which leads to a stronger dilution effect. Finally, note that with current measurements of the primordial helium abundance (grey band in the plot) the constraints on the HNL lifetime can\hfill
\enlargethispage{0.8cm}

\newpage
be notably improved, compared to the commonly used bound of 0.1 s. Currently, the theoretical error in the determination of $D/H|_\mathrm{P}$ is substantial, which leads to $Y_\mathrm{P}$ being the driving power behind the bounds on HNLs.
\enlargethispage{0.8cm}

To summarize, the influence of HNLs on BBN depends on their mass, lifetime and mixing pattern. For lifetimes smaller than the time of neutrino decoupling (${\sim}0.1$ s), the SM plasma is in thermal equilibrium and any deviation from this condition will be efficiently washed out. Therefore, BBN will not be able to constrain such HNLs. For lifetimes larger, however, there are effects that both increase and decrease the primordial abundances, making a numerical modelling of these processes a requisite. The name of the game is to accurately determine the expansion rate and the rates of the neutron-proton conversion reactions in Eq.~\eqref{eq:np_reactions}.

We note that the inclusion of neutron-proton conversions due to interactions with mesons requires an accuracy of order $10^{-7}$ in the determination of the HNL number density, which is beyond the framework established here and is therefore not taken into account. We refer the interested reader to~\cite{Boyarsky:2020dzc, MV_paper}, where the impact of HNLs on BBN is studied through a semi-analytical treatment of the effects mentioned before.

\subsection{Relevant Domain of Parameter Space}
There must be at least two HNLs in order to comply with the observed pattern of neutrino oscillations. In the case of two HNLs, there are 11 free model parameters, while in the case of three HNLs there are 18 parameters \cite{Drewes:2013gca}. For the BBN phenomenology, however, only the masses and mixing angles are relevant, since they determine when the HNLs decay and how much entropy is injected into each sector. The requirement that HNLs are responsible for neutrino mass generation provides a constraint on the HNL mixing angles -- the so-called seesaw bound. A convenient expression is given by~\cite{Asaka:2011pb}:
\begin{align}
   |\theta|^{2}\gtrsim |\theta|^{2}_{\text{see-saw}} \simeq \frac{1}{m_N}\sqrt{\Delta m_{\text{atm}}^{2}} \approx 5\cdot 10^{-11}\left(\frac{1\text{ GeV}}{m_{N}}\right)\, .
   \label{eq:see-saw}
\end{align}
Note that this is only a simplified limit (accurate within an order of magnitude or so) and that a more sophisticated estimation of the seesaw bound can be found in e.g. \cite{Asaka:2011pb} (see also the discussion in \cite{Drewes:2019mhg}). Nevertheless, it enables us to roughly estimate the part of parameter space where BBN can improve upon the seesaw bound. Using the approximate BBN bound of $\tau_N \sim 0.1\,\mathrm{s}$, we find that it intersects the seesaw bound at $m_N \approx 1\,\mathrm{GeV}$. For smaller masses, BBN provides stronger constraints than the seesaw bound. Vice versa, for masses well above $1\,\mathrm{GeV}$ and $\tau_N \gg 0.1\,\mathrm{s}$ HNLs cannot explain the pattern of neutrino oscillations, while for $\tau_N \ll 0.1\,\mathrm{s}$ BBN is more or less insensitive to such particles. Hence, in this work we focus on the influence of HNLs on BBN in the most relevant mass region $m_N<1\,\mathrm{GeV}$.


\section{Methodology}
\label{sec:methodology}
The contribution of baryons to the cosmological evolution during BBN is suppressed by the baryon-to-photon ratio $\eta \approx 6.09\times 10^{-10}$ \cite{Aghanim:2018eyx}. Therefore, we can first compute the evolution of the background cosmology in the presence of HNLs and subsequently pass on the relevant quantities\footnote{This includes the time, scale factor, temperature, time derivative of temperature, Hubble rate, total energy density and the neutron-proton conversion rates.} to an external code that takes care of the nuclear reaction network and solves for the primordial nuclei abundances. To this end, we have written a Boltzmann code, \texttt{pyBBN}\footnote{\href{https://github.com/ckald/pyBBN}{https://github.com/ckald/pyBBN}}, that computes the background cosmology and communicates with an external BBN code, the modified \texttt{KAWANO} code \cite{Kawano:1992ua, Ruchayskiy:2012si}, to solve for the nuclear reaction network and return the primordial abundances. The \texttt{pyBBN} code accurately simulates BBN in the presence of HNLs with masses up to $m_N \sim1$ GeV and includes all relevant interactions discussed in Section~\ref{subsec:HNL_interactions_plasma}, as well as a more refined computation of the neutron-proton conversion rates (see Appendix~\ref{app:weak_rates}). We note that the modified \texttt{KAWANO} code itself is rather outdated, since it does not include the most recent measurements of the nuclear reaction rates as compared to modern BBN codes \cite{Consiglio:2017pot, Pitrou:2018cgg, Arbey:2018zfh}. However, we make a direct comparison between the modified \texttt{KAWANO} code and \texttt{PArthENoPE2.\hspace{-0.5mm}0} \cite{Consiglio:2017pot} in Standard Model BBN (see Appendix \ref{app:Abundances_comparison}) and show that any deviations between the two are well below the percent level. Our code is subjected to many tests, of which a representative selection is summarized in Appendix \ref{app:pyBBN_tests}.
\vspace{2mm}

In this work, a number of well-justified approximations are made in order to simplify the computation of the time evolution of the system:
\begin{itemize}
    \item HNLs are treated as Dirac fermions with 4 degrees of freedom. Realistic low-scale seesaw models introduce pseudo-Dirac fermions with an approximate $U(1)$ `lepton number'-like symmetry, where the small symmetry-breaking terms have only negligible consequences for the BBN phenomenology~\cite{Shaposhnikov:2006nn, Gavela:2009cd}. The influence of Dirac HNLs on BBN is the same as two degenerate Majorana HNLs, as long as the mixing pattern, lifetime and spectrum are the same. This means that the mixing angle of the Dirac HNL $\theta_\mathrm{D}$ is related to that of the Majorana HNLs $\theta_\mathrm{M}$ by $\theta_\mathrm{D} = \sqrt{2}\theta_\mathrm{M}$. 
    
    \item Matter effects on the mixing angle of HNLs are neglected. The deviation of the HNL mixing angle in a medium $\theta_\mathrm{M}$ from the one in vacuum $\theta_\mathrm{V}$ is given by~\cite{Notzold:1987ik, MV_paper}:
    \begin{align}
        \label{eq:mediumeffectsmixing}
        \theta_\mathrm{M}(T) \approx \frac{\theta_\mathrm{V}}{1+2.2\cdot 10^{-7}\left(\frac{T}{1\,\text{GeV}}\right)^6\left(\frac{1\,\text{GeV}}{m_N}\right)^2} \ .
    \end{align} 
    For the mass and lifetime range considered in this work, medium effects alter the HNL mixing angles well below the sub-percent level and are thus not relevant.
    \item HNL decays with more than one meson in the final state are not considered. A comparison of the HNL decay widths of channels involving single mesons with the total decay width of HNLs into quarks, shows that multi-meson final states are only relevant for masses $m_N \gtrsim 1$ GeV \cite{Bondarenko:2018ptm}.
    \item Chemical potentials are neglected.
    Electrons and positrons strongly scatter and annihilate within the electromagnetic sector and quickly erase chemical potentials. A similar argument can be given for HNLs and neutrinos while still in equilibrium, since their interactions with the quark, neutrino and electromagnetic sectors wash out excessive chemical potentials to a sufficient degree. Right before they decouple, we evolve their distribution functions manually and therefore no information on the chemical potential is required afterwards. In addition, this work assumes a negligible primordial lepton asymmetry, which sets the chemical potential of a particle equal to that of its anti-particle  (see~\ref{subsec:applicability_bounds} for a discussion of the opposite case of non-zero lepton asymmetry). This leads to the convenient property that the distribution function of a particle of each species is the same as that of the corresponding anti-particle.
    \item Neutrino oscillations are taken into account by following a similar procedure as in \cite{Ruchayskiy:2012si}. A simple comparison of the characteristic timescale of oscillations between two flavours and the Hubble time at $T\sim 1$ MeV, shows that the former is much smaller than the latter. Since the weak reaction rates involving neutrinos are of the same order as the Hubble rate at this temperature, this means that active neutrinos will oscillate many times between subsequent interactions. Therefore, it is possible to describe the oscillation phenomenon by means of time-averaged transition probabilities $P_{\alpha\beta}$, which read \cite{Strumia:2006db}:
    \begin{align}
        \label{eq:nuPee}
       P(\nu_\alpha\rightarrow\nu_\beta) = \begin{cases}
       1 - 2V_{\alpha 1}^2V_{\alpha 2}^2 - 2V_{\alpha 3}^2(1 - V_{\alpha 3}^2)\, ,& \alpha = \beta\\
       -2V_{\alpha 1}V_{\alpha 2}V_{\beta 1}V_{\beta 2} + 2V_{\alpha 3}^2V_{\beta 3}^2\, , & \alpha \neq \beta
       \end{cases}
    \end{align}
    where $V$ is the unitary Pontecorvo-Maki-Nakagawa-Sakata (PMNS) matrix,
    \begin{align}
        V = \begin{pmatrix}
        c_{12}c_{13} & s_{12}c_{13} & s_{13}\\
        -s_{12}c_{23}-c_{12}s_{13}s_{23} & c_{12}c_{23}-s_{12}s_{13}s_{23} & c_{13}s_{23}\\
        s_{12}s_{23}-c_{12}s_{13}c_{23} & -c_{12}s_{23}-s_{12}s_{13}c_{23} & c_{13}c_{23}
        \end{pmatrix}\, ,
    \end{align}
    with $c_{ij} = \cos(\theta_{ij})$, $s_{ij} = \sin(\theta_{ij})$ and $\theta_{ij}$ the active neutrino mixing angles in vacuum. We use the best-fit values $\sin^2(\theta_{12}) = 0.297,\, \sin^2(\theta_{13}) = 0.0215\, \mathrm{and}\, \sin^2(\theta_{23}) = 0.425\, \text{(NH) or }\sin^2(\theta_{23}) = 0.589\, (\mathrm{IH})$, that are derived from a global fit of neutrino oscillation data \cite{Capozzi:2017ipn, Capozzi:2020qhw}. Note that the CP-violating phase usually present in the PMNS-matrix is neglected here, as in this work a zero lepton asymmetry is assumed.\\ At temperatures of a few MeV, electron neutrinos interact through both neutral \emph{and} charged currents, which is not the case for muon and tau neutrinos, due to a negligible abundance of muons and tau leptons available in the plasma. This will lead to a non-diagonal charged current contribution to the neutrino self-energy (present only for electron neutrinos), which cannot be absorbed in a redefinition of the fields. It means that, when the neutrino mass matrix is diagonalized, the mixing angles of the PMNS-matrix will now depend on properties of the medium. Following the approach 
    
    presented in \cite{Strumia:2006db}, the mixing angles in medium are given by:
    \begin{align}
      \theta^\mathrm{m}_{12} &= \frac{1}{2}\arctan\left[\frac{\sin(2\theta_{12})}{\cos(2\theta_{12})+\frac{A_\mathrm{MSW}}{\Delta m^2_\mathrm{sol}}}\right]\\
      \label{eq:oscillation_angle_1323}
      \theta^\mathrm{m}_{13(23)} &= \frac{1}{2}\arctan\left[\frac{\sin(2\theta_{13(23)})}{\cos(2\theta_{13(23)})\pm\frac{A_\mathrm{MSW}}{\Delta m^2_\mathrm{atm}}}\right]\\
      A_\mathrm{MSW} &= \frac{16\zeta(3)\sqrt{2}G_\mathrm{F}E^2T^4}{\pi M_\mathrm{W}^2}\ ,
    \end{align}
    where in the denominator of Eq.~\eqref{eq:oscillation_angle_1323} the '+'-sign is used for the normal neutrino hierarchy and the '$-$'-sign for the inverted neutrino hierarchy. Here, $\Delta m^2_\mathrm{sol} = 7.37\times 10^{-5}\, \mathrm{eV}^2$ and $\Delta m^2_\mathrm{atm} = 2.56\times 10^{-3}\, \mathrm{eV}^2$ denote the measured solar and atmospheric neutrino mass-squared differences respectively~\cite{Capozzi:2017ipn}. In our analysis we assume a normal neutrino hierarchy, but we have checked that the results do not change significantly for the inverted hierarchy.
    \item The QCD phase transition is modelled as an instantaneous event that takes place at temperature $T = \Lambda_\mathrm{QCD} = 150\,\mathrm{MeV}$. At this point, quarks and gluons are replaced by mesons and consequently the system undergoes a sudden jump in its expansion. The entropic degrees of freedom $g_{s}$ of relativistic particles before and after the phase transition are computed, after which entropy conservation is used to estimate the jump in the quantity $aT$:
    \begin{align}
        (aT)_\mathrm{after} = (aT)_\mathrm{before}\left(\frac{g_{s,\mathrm{before}}}{g_{s,\mathrm{after}}}\right)^\frac{1}{3}\ .
    \end{align}
    Under the assumption that the temperature remains constant during the phase transition, the scale factor will increase by a similar factor:
    \begin{align}
        a_\mathrm{after} = a_\mathrm{before}\left(\frac{g_{s,\mathrm{before}}}{g_{s,\mathrm{after}}}\right)^\frac{1}{3}\ .
    \end{align}
    This means that decoupled particles (in this case HNLs if the mixing angle is small enough) will experience a dilution effect, while particles in equilibrium are unaffected.
\end{itemize}
Some of these approximations make it possible to simulate the BBN epoch by only considering the evolution of the particle state of each species, i.e., the evolution of the anti-particle state is consequently set at the same time. The gravitational contribution of both particle and anti-particle of each species is then captured by the number of degrees of freedom, which enters in the definition of the energy density. In what follows, the system of equations used to simulate the BBN epoch is extensively laid out.

\subsection{System of Equations}
\label{sec:system_of_equations}
The behaviour of the homogeneous and isotropic Universe at large is governed by the Friedmann equations that connect its expansion rate to the energy density of the primordial plasma:
\begin{align}
    \label{eq:Friedmann_equation}
    H^2  &= \frac{8 \pi G \rho}{3}\\
    \label{eq:Energy_conservation}
    \frac{\mathrm{d} \rho}{\mathrm{d} t} &= -3H(\rho + p)\ ,
\end{align}
where $H \equiv \frac{\dot{a}}{a}$ is the Hubble expansion rate, $G$ is the gravitational constant, $\rho$ and $p$ are the total energy density and pressure of the primordial plasma respectively.

Writing the energy density as the sum of contributions from equilibrium and non-equilibrium species, $\rho = \rho_\mathrm{eq} + \rho_\mathrm{noneq}$, the evolution of the plasma temperature can be obtained from Eq.\eqref{eq:Energy_conservation} and reads:
 \begin{align}
    \label{eq:Temperature_evolution}
     \frac{\mathrm{d}T}{\mathrm{d}t} = -\frac{3H(\rho+p)+\frac{\mathrm{d}\rho_\mathrm{noneq}}{\mathrm{d}t}}{\frac{\mathrm{d}\rho_\mathrm{eq}}{\mathrm{d}T}}\ .
 \end{align}
Expressions for $\rho_\mathrm{eq}$ and $\rho_\mathrm{noneq}$ can be substituted in this equation to obtain an explicit formula
for the temperature evolution (see appendix \ref{app:TemperatureEvolution} for a full derivation).
\vspace{2mm}

Since the early Universe is homogeneous and isotropic, the distribution function of a particle $\alpha$ depends only on time and the absolute momentum $p$: $f_\alpha(t, \mathbf{x}, \mathbf{p}) = f_\alpha(t, p)$. 
A separate distribution function can be assigned to every quantum degree of freedom, but frequently the processes in the system will wash out the differences between some of them. For example, electromagnetic interactions of electrons easily flip the helicity, keeping their left- and right-helical distribution functions equal to each other. Together with the approximations of no lepton asymmetry and zero chemical potential, this means that $f_\alpha(t, p)$ is the 1-particle distribution function that we assign to each degree of freedom of a specie. Therefore, it is sufficient to track the evolution of only one $f_\alpha(t, p)$ and include the cosmological impact of a specie through the number of internal degrees of freedom $g_\alpha$. The density and pressure of a particle specie $\alpha$ are then defined by:
\begin{align}
    \rho_\alpha &= \frac{g_\alpha}{2\pi^2} \int\mathrm{d}p  \,p^2 E_\alpha f_\alpha(t,p) \\
    p_\alpha &= \frac{g_\alpha}{6\pi^2} \int\mathrm{d}p\,\frac{p^4}{E_\alpha}f_\alpha(t,p)\, ,
\end{align}
where $g_\alpha$ is the number of degrees of freedom. At temperatures around 1 MeV, weak interactions decouple and any spectral distortions in neutrino distribution functions are not washed out efficiently any longer. Distortions in the distribution functions of neutrinos can originate from the decay of HNLs into high-energy neutrinos. While spectral distortions are not relevant in SBBN \cite{Pitrou:2018cgg}, they can play a significant role when HNLs are present and affect the neutron-proton conversion rates. One way to obtain the evolution of distribution functions and to keep track of distortions due to non-equilibrium dynamics is by means of the unintegrated Boltzmann equation:
\begin{align}
\label{eq:BoltzmannEquation}
\frac{\mathrm{d}f_{\alpha}}{\mathrm{d}t} = \frac{\partial f_{\alpha}}{\partial t} - Hp\frac{\partial f_\alpha}{\partial p} = I_\alpha\,,
\end{align}
where $I_\alpha$ is the collision term that encodes the details of interactions. For a reaction involving specie $\alpha$ and $(\gamma-1)$ other particles of the form
\begin{align*}
\alpha + 2 + 3 + ... + \beta \leftrightarrow (\beta+1) + (\beta+2) + ... + \gamma\ \ .
\end{align*}
$I_\alpha$ is equal to \cite{Kolb:1990vq}:
\begin{align}
\label{eq:CollisionIntegral}
I_\alpha = \frac{N_W}{2g_\alpha E_\alpha}\sum_{\mathrm{ini, fin}}\int\prod_{i=2}^{\gamma}\frac{\mathrm{d}^3p_i}{(2\pi)^32E_i}S|\mathcal{M}|^2F[f](2\pi)^4\delta^4(P_\mathrm{in}-P_\mathrm{out})\ ,
\end{align}  
with $g_\alpha$ the degrees of freedom of particle $\alpha$, $P_i$ the four momentum of a state, $S$ the symmetry factor and $|\mathcal{M}|^2$ the \emph{unaveraged}, squared matrix element summed over helicities of initial and final states (see Appendix~\ref{app:MatrixElements} for details). The factor $N_W$ accounts for additional degrees of freedom in final-state particles other than helicities (for example, $N_W = N_c = 3$ for two quarks in the final state). The sum runs over all possible initial states `ini' = $\{\alpha,\, 2,\, \dots,\, \beta\}$ and final states `fin' = $\{(\beta+1),\, (\beta+2),\, \dots,\, \gamma\}$ involving particle $\alpha$. $F[f]$ is the functional that captures the influence of the medium on the interactions and is given by
\begin{align}
\label{eq:FunctionalF}
F[f] = (1\pm f_{\alpha})\dots(1\pm f_\beta)f_{\beta+1}\dots f_\gamma - f_{\alpha}\dots f_\beta(1\pm f_{\beta+1})\dots(1\pm f_\gamma)\ .
\end{align}
Here $(1 - f)$ is the Pauli blocking factor used for fermions and $(1 + f)$ the Bose enhancement factor used for bosons. For the interactions considered in this work, the collision term can be simplified considerably to obtain a 1- and 2-dimensional integral for three-particle and four-particle reactions respectively (see Appendix \ref{app:CollisionIntegral} for more details).
\vspace{2mm}

The collision term describes how the distribution function changes through both the direct and reverse reactions. High rates of interactions in the early Universe thermalize the distributions of most particles, bringing them to Fermi-Dirac or Bose-Einstein form (for fermions and boson respectively) and which makes the collision integral vanish. This provides a convenient initial condition for the system of Boltzmann equations considered, as we assume that \textit{all} particles were in thermal equilibrium at some high temperature. The equilibrium particle distribution functions are determined only by the temperature $T$ and read:
\begin{align}
    f(t, p) = \frac{1}{e^{E(p)/T(t)} \pm 1}\, ,
\end{align}
where $E(p) = \sqrt{p^2 + m^2}$ is the energy of the particle and the plus-sign and minus-sign in the denominator are used for fermions and bosons respectively.
At the temperatures considered in this work, the primordial background plasma consists of:
\begin{itemize}
    \item Photons and electrons/positrons, which are in equilibrium at all relevant times and have distribution functions of the Bose-Einstein and Fermi-Dirac form.
    \item Quarks and gluons above QCD phase transition, which are also in thermal equilibrium down to $T = \Lambda_\mathrm{QCD}$.
    \item Active neutrinos, which decouple at a temperature of few MeV. Slightly before decoupling, the Boltzmann equation is solved in order to accurately account for spectral distortions in their distribution functions. Neutrino oscillations are included by means of the time-averaged transition probabilities in Eq.~\eqref{eq:nuPee}. The set of three Boltzmann equations for active neutrinos then reads:
        \begin{align}
        \label{eq:BoltzmannEquationActiveNeutrinos} \frac{\partial f_{\alpha}}{\partial t} - Hp\frac{\partial f_\alpha}{\partial p} = \underset{\beta}{\sum} P_{\alpha\beta}I_\beta\,,
        \end{align}
    where $\alpha,\beta \in \{e,\mu,\tau\}$.
    \item Heavy Neutral Leptons. The condition that high-mass HNLs should survive down to temperatures of few MeV in order to affect BBN, restricts the mixing angle to small values. Consequently, this means that such HNLs decouple (semi-)relativistically. At high temperatures the SM plasma is in thermal equilibrium and medium effects can have a strong influence on the decoupling process of HNLs. Therefore, the full unintegrated Boltzmann equation for HNLs is also solved, to simulate their decoupling properly.
    \item Unstable decay products of HNLs. Some of these decay products -- muons and pions -- are in equilibrium until their decay rate starts to dominate the interaction rates that keep them in equilibrium. At temperatures of interest for BBN, all muons and mesons in the plasma are assumed to originate from HNL decays. Their subsequent decay proceeds through various decay channels, of which some can have yet again unstable particles in the final state. A Boltzmann equation for each of these unstable particles must be solved accordingly.
\end{itemize}

To summarize, the system of equations consists of: The Friedmann equation~\eqref{eq:Friedmann_equation} for the scale factor $a(t)$, the temperature evolution equation~\eqref{eq:Temperature_evolution} for the temperature $T(a)$, three Boltzmann equations~\eqref{eq:BoltzmannEquationActiveNeutrinos} for the active neutrino distribution functions $f_{\nu_\alpha}$, a Boltzmann equation~\eqref{eq:BoltzmannEquation} for the Dirac HNL distribution function $f_N$ and $X$ Boltzmann equations~\eqref{eq:BoltzmannEquation} for the distribution functions of $X$ unstable HNL decay products $f_{X_i}$. Initial conditions for Boltzmann equations are taken as equilibrium distributions at time slightly before decoupling. There is an equal number of evolution equations as there are unknowns. This system of equations is therefore closed and can be solved numerically.

\subsection{Cosmological Data and Analysis}
\label{subsec:bbn_data_analysis}
Common methods to measure the primordial helium abundance $Y_\mathrm{P}$ are based on \emph{1)} observations of the CMB damping tail \cite{Aghanim:2018eyx} and \emph{2)} observations of hydrogen and helium recombination lines in low-metallicity regions \cite{Izotov:2014fga,Aver:2015iza, Peimbert:2016bdg, Fern_ndez_2018, Valerdi:2019beb}. The former method gives a determination of $Y_\mathrm{P}$ with an error of ${\sim}10\%$ \cite{Aghanim:2018eyx}, while the latter method is reportedly able to provide a measurement of $Y_\mathrm{P}$ with percent-level accuracy. Likewise, the primordial deuterium abundance is commonly determined by observations of absorption features in quasar spectra due to metal-poor gas clouds~\cite{Cooke:2017cwo, Riemer-Sorensen:2017pey, Balashev:2015hoe, zavarygin_mnras}, which also is reported to yield an accuracy around one percent.

Our main analysis is performed with the means and errors for the primordial helium and deuterium abundances as recommended by the PDG \cite{pdg}. At $1\sigma$ they read:
\begin{align}
Y_\mathrm{P} 	            &= 0.245 \pm 0.003 \label{eq:Yp}\, , \\
D/H|_\mathrm{P}   	&= (2.547  \pm 0.025) \times 10^{-5} \label{eq:DH} \,.
\end{align}

The prediction of the primordial abundances is subject to errors originating from uncertainties in the neutron lifetime and the nuclear reaction rates \cite{Pitrou:2018cgg, Fields:2019pfx}.
They are given by:
\begin{align}
\sigma(Y_\mathrm{P})^\mathrm{Theo} 	        		&= 0.00018 \,, \\
\sigma(D/H|_\mathrm{P} )^\mathrm{Theo} &= 0.05 \times D/H|_\mathrm{P}^\mathrm{Theo} \label{eq:sigmaDH} \, .
\end{align}
Since the uncertainty in the neutron lifetime is directly accounted for, we fix it to the PDG recommended value of $\tau_n = 880.2\,\mathrm{s}$.
\vspace{2mm}

Boltzmann simulations can be computationally expensive and therefore it would not be practical to vary the baryon-to-photon ratio besides the HNL mass and lifetime. Therefore, we compute the primordial abundances for a given value of the baryon-to-photon ratio, which is taken as $\eta_0 = 6.09\times 10^{-10}$ \cite{Aghanim:2018eyx}, and then obtain the abundances for other values of $\eta$ by using general scaling relations in a similar fashion as in~\cite{Fields:2019pfx}:
\begin{align}
    Y_\mathrm{P} \propto \left(\frac{\eta}{\eta_0}\right)^{0.039} \qquad D/H|_\mathrm{P} \propto \left(\frac{\eta}{\eta_0}\right)^{\hspace{-0.7mm}-1.62}\, .
\end{align}
We have explicitly verified that these scaling relations apply independently of HNL mass, mixing angle and mixing pattern. 
The inclusion of HNLs in the system induces deviations of the primordial abundances compared to the SBBN case. In order to quantify these deviations, we define the following $\chi^2$: 
\begin{align}\label{eq:chiBBN}
\chi_{\rm BBN}^2 =\, \frac{\left[Y_{\rm P} - Y_{\rm P}^{\rm Obs}\right]^2}{\sigma^2(Y_{\rm P})^{\rm Theo} + \sigma^2(Y_{\rm P})^{\rm Obs}} + \frac{\left[{\rm D/H}|_{\rm P} - {\rm D/H}|_{\rm P}^{\rm Obs}\right]^2}{\sigma^2({\rm D/H}|_{\rm P} )^{\rm Theo} + \sigma^2({\rm D/H}|_{\rm P} )^{\rm Obs}} \, .
\end{align}
The statistical analysis is performed by computing the quantity $\chi^2_\mathrm{BBN}$ for different combinations of $(m_N, \tau_N, \eta)$ and then by marginalizing over $\eta$.
Next, a minimum $\chi^2_{\rm min}$, which in this work is taken as the $\chi^2$ in SBBN, is subtracted from all $\chi^2_\mathrm{BBN}$. This is a reasonable approximation of $\chi^2_{\rm min}$, since BBN predictions in the SM already give a good fit to the data and this approach will result in conservative bounds. A combination $(m_N, \tau_N)$ is then ruled out at $2\sigma$ when $\Delta \chi^2 \equiv \chi^2_\mathrm{BBN} - \chi^2_\mathrm{min} = 6.18$, where this last value is obtained by evaluating the quantile function of the $\chi^2$-distribution with 2 degrees of freedom $Q(p,2)$ at the point $p = 0.954$.


\section{Results}
\label{sec:results}
Here we present our main findings and bounds on the lifetime and mixing angles of HNLs.
\subsection{Overview of Relevant Effects}
\label{subsec:spectral_dilution}

\begin{enumerate}
    \item \textbf{Neutrino spectral distortions}\\
    The impact of neutrinos on BBN is defined in terms of their contribution to the expansion rate and their participation in the neutron-proton conversion reactions. SBBN is characterized by the fact that neutrino decoupling and neutron decoupling happen around the same time. This means that there is an absence of high-energy neutrinos around the time neutrons go out of equilibrium, as their distribution function is very close to that of a thermal one.
    The current generation of high-precision BBN codes (see e.g. \cite{Consiglio:2017pot, Pitrou:2018cgg, Arbey:2018zfh}) takes advantage of this mechanism and treat active neutrinos as thermal-like particles with an effective temperature at all times, i.e., they account for the gravitational effect of incomplete neutrino decoupling, but they do not fully consider the effect of spectral distortions on the weak rates. Indeed, in SBBN these distortions are negligible, as they only add a correction to the primordial helium abundance at the level of $10^{-3}$ (see~\cite{Froustey:2019owm} and Table~\ref{tab:eq_vs_noneq}).
    \vspace{2mm}
    
    Once HNLs enter the system, this picture changes significantly. If HNLs decay around time of neutrino decoupling, SM reactions will not be efficient enough to \emph{completely} bring high-energy neutrinos back in equilibrium (and thus restore the balance in the rates as in SBBN). Their energy loss rate is sufficiently high to distribute the excessive energy over the neutrinos in the thermal bath, but falls short of erasing non-equilibrium signatures in their spectra. We show that these spectral distortions in the active neutrino distribution functions are one of the main driving powers behind changes in the primordial helium abundance. High-energy neutrinos from HNL decays will dominate the neutron-proton conversion rates, moving their ratio away from $\Gamma_{p\rightarrow n}/\Gamma_{n\rightarrow p} = e^{-(m_n-m_p)/T}$ to higher values. Since the neutron-to-proton ratio is
    determined by the aforementioned ratio of rates, the presence of high-energy neutrinos leads to an \emph{increase} of the primordial helium abundance as compared to the SBBN case.
    
    \newpage
    We observe this trend of increasing $Y_\mathrm{P}$ due to spectral distortions irrespective of mass, lifetime and mixing pattern.
    The primordial deuterium abundance, on the other hand, is not so sensitive to the dynamics that determine the neutron-to-proton ratio, but rather to the baryon-to-photon ratio.
    \vspace{2mm}

    In Figure~\ref{fig:thermalized_neutrinos} we demonstrate the influence of spectral distortions on the neutron-proton conversion rates for a 200 MeV HNL of lifetime 0.08 s. From the left panel it is clear that the conversion reactions decouple around $T\approx1$ MeV, which marks the temperature range of interest we will focus on. In the right panel the ratio $\Gamma_{p\rightarrow n}/\Gamma_{n\rightarrow p}$ obtained by using non-equilibrium neutrino distributions and thermal-like neutrino distributions is compared to the same quantity in SBBN. We find that spectral distortions increase this ratio to higher values (dashed curve). More specifically, they increase the rates of the reactions $p\overline{\nu_e}\rightarrow ne^+$ and $n\nu_e\rightarrow pe^-$, since their cross-section is proportional to the squared neutrino energy. Whereas in SBBN the ratio of these two rates is mostly determined by the neutron-proton mass difference, this barrier is easily overcome in the presence of high-energy neutrinos. A larger value of the ratio $\Gamma_{p\rightarrow n}/\Gamma_{n\rightarrow p}$ naturally leads to a higher neutron-to-proton ratio and thus a higher primordial helium abundance. Note that the negligence of spectral distortions leads to the opposite effect on the ratio of proton-neutron conversion rates (solid curve) and therefore also on the primordial helium abundance. In Table~\ref{tab:eq_vs_noneq} we quantify this effect. We see that using non-equilibrium neutrino spectra as opposed to thermal-like spectra increases the primordial helium abundance with a relative difference higher than the $2\sigma$ error in $Y_\mathrm{P}$ (Eq.~\ref{eq:Yp}). As an example, for a 200 MeV HNL of lifetime 0.08 s, this makes the difference between being excluded by current data or not.

    \begin{figure}[t!]
    \vspace{-0.18cm}
    \begin{center}
    \includegraphics[width=\textwidth]{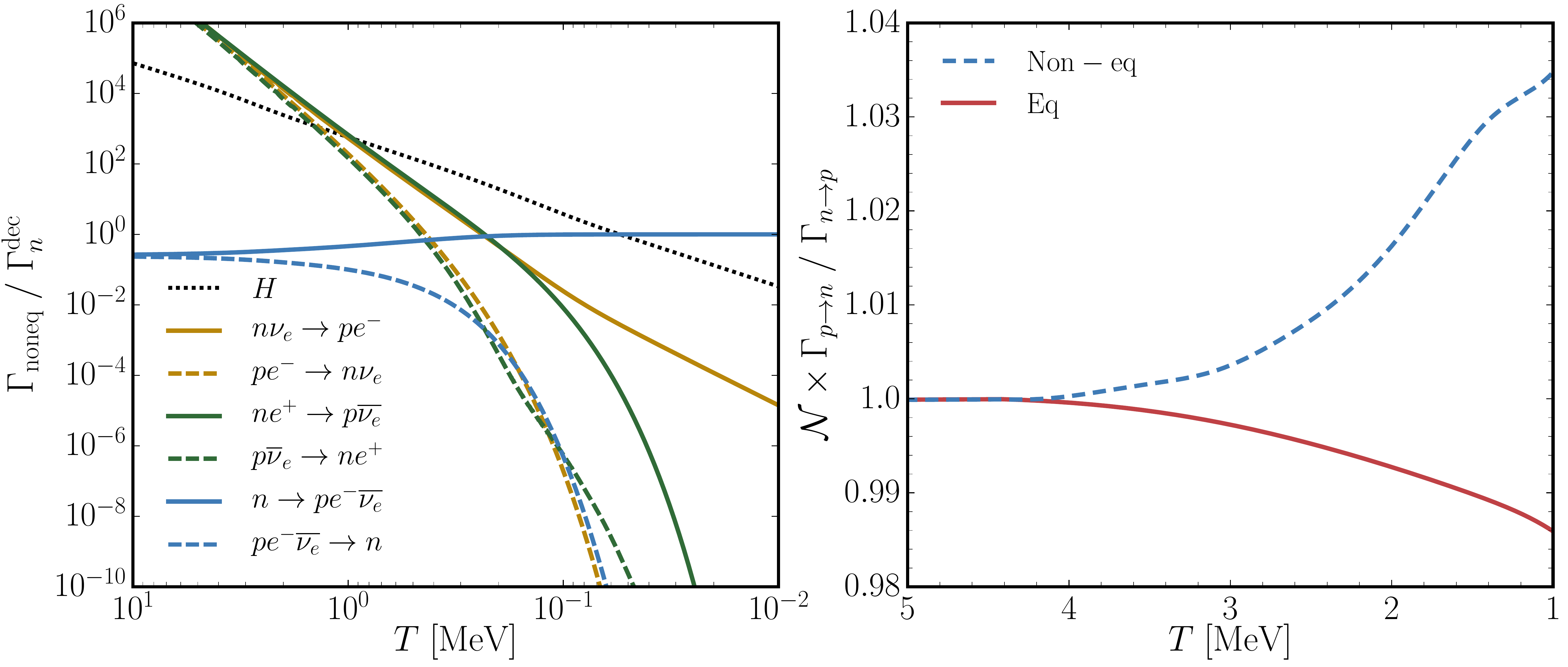}
    \caption{Evolution of the neutron-proton conversion rates and their dependence on neutrino spectral distortions in the presence of an HNL of mass 200 MeV and lifetime 0.08 s that mixes with electron neutrinos only. \emph{Left:} The neutron-proton conversion rates $\Gamma_\mathrm{noneq}$ and the Hubble expansion rate $H$, normalized by the neutron decay rate in vacuum $\Gamma_n^\mathrm{dec}$. \emph{Right:} Ratio of proton-to-neutron and neutron-to-proton conversion rates using non-equilibrium neutrino distributions (`Non-eq') and thermal-like neutrino distributions (`Eq'). Here only scattering reactions are considered. The normalization factor $\mathcal{N}$ is equal to $(\Gamma_{p\rightarrow n}/\Gamma_{n\rightarrow p})^{-1}$ in SBBN.}
    \label{fig:thermalized_neutrinos}
    \end{center}
    \end{figure}

\begin{table}[h!]
    \vspace{-0.4cm}
    \centering
    {\def\arraystretch{1.35}
    \begin{tabular}{c|ccc|ccc}
    \hline\hline
        \multirow{2}{*}{\textbf{Model}} & \multicolumn{3}{c|}{$\boldsymbol{Y_\mathrm{P}}$} & \multicolumn{3}{c}{$\boldsymbol{10^5\times D/H|_\mathrm{P}}$}\\
        & Non-eq. & Eq. & Rell. diff. & Non-eq. & Eq. & Rell. diff.\\
        \hline\hline
        SBBN & 0.24657 & 0.24642 & 0.06\% & 2.6082 & 2.5913 &0.6\% \\
        \hline
        \multirow{2}{*}{\shortstack{$m_N = 200$ MeV\\ $\tau_N = 0.08\,\mathrm{s}$}} & \multirow{2}{*}{0.25498} & \multirow{2}{*}{0.24268} &  \multirow{2}{*}{4.8\%} & \multirow{2}{*}{2.4373} & \multirow{2}{*}{2.4458} & \multirow{2}{*}{0.3\%}\\
        & & & & & &\\
        \hline\hline
    \end{tabular}
    }
    \caption{Primordial helium and deuterium abundances in SBBN and in the presence of HNLs when non-equilibrium neutrino distributions are used (`Non-eq.') or when thermal-like neutrino distributions are used (`Eq.'). Non-equilibrium distributions take into account neutrino spectral distortions, while equilibrium distributions only account for the gravitational effects of HNLs. HNL mixing with electron neutrinos only is considered here. These values are obtained using $\tau_n = 880.2$ s and $\eta = 6.09\times10^{-10}$.}
    \label{tab:eq_vs_noneq}
\end{table}
\newpage
    \item \textbf{Modified expansion history}\\
    While HNLs add to the energy density of the Universe and thus the expansion rate, their decays into the electromagnetic sector of the plasma can strongly dilute the abundance of decoupled species and consequently decrease the expansion rate. Moreover, the injection of non-equilibrium neutrinos further heats up the EM plasma by transferring energy via $\nu-e$ scatterings and by shifting the balance of the reactions $\nu\overline{\nu}\leftrightarrow e^+e^-$ to the right. The impact of these effects on active neutrinos is shown in Figure~\ref{fig:dilution}. In the left panel the change in the effective number of extra relativistic species $N_\mathrm{eff}$ is plotted against the lifetime of a 200 MeV HNL. The grey band in the plot is the measured value of $N_\mathrm{eff}$ by Planck at 2$\sigma$ and reads $N_\mathrm{eff}^\mathrm{CMB} = 2.89 \pm 0.62$~\cite{Aghanim:2018eyx, Sabti:2019mhn}. This plot shows that HNLs can significantly diminish $N_\mathrm{eff}$, while the SBBN prediction is obtained back again when the lifetime is short enough and neutrinos are still in equilibrium. We emphasize that this is only the case for HNL masses above ${\sim}50$ MeV and lifetimes below ${\sim}0.1\,\mathrm{s}$~\cite{HNLs_Neff}. In the right panels of Figure~\ref{fig:dilution} the Hubble rate and the total active neutrino energy density in the presence of 200~MeV HNLs are compared to the same quantities in SBBN. The dilution of neutrinos can be so severe, that the Hubble rate becomes smaller than the one in SBBN at a given temperature. If the injection of energy into neutrinos does not compensate this dilution, the net result will be a decrease of the expansion rate and a later decoupling of neutron-proton reactions.
    Note that while the dilution of neutrinos also leads to lower neutron-proton conversion rates, this effect only partially compensates for the decrease in $Y_\mathrm{P}$ as implied by a lower expansion rate.
    \vspace{2mm}

    In Table~\ref{tab:eq_vs_noneq} we see the consequences of this effect explicitly for $Y_\mathrm{P}$ and $D/H|_\mathrm{P}$ by comparing the middle columns (`Eq') between the different cases. In our benchmark example of an HNL of mass 200 MeV and lifetime 0.08 s, this causes a decrease in $Y_\mathrm{P}$ due to later decoupling of neutron-proton conversion reactions and a decrease in $D/H|_\mathrm{P}$ due to dilution of baryons (which therefore requires a higher initial value). As expected, the gravitational effect is the dominant effect for the primordial deuterium abundance.

\end{enumerate}
In summary, for the primordial helium abundance this dilution effect is sub-dominant and the net effect is an increase of the primordial abundances, driven by spectral distortions in the active neutrino distribution functions. For the primordial deuterium abundances, on the other hand, the gravitational effect is of most importance, with the dilution effect dominating at high masses.
\enlargethispage{1.2cm}

\newpage

\begin{figure}[t]
\vspace{-0.15cm}
\begin{center}
\includegraphics[width=\textwidth]{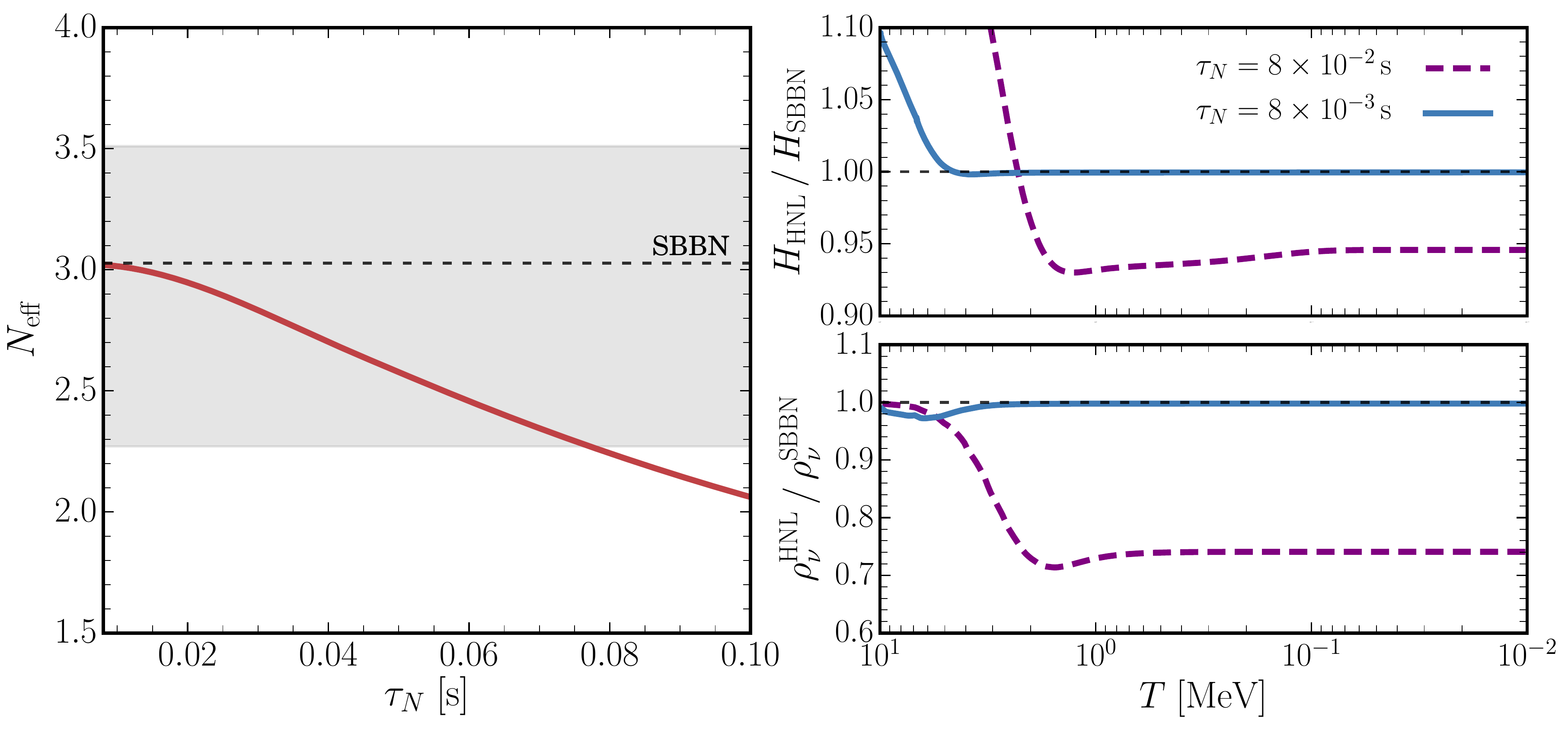}
\end{center}
\vspace{-0.1cm}
\caption{\emph{Left:} Impact of a 200 MeV HNL on the effective number of extra relativistic species $N_\mathrm{eff}$ as a function of its lifetime. The dashed line corresponds the value of $N_\mathrm{eff}$ in SBBN and the grey contour corresponds to the mean $\pm2\sigma$ measurement by Planck~\cite{Aghanim:2018eyx}. \emph{Right:} Influence of a 200 MeV HNL on the Hubble rate (\emph{top} panel) and the active neutrino energy density (\emph{bottom} panel), both normalized by the same quantities in SBBN.}
\label{fig:dilution}
\end{figure}

\vspace{-0.1cm}
\subsection{Bounds on the Lifetime and Mixing Angles of HNLs}
\label{subsec:bounds_mixing_lifetime}
We present here the BBN bounds for HNLs with masses up to 1 GeV. The bounds are shown for three mixing patterns: those that mix purely with electron, muon or tau neutrinos. A direct comparison between the bounds obtained with \texttt{pyBBN} and those in~\cite{Dolgov:2000jw, Ruchayskiy:2012si, MV_paper} is provided in Appendix~\ref{app:HNL_bounds_theta}. In Figure \ref{fig:bounds_tau} the bounds on the HNL lifetime are shown. We note that these bounds are independent of the Dirac or Majorana nature of the HNLs. The trend of the bounds can be categorised into three mass regions:
\begin{enumerate}
    \item[A.] $m_N \lesssim 135 \, \mathrm{MeV}$\\
    In this mass range only three-body decays into leptons occur. The reason for the weakening of the bounds at lower masses is twofold: \emph{1)} such HNLs decay into neutrinos with energies low enough to be sufficiently washed out before they can affect the neutron-proton conversion reactions and \emph{2)} the HNLs are still in thermal equilibrium, which leads to a Boltzmann suppression of their number density at temperatures lower than their mass.
    \item[B.] $135\, \mathrm{MeV} \lesssim m_N \lesssim 200 \, \mathrm{MeV}$\\
    In this region the main decays of HNLs transition from three-body decays into leptons to two-body decays into pions and leptons. Since HNLs in this range have masses that are close to the pion mass, the created neutrinos have relatively lower energies and affect BBN less severely. Moreover, this region also marks the transition between non-relativistic HNL decoupling and relativistic HNL decoupling. The energy density of HNLs reaches a (local) maximum at $m_N \sim 140\,\mathrm{MeV}$ and then decreases for higher masses, until it becomes more or less mass-independent around $m_N \gtrsim 200\,\mathrm{MeV}$. This decrease of the HNL density also diminishes their effect on BBN.
    \item[C.] $m_N \gtrsim 200 \, \mathrm{MeV}$\\
    The bounds for such masses become more or less constant. HNLs in this mass range decouple while they are in the (ultra-)relativistic regime, which results in their abundance at low temperatures to become roughly mass independent. Moreover, high-energy neutrinos from HNL decays quickly lose their energy due to subsequent interactions, leaving eventually a non-equilibrium neutrino population with an energy range that is only weakly sensitive to the total amount of energy initially injected into the neutrino sector. The reason that $\tau$-mixing gives stronger bounds than the other cases is simply due to a larger branching fraction of HNL decays into neutrinos (see Figure~\ref{fig:HNL_BR}). Also note the deviation of the muon mixing bound from tau mixing bound starting at $m_N\approx 250$ MeV, which corresponds to the threshold of the decay channel $N\rightarrow\mu^\pm\pi^\mp$.
\end{enumerate}

\begin{figure}[t!]
\begin{center}
\includegraphics[width=0.9\textwidth]{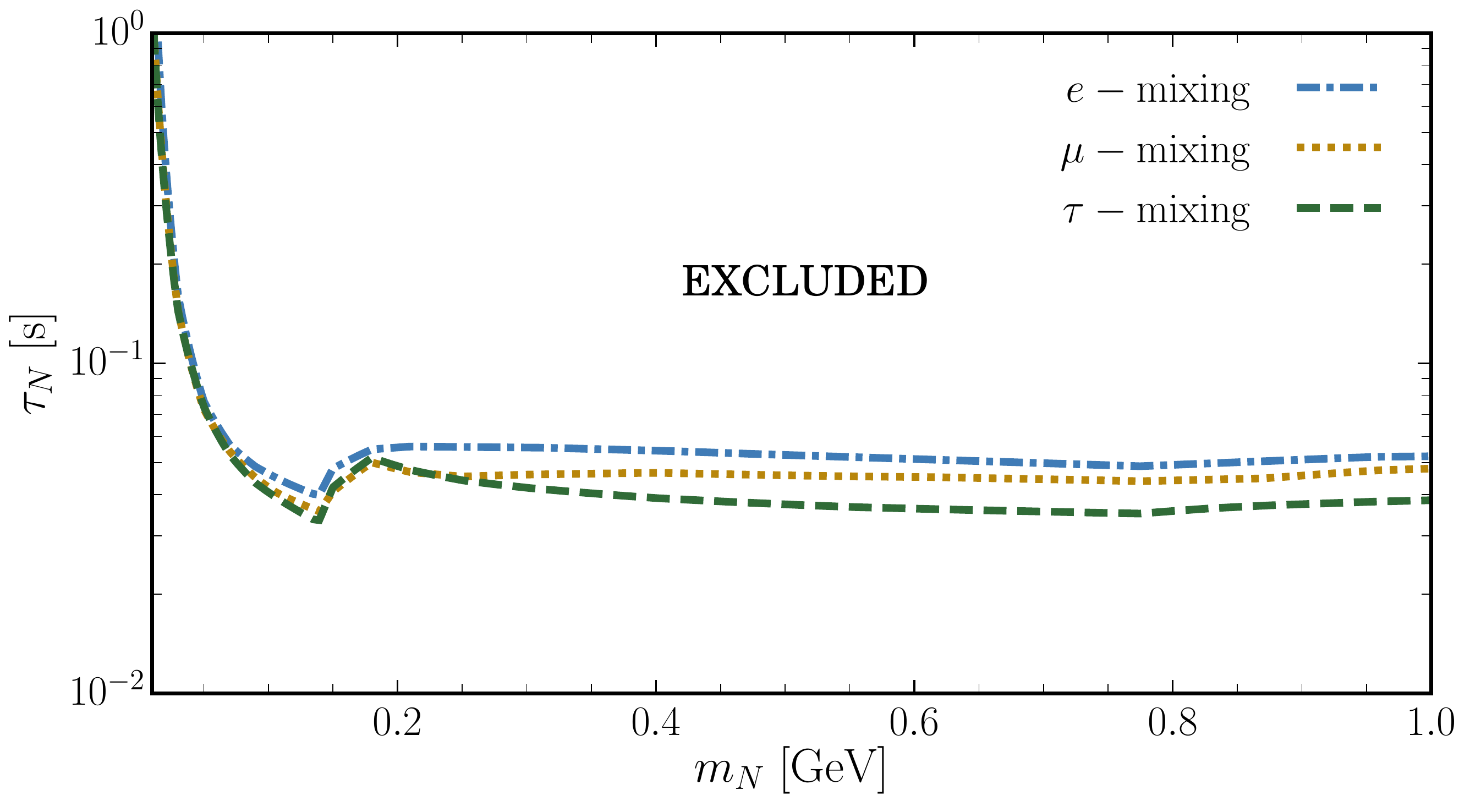}
\end{center}
\caption{Constraints at 2$\sigma$ from BBN on the lifetime of HNLs that mix purely with electron, muon or tau neutrinos. See also Figure~\ref{fig:bounds_angles} for BBN bounds for masses down to few MeV.}
\label{fig:bounds_tau}
\end{figure}

\begin{figure}[H]
\vspace{-0.7cm}
\begin{center}
\includegraphics[width=0.9\textwidth]{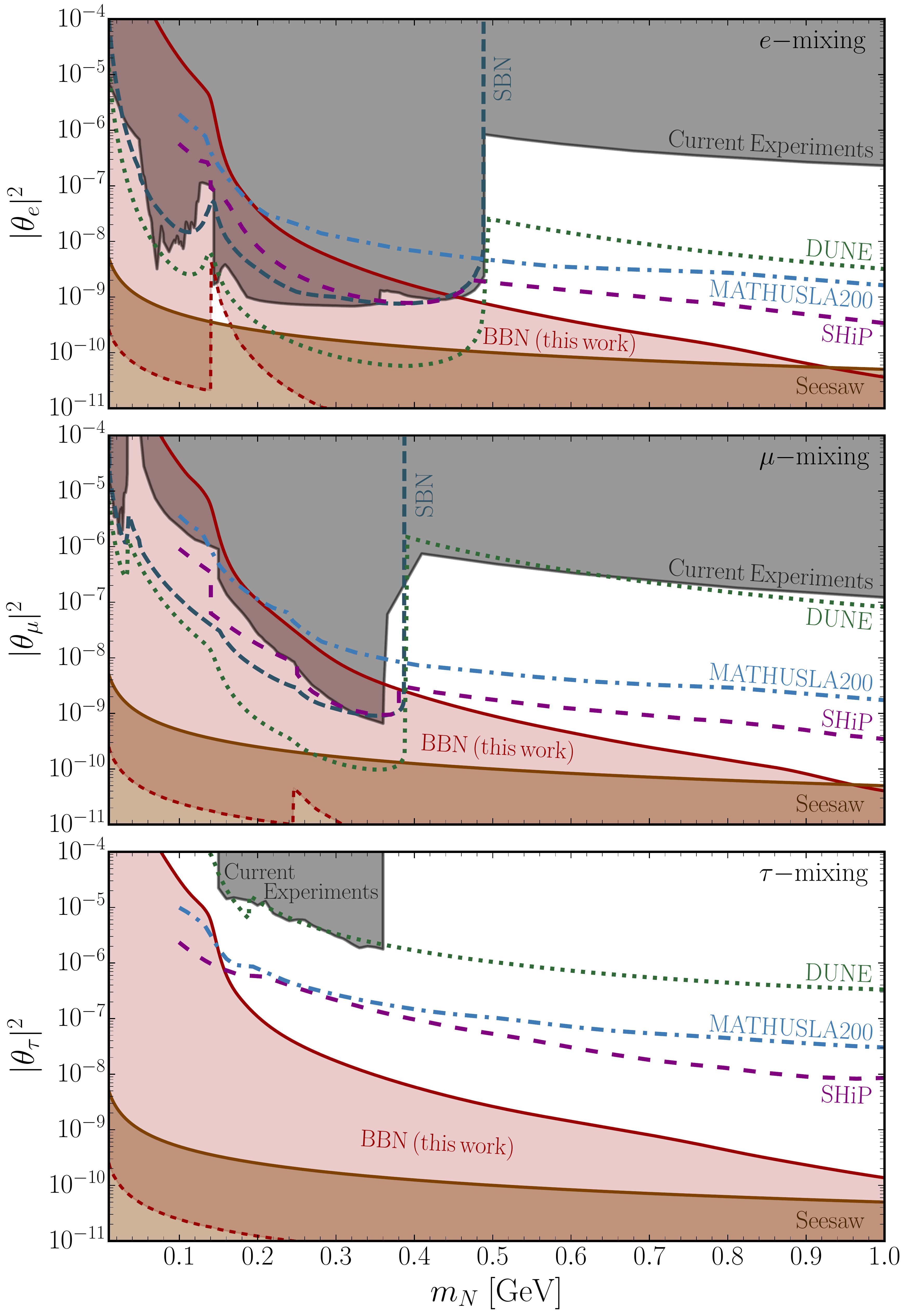}
\end{center}
\vspace{-0.3cm}
\caption{Constraints on the mixing angles of HNLs that mix purely with electron (\emph{top}), muon (\emph{middle}) or tau (\emph{bottom}) neutrinos. The grey regions indicate currently existing limits from experiments (see~\cite{deGouvea:2015euy,Alekhin:2015byh, beacham2019physics, Abe:2019kgx, NA62:2020mcv}), the brown regions (Seesaw) are based on Eq.~\eqref{eq:see-saw} and the red regions (BBN) are the 2$\sigma$ results of this work (see also Figure~\ref{fig:bounds_angles} for BBN bounds for masses lower than displayed here). The lower red, dashed curves roughly denote the boundaries below which our approach is not valid anymore (see text for details and Figure~\ref{fig:bounds_angles} for the full curves). The sensitivity estimates for upcoming and proposed experiments are based on~\cite{Ballett:2016opr} (SBN), \cite{Ballett:2019bgd, Abi:2020evt} (DUNE), \cite{Curtin:2018mvb} (MATHUSLA200) and \cite{SHiP:2018xqw,Gorbunov:2020rjx} (SHiP).}
\label{fig:full_theta}
\end{figure}

For $m_N > m_\pi$, BBN is able to constrain HNLs with lifetimes down to $0.03 - 0.05$ s, depending on their mixing pattern. For example, the lifetime of HNLs that mix with tau neutrinos is constrained almost twice as strong as compared to the electron neutrino mixing case. This analysis improves upon the commonly used bound of $0.1\,\mathrm{s}$ in previous literature by a factor of $2 - 3$. The bounds on the lifetimes can be translated to bounds on the mixing angles by using the definition of the decay width $\tau_N = \Gamma_N^{-1}(m_N, \theta^2)$. We put the obtained bounds on the mixing angles of two degenerate Majorana HNLs in a wider context in Figure~\ref{fig:full_theta} and compare them to currently existing bounds from experiments, as well as forecasts for upcoming and proposed experiments. The forecasted bounds are added for a number of experiments that have the potential to probe the parameter space close to the BBN bound, which in this context are SBN~\cite{Antonello:2015lea}, DUNE~\cite{Acciarri:2015uup}, MATHUSLA200~\cite{Alpigiani:2018fgd} and SHiP~\cite{Alekhin:2015byh}. Note that some of the experimental bounds shown here are obtained for a single Majorana HNL and are therefore an underestimation within the context of the two Majorana HNLs considered here (usually must be scaled down by a factor $\sqrt{2}$).{\parfillskip=0pt\par}

\pagebreak

\newpage
\vspace*{-0.35cm}
\enlargethispage{0.25cm}

The dashed, red curves in Figure~\ref{fig:full_theta} roughly indicate the boundaries below which our approach starts to break down. This is because we have considered HNLs with short lifetimes in this work, while for much longer lifetimes a number of new effects become relevant (see also~\cite{MV_paper}):
\begin{itemize}
    \item HNLs with too small mixing angles are no longer able to enter thermal equilibrium during the Universe's evolution, which eventually results in a different initial condition for their abundance. This limit can be estimated by evaluating for which lifetimes the HNL-SM scattering rate (medium effects in Eq.~\eqref{eq:mediumeffectsmixing} included) is smaller than the Hubble rate, $\Gamma_{N\leftrightarrow \mathrm{SM}}(T_\mathrm{max}) \lesssim 3H(T_\mathrm{max})$, where $T_\mathrm{max}$ is the temperature at which $\Gamma_{N\leftrightarrow \mathrm{SM}}$ peaks. For $m_N = 0.1\,\mathrm{GeV}$ this gives $\tau_N\gtrsim\mathcal{O}(10^4)\,\mathrm{s}$ and goes down to $\tau_N\gtrsim\mathcal{O}(1)\,\mathrm{s}$ for $m_N = 1\,\mathrm{GeV}$.
    \item If charged mesons originating from HNL decays are sufficiently present in the plasma when the nuclear reaction network sets on (around $t\,{\sim}\,200\,\mathrm{s}$), then they will be able to destroy light nuclei and decrease their abundances~\cite{Kawasaki:2004qu, Pospelov:2010cw}. An estimate for when this effect becomes relevant can be obtained by determining the lifetimes for which the number density of mesons exceeds the number density of helium, $n_\mathrm{mesons}(T_\mathrm{BBN}) \gtrsim \frac{1}{4}n_\mathrm{baryons}(T_\mathrm{BBN})$, where $T_\mathrm{BBN}$ is the temperature of start of primordial nucleosynthesis. This gives a limit of $\tau_N\gtrsim\mathcal{O}(30)\,\mathrm{s}$.
\end{itemize}
The lower boundaries are a combination of these two points, where the second one only comes into effect when HNLs can decay into charged mesons (pions or kaons).
\vspace{2mm}

In this figure we see that the combination of current experimental bounds and bounds from BBN allows to exclude short-lived HNLs with masses up to ${\sim}450\,\mathrm{MeV}$ for electron neutrino mixing and up to ${\sim}\mathrm{360}\,\mathrm{MeV}$ for muon neutrino mixing. Given the range of validity of our bounds at these masses (see also Figure~\ref{fig:bounds_angles}), this holds for lifetimes up to at least a few tens of seconds. For tau neutrino mixing there are currently no experimental bounds that overlap with the BBN bounds in the mass range $0.1 - 1$ GeV and, therefore, no such exclusions can be made. Nevertheless, SBN, DUNE and proposed experiments such as SHiP and MATHUSLA200 will be able to cover a significant part of the parameter space for all three mixing cases and, together with the bounds from BBN, greatly reduce the size of the remaining gaps for masses up to 1 GeV.

\subsection{Fitting Functions}
We summarize the output of \texttt{pyBBN} in the form of fitting functions for the primordial helium abundance as a function of the HNL lifetime. These fitting functions may be useful as a quick way to obtain the results of this work (without the meson effect) and can be used in a fast check when, for example, improved data will be available. For the three mixing patterns considered before, they read:
\begin{align}
    Y_\mathrm{P}^\mathrm{Fit}\Big|_\mathrm{e-mixing} &= 0.24657 + 2.2\,\tau_N^2\exp(2.1\tau_N)\\
    Y_\mathrm{P}^\mathrm{Fit}\Big|_\mathrm{\mu-mixing} &= 0.24657 + 2.6\,\tau_N^2\exp(2.5\tau_N)\\
    Y_\mathrm{P}^\mathrm{Fit}\Big|_\mathrm{\tau-mixing} &= 0.24657 + 3.7\,\tau_N^2\exp(2.4\tau_N)\ ,
\end{align}
where $\tau_N$ is the HNL lifetime in seconds. These fitting functions are tested for lifetimes $0.01\,\mathrm{s}\leq\tau_N\leq0.05\,\mathrm{s}$ and masses $150\,\mathrm{MeV}\leq m_N \leq 1\,\mathrm{GeV}$ ($e$-mixing and $\mu$-mixing) and $250\,\mathrm{MeV}\leq m_N \leq 1\,\mathrm{GeV}$ ($\tau$-mixing). They have a maximum deviation from the simulated data of ${\sim}0.5$\% (for masses in the lower range), which decreases with increasing mass or decreasing lifetime (already down to $\ll 0.1$\% for $m_N \approx 1$ GeV).

\newpage
\vspace{-0.4cm}
\section{Discussion}
\label{sec:discussion}
In this section we comment on the effects neglected in our analysis and the sensitivity of the obtained bounds to the joint theoretical and observational errors in the primordial abundances. Next, we elaborate on the implications and relevancy of the bounds for several particle physics and cosmological scenarios. Lastly, we briefly compare with results in previous literature and discuss the impact of more accurate determinations of the primordial abundances.

\subsection{Robustness of the Bounds}
\label{subsec:robustness}
\paragraph{Neglected processes.} Despite of having the rigorousness and thoroughness of the Boltzmann approach at hand to track down non-equilibrium dynamics, there are still some processes left that require an accuracy that is nearly unattainable when using this method. The most relevant example of this is the treatment of neutron-proton conversion reactions due to mesons $h^\pm = \{K^\pm,\pi^\pm\}$ originating from HNL decays $N\rightarrow \ldots\rightarrow h^\pm + X$. These reactions are mediated by the strong force and have higher rates than the ones in Eq.~\eqref{eq:np_reactions}. The study in~\cite{Boyarsky:2020dzc} has found that in order to accurately model this process, a precision of ${\sim}10^{-7}$ in the determination of the HNL number density is necessary. In this same reference it has been shown that this effect \emph{increases} the derived bounds for HNL mass $m_N \geq m_h + m_\ell$ 
in the case of electron- and muon-mixing, while for tau-mixing this only happens when $m_N$ exceeds the $\eta$-meson mass. Given the current precision in measurements of the primordial abundances, this effect improves upon our bounds by a factor of {$\sim$}2, but becomes less dominant with more precise measurements of $Y_\mathrm{P}$ (see Figure~\ref{fig:bounds_tau_future}).
Other effects neglected in this work (besides those occurring at long lifetimes, see Section~\ref{subsec:bounds_mixing_lifetime}) are muon neutrino-driven neutron-proton conversions and some minor radiative corrections to the weak reactions (see, however, Appendix~\ref{app:corrections_to_rates} for a comment on the latter). Nevertheless, these effects either make the bounds slightly stronger (former) or affect the primordial abundances in a way well below the sensitivity of our analysis (latter).
\enlargethispage{1cm}

\paragraph{Errors in $\boldsymbol{Y_\mathrm{P}}$, $\boldsymbol{D/H|_\mathrm{P}}$ and $\boldsymbol{\tau_n}$.} The bounds also depend on the measured and theoretical errors in the primordial abundances. From Eqs.~\eqref{eq:Yp} -- \eqref{eq:sigmaDH} we see that the observational error in $Y_\mathrm{P}$ and the theoretical error in $D/H|_\mathrm{P}$ are the dominating errors (see, however,~\cite{Mossa:2020qgj} for recent developments in relation to the latter). In fact, since the latter is considerably large, it makes $Y_\mathrm{P}$ the driving power behind the constraints. We find that using a smaller theoretical uncertainty for $D/H|_\mathrm{P}$, such as the one in~\cite{Pitrou:2018cgg}, changes our lower mass bounds only, with a maximum improvement of a factor {$\sim$}2 at $m_N = 10\; \mathrm{MeV}$ and already down to $\ll$ 1\% at $m_N = 100\; \mathrm{MeV}$. This improvement happens mainly in the lower mass region, since such light HNLs have lifetimes $\tau_N \gtrsim 0.1$~s and can alter the Hubble expansion and thus the baryon-to-photon ratio. Consequently, this latter quantity affects the primordial deuterium abundance the most. In this work the $2\sigma$ allowance roughly corresponds to an error in $Y_\mathrm{P}$ of about $2.5\%$. Taking into account that recent measurements of $Y_\mathrm{P}$~\cite{Aver:2015iza, Peimbert:2016bdg, Fern_ndez_2018, Valerdi:2019beb} are all consistent with each other within $1\sigma$\hspace{0.3mm}\footnote{The study~\cite{Izotov:2014fga} reports a primordial helium abundance with a relative difference of approximately 4\% from the PDG recommended value~\cite{pdg}. Using such error in our analysis relaxes our bounds roughly by a factor ${\sim}1.5$.}, we find that the 2$\sigma$ bounds presented here correspond to a modest error in the primordial helium abundance. In addition, since $Y_\mathrm{P}$ always increases in the presence of short-lived HNLs, using a neutron lifetime of $\tau_n = 880.2$ s rather than $\tau_n = 887.7$ s~\cite{pdg} in our analysis results in weaker bounds.
\vspace{2mm}

Based on these considerations, we find that the constraints presented in this work are \emph{conservative} and therefore provide a rigorous complement to bounds from laboratory experiments.
\newpage

\vspace*{-0.9cm}
\enlargethispage{0.7cm}
\subsection{Applicability of the Bounds}
\label{subsec:applicability_bounds}
\paragraph{Range of validity.} In this work we studied the impact of short-lived HNLs on BBN and therefore did not include the impact of a couple of effects that come into play at much longer lifetimes, as detailed at the end of Section~\ref{subsec:bounds_mixing_lifetime}. This provides the range of validity of our bounds, which is marked by the red, dashed lines in Figure~\ref{fig:full_theta} or, equivalently, the dotted lines in Figure~\ref{fig:bounds_angles}.

\paragraph{Generic mixing patterns.} We have considered cases where HNLs mix with one neutrino flavour only. While the mixing pattern can affect the initial abundance of a certain neutrino flavour, subsequent SM interactions lead to a quick energy loss and redistribution of the injected energy among the other flavours. Moreover, neutrino oscillations are still relevant at temperatures when HNLs decay and further distribute the injected energy over the different neutrino flavours. Therefore, it is the total amount of neutrinos injected as a whole that matters most and -- to lesser extent -- their energy range. For example, HNLs that mix with tau neutrinos decay more often into neutrinos than the other two mixing cases (see Figure~\ref{fig:HNL_BR}), which leads to this type of mixing to be the strongest constrained. As such, for a given mass, any mixing pattern will give a bound on the lifetime that is in between those in Figure~\ref{fig:bounds_tau}.
\vspace{-0mm}

\paragraph{Relation to realistic models.} Next, in the simulations we modelled HNLs as Dirac particles with 4 degrees of freedom. This is equivalent to dealing with two Majorana particles, each with 2 degrees of freedom, that are degenerate in mass and have the same set of mixing angles. The mixing angles between the two cases are related by $\theta_D = \sqrt{2}\theta_M$. Such degeneracy approximately appears in so-called `symmetry protected' scenarios which impose a `lepton number'-like symmetry \cite{Drewes:2015iva, Drewes:2016jae} and are motivated by low-scale leptogenesis and seesaw scenarios for large mixing angles (significantly above the estimate in Eq. \eqref{eq:see-saw}). While the Dirac sterile neutrinos considered in our simulations are not able to account for the non-zero neutrino masses, since the mass generation mechanism requires small symmetry breaking terms (see also discussion in \cite{Kersten:2007vk}), the limit of approximate `lepton number'-like symmetry is of phenomenological interest because it operates with the largest experimentally allowed mixing angles \cite{Drewes:2015iva, Antusch:2017hhu}. To successfully produce non-zero active neutrino masses \emph{and} generate a baryon asymmetry, the masses $m_{Ni}$ and mixing angles $\theta_{\alpha i}$ of the two heavy Majorana HNLs $i = \{1,2\}$ need to have a small but finite difference, forming a pseudo-Dirac fermion~\cite{Kersten:2007vk}. From an experimental point of view, as the mass splitting between the two Majorana HNLs $|m_{N1} - m_{N2}| \ll m_N$ is too small to be resolved, this has as a consequence that experiments are not sensitive to the individual mixing angles, but rather to their sum |$\theta_{\alpha}|^2=\sum_i |\theta_{\alpha i}|^2$~\cite{Antusch:2017pkq}. A similar reasoning also applies for the BBN phenomenology. Therefore, the results presented here can be used as bounds on such very degenerate Majorana HNLs/pseudo-Dirac HNLs.
\vspace{2mm}

Note that the bounds do not apply to models with single Majorana HNLs, since this implies a different cosmological evolution than the one studied here. Moreover, in many cases adding a viable dark matter candidate does not change the calculations presented here due to very weak interactions with SM species or a low abundance (see e.g. \cite{Boyarsky:2009ix} for the case of the $\nu$MSM).
\vspace{-1mm}

\paragraph{Alternative models of HNLs.} Here we considered minimal type-I seesaw models, that involve HNLs interacting with SM particles through the neutrino portal. We assumed a negligible primordial lepton-asymmetry. Generally, a large lepton asymmetry causes a suppression of the mixing angle in a medium, diminishing active-sterile neutrino conversions. This affects the thermalisation and decoupling of HNLs at high temperatures (see e.g.~\cite{Hannestad:2012ky,Saviano:2013ktj} for the case of light sterile neutrinos). For higher masses a proper calculation of this effect should be considered to see for which part of parameter space it is relevant (see~\cite{Gelmini:2020ekg} for a study at very small mixing angles and where this effect is considered).
\vspace{2mm}

A recent example of an other model that incorporates HNLs is the dipole portal~\cite{Magill:2018jla}, where HNLs couple to the electromagnetic field strength tensor and the SM neutrino fields. They decay mainly via the channel $N\rightarrow\nu\gamma$, which resembles the decay of HNLs into neutral pions $N\rightarrow \nu\pi^0\rightarrow \nu\gamma\gamma$ studied here. Therefore, we expect that this dipole interaction has similar consequences for BBN as studied in this work (neutrino spectral distortions and late reheating). Consequently, our constraints on the HNL lifetime for masses higher than the pion mass can be -- very roughly and with caution -- mapped on this model's free parameters, simply by using the lifetime formula $\tau_N^{-1} \propto |d|^2m_N^3$~\cite{Magill:2018jla}. 
\vspace{2mm}

Any other scenario that introduces additional long-lived (non-DM) companions to HNLs will alter the evolution of the Universe and affect BBN in a different way than studied here. In many cases this means a relaxation of the bounds presented in this work.

\enlargethispage{0.4cm}
\subsection{Comparison with Previous Literature and Bounds from the CMB}
\label{subsec:comparison_previous_CMB}
The impact of HNLs on Big Bang Nucleosynthesis has been previously studied in~\cite{Ruchayskiy:2012si} for masses below the pion mass and in~\cite{Dolgov:2000jw} for masses below 200 MeV. The former employed a similar Boltzmann approach as the one here, while in the latter several assumptions were made to simplify the computation of the collision integral and kinetic equations, which lead to a weaker impact of HNLs on BBN~\cite{Dolgov:2000pj} (see Appendix~\ref{app:HNL_bounds_theta} for details). A more recent study presented in~\cite{MV_paper} estimates the impact of HNLs on BBN for masses $50\,\mathrm{MeV}\leq m_N\leq 1\,\mathrm{GeV}$ in a semi-analytical way, taking into account all relevant effects. This study treats high-energy neutrinos from HNL decays as separate species that are suspended in a thermal bath of equilibrium particles with which they also interact and imposes several simplifications to avoid using the unintegrated Boltzmann equation in Eq.~\eqref{eq:BoltzmannEquation}. In Appendix~\ref{app:HNL_bounds_theta} we make a comparison with~\cite{Dolgov:2000jw, Ruchayskiy:2012si}, where we use the same statistical analysis as~\cite{Ruchayskiy:2012si} to obtain the BBN bound and find excellent agreement. Furthermore, in the same Appendix we compare the results between~\cite{MV_paper} and this work. We have extensively tested both codes under numerous scenarios and find good agreement in the bounds, with a deviation of at most ${\sim}$30\% for masses below 200 MeV. A plausible explanation for this discrepancy is that in~\cite{MV_paper} the contribution of HNLs to the total energy density of the Universe is neglected during their decoupling. Another recent study in~\cite{Gelmini:2020ekg} presents BBN bounds for masses $150\,\mathrm{MeV}\leq m_N\leq 450\,\mathrm{MeV}$ and mixing angles below the seesaw limit, i.e., when HNLs have a very long lifetimes and possibly never enter thermal equilibrium. Their constraints are therefore \emph{complementary} to the results obtained in this work.
\vspace{2mm}

Lastly, we note that CMB measurements by Planck can also constrain HNLs, albeit weaker than BBN for the mass and lifetime ranges of interest here. For masses below {$\sim$}20 MeV, however, the CMB does provide stronger limits and has been previously used to probe light sterile neutrinos (see e.g.~\cite{Ruchayskiy:2012si}). We show our CMB bounds and compare them to the BBN bounds in Appendix~\ref{app:CMB_constraints}.

\begin{figure}[t!]
\vspace{-0.3cm}
\begin{center}
\includegraphics[width=0.88\textwidth]{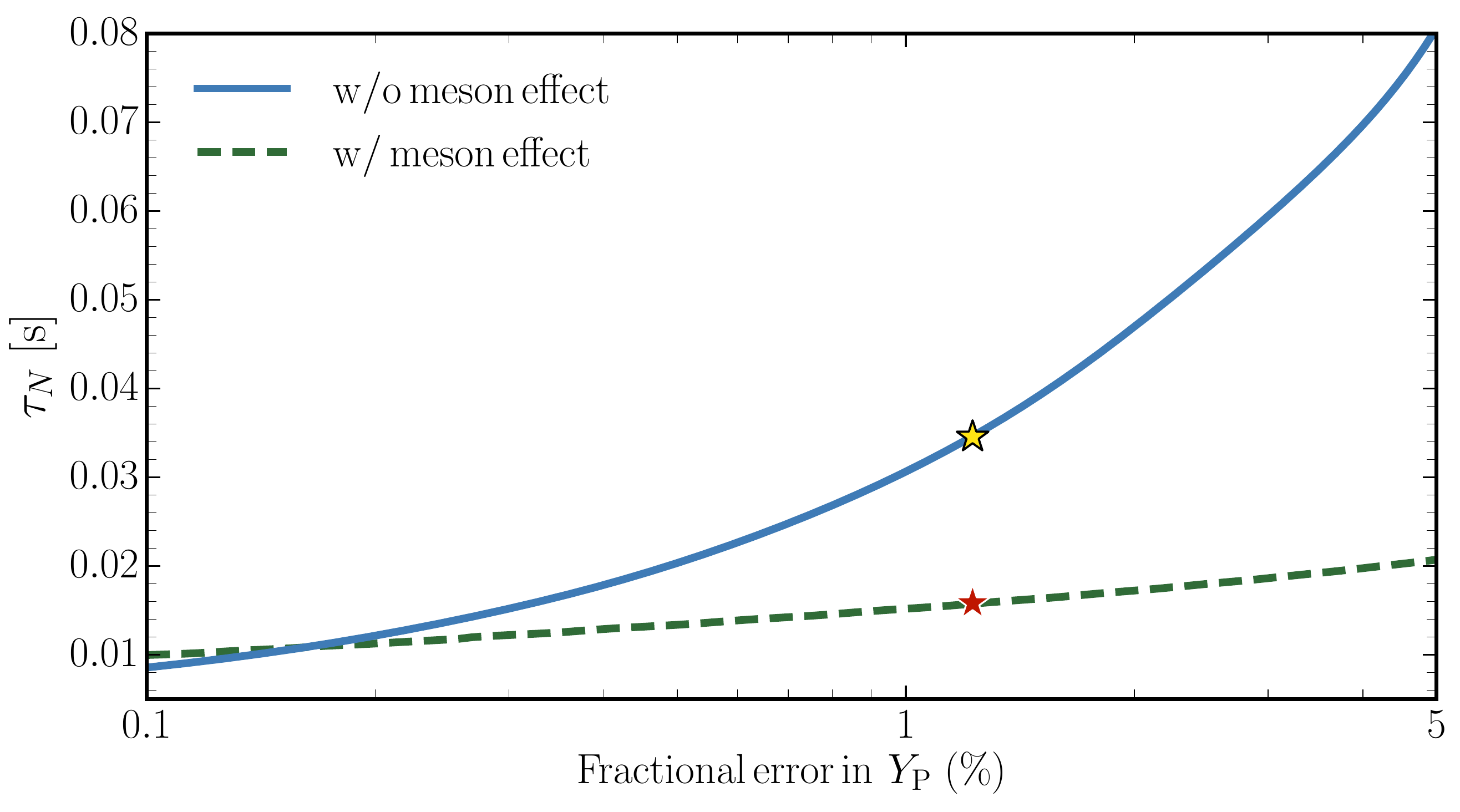}
\end{center}
\vspace*{-0.2cm}
\caption{Projected constraints at 95.4\% CL on the lifetime of a 775 MeV HNL that mixes with tau neutrinos only. The bounds are shown as a function of the fractional error in the primordial helium abundance (joint theoretical and observational). The green dashed line includes the meson effect from~\cite{Boyarsky:2020dzc}. The stars indicate the current $1\sigma$ precision. These bounds are obtained using $\tau_n = 880.2$ s, $\eta = 6.09\times10^{-10}$ and the central value of $Y_\mathrm{P}$ in Eq.~\eqref{eq:Yp}.}
\label{fig:bounds_tau_future}
\end{figure}
\subsection{Future Cosmological Constraints}
The two main errors in the BBN analysis are the theoretical error in $D/H|_\mathrm{P}$ and the observational error in $Y_\mathrm{P}$. Future advancements in the determination of the primordial abundances would involve improvements from both a theoretical and an observational perspective. As mentioned in Subsection \ref{subsec:robustness}, a more accurate prediction of $D/H|_\mathrm{P}$ will improve our bounds at lower masses. For $Y_\mathrm{P}$ the situation is less clear. We attempt to obtain a quantitative picture of the sensitivity of the bounds in Figure~\ref{fig:bounds_tau_future} by showing which lifetimes of a 775~MeV HNL could be constrained when the joint theoretical and observational error in $Y_\mathrm{P}$ is altered. From Figure~\ref{fig:full_theta} it is clear that in order to close the remaining gap between BBN and SHiP, lifetimes $\tau_N \lesssim 0.01$ s must be probed. In Figure~\ref{fig:bounds_tau_future} we see that this would correspond to a per mille accuracy in the determination of $Y_\mathrm{P}$. We have explicitly checked that this is also the case when the meson effect as described in~\cite{Boyarsky:2020dzc} is included. Such precision with accurate determinations of systematic errors seems unlikely to be attainable in the foreseeable future.
\vspace{2mm}

Finally, the next generation of CMB experiments, such as the Simons Observatory~\cite{Ade:2018sbj} and CMB-S4~\cite{Abazajian:2019eic}, aim to measure $N_\mathrm{eff}$ down to percent-level precision. This would bring the CMB bounds more in line to those from BBN for high HNL masses. For masses of a few tens of MeV and lower it would improve upon the current CMB bounds from Planck (see Appendix~\ref{app:CMB_constraints}) and become the strongest probe.

\section{Conclusion}
\label{sec:conclusion}
Big Bang Nucleosynthesis provides rigorous constraints on the lifetime and mixing angles of Heavy Neutral Leptons that are complementary to those from terrestrial experiments and therefore of great use when defining goals for future experiments. We have developed a Boltzmann code, \texttt{pyBBN}, that simulates the BBN epoch in the presence of short-lived, thermally decoupled HNLs with masses up to 1 GeV. We take into account all relevant HNL decay channels, as well as subsequent interactions of decay products (thermalization and decay showers), dilution due to QCD phase transition, SM neutrino oscillations and matter effects, improved calculations of weak reaction rates and proper evolution of the baryon abundance due to non-standard thermal history. Moreover, we marginalize over the baryon-to-photon ratio to obtain pure BBN bounds and make use of the latest measurements of the primordial helium and deuterium abundances~\cite{pdg}. We summarize the impact of HNLs on BBN in Section~\ref{sec:new-physics-and-BBN}, lay out our methodology in Section~\ref{sec:methodology}, present the main results in Section~\ref{sec:results} and scrutinize the results in Section~\ref{sec:discussion}. Figures~\ref{fig:bounds_tau}, \ref{fig:full_theta} and \ref{fig:bounds_angles} show the constraints obtained in this work. We conclude that:
\begin{itemize}
    \item HNLs affect BBN mainly by their decay into active neutrinos and mesons, which increases the neutron-to-proton ratio and consequently the primordial helium abundance. While the primordial deuterium abundance is also affected due to late reheating, the theoretical error in its determination is currently too large to provide any constraining power.
    \item BBN is able to constrain HNLs with lifetimes down to $0.03 - 0.05$ s, depending on the mixing pattern. The strongest bounds are seen in the tau-mixing case, followed by muon-mixing and finally electron-mixing. This is a factor $2 - 3$ improvement compared to the commonly used bound of 0.1 s. Where applicable, the inclusion of the meson effect as described in~\cite{Boyarsky:2020dzc} further improves these bounds by a factor ${\sim}2$ (see Figure~\ref{fig:bounds_angles}).
    \item The combination of current bounds from collider, collider-based and neutrino experiments together with BBN bounds excludes HNLs that mix with electron neutrinos up to a mass of about 450 MeV and up to 360 MeV for muon neutrino mixing (see Figure~\ref{fig:full_theta}), in both cases for lifetimes up to at least a few tens of seconds. For mixing with tau neutrinos no such mass and lifetime ranges can be excluded, because currently there are only few laboratory experiments that probe this part of parameter space and there is no overlap with the BBN bounds.
    \item Upcoming and proposed experiments, such as SBN~\cite{Antonello:2015lea}, DUNE~\cite{Acciarri:2015uup}, MATHUSLA~\cite{Alpigiani:2018fgd} and SHiP~\cite{Alekhin:2015byh}, will further probe the parameter space of HNLs and, together with BBN, close down a significant part for HNL masses up to 1 GeV.
    \item Improved measurements of the primordial helium abundance will strengthen the current bounds, while refined predictions of the primordial deuterium abundance will mainly allow to further exclude HNLs in the lower mass regions.
\end{itemize}

\section*{Acknowledgements}
We are grateful to Maksym Ovchynnikov and Vsevolod Syvolap for the collaboration on this topic. We also thank Miguel Escudero, Kyrylo Bondarenko, Oleg Ruchayskiy, Alexey Boyarsky, Shintaro Eijima, James Alvey, Marco Drewes and Artem Ivashko for helpful discussions and comments on the draft of this work.
NS is a recipient of a King's College London NMS Faculty Studentship. AF acknowledges support of the DFG (German Research Foundation) through the research training group \emph{Particle physics beyond the Standard
Model} (GRK 1940). AM is supported by the
Netherlands Organization for Scientific Research (NWO)
under the program ``Observing the Big Bang'' of the Foundation for Fundamental Research on Matter (FOM).


\newpage
\bibliographystyle{JHEP}
\bibliography{main}
\newpage


\appendix

\include{appendices}

\end{document}

%% file: appendices.tex
%
%

\section{Summary of BBN Bounds}
\label{app:HNL_bounds_theta}
\paragraph{Previous Bounds.} In this section we compare our results with the previous bounds in the literature on the lifetime of HNLs up to 100 MeV. For these masses, there are only four decay channels (see Eq.~\eqref{eq:HNLnuemixingdecays4p}). We compare with~\cite{Dolgov:2000jw, Ruchayskiy:2012si}. The study in~\cite{Dolgov:2000jw} made a number of assumptions to simplify their calculations: HNLs were assumed to be non-relativistic at all times, Boltzmann distribution functions were used inside collision integrals and thermal-like distributions in all other places, and neutrino spectral distortions and temperature distortions were assumed to be relatively small. Since our approach is similar to the one in~\cite{Ruchayskiy:2012si}, we adapt the same methodology for the evolution of the system in this comparison.

\begin{figure}[h!]
\begin{center}
\includegraphics[width=0.9\textwidth]{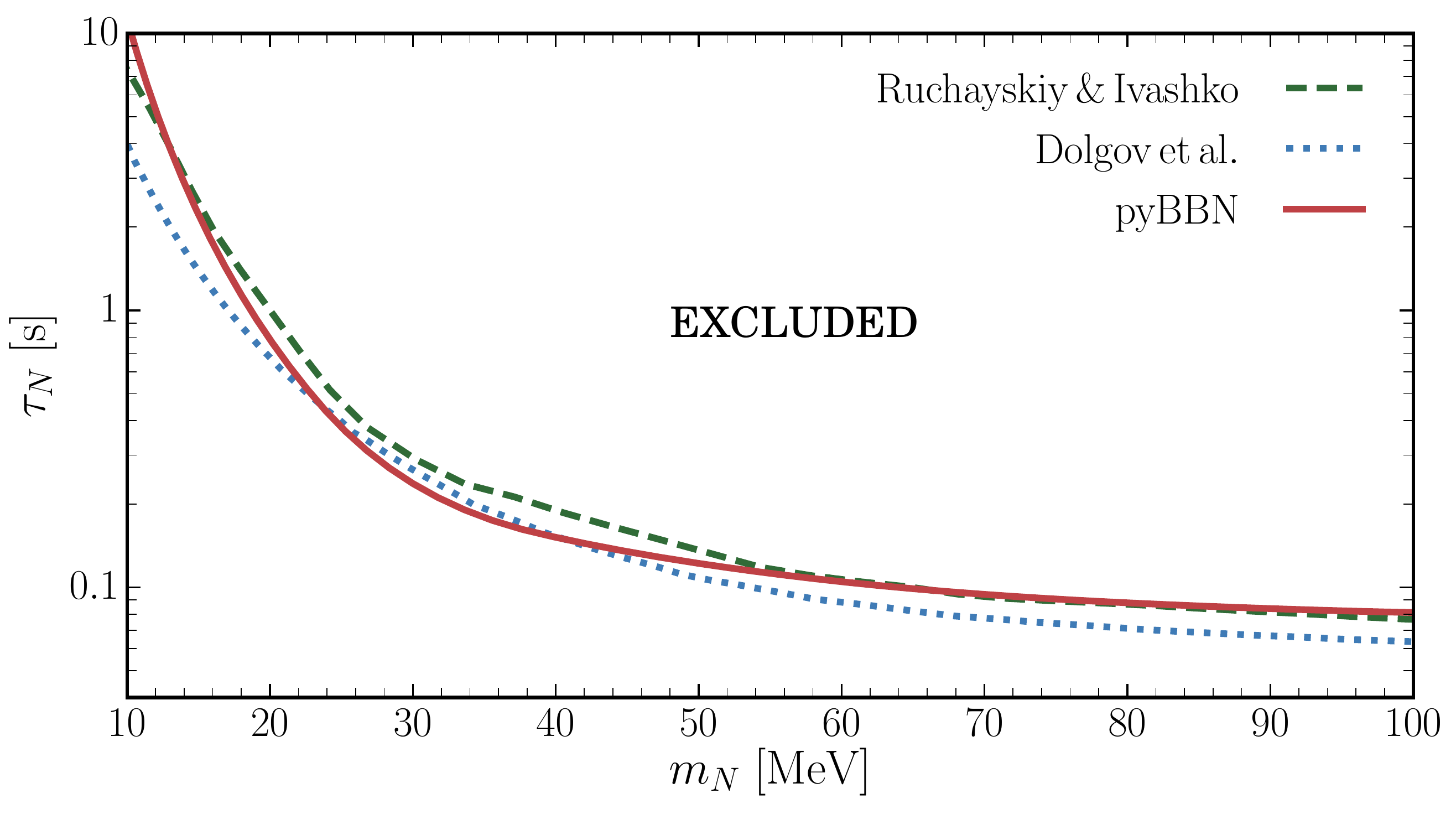}
\end{center}
\vspace{-0.1cm}
\caption{Constraints from BBN on the lifetime of HNLs up to mass of 100 MeV. Mixing with electron neutrinos only is assumed here. Dashed curve is from \cite{Ruchayskiy:2012si}, dotted from \cite{Dolgov:2000jw}.}
\label{fig:low_mass_bounds}
\end{figure}

In what follows, we use for the neutron lifetime $\tau_n = 885.7\,\mathrm{s}$ and for the baryon-to-photon ratio $\eta = 6\times10^{-10}$. Moreover, we apply the same exclusion condition as in~\cite{Ruchayskiy:2012si} that is based on the measured primordial helium abundance in~\cite{Aver:2011bw}. Results are shown in Figure~\ref{fig:low_mass_bounds}, where mixing with only electron neutrinos is considered. We find good agreement between the bounds obtained from \texttt{pyBBN} and those from~\cite{Dolgov:2000jw, Ruchayskiy:2012si}.

\begin{figure}[h!]
\vspace*{-0.2cm}
\begin{center}
\includegraphics[width=\textwidth]{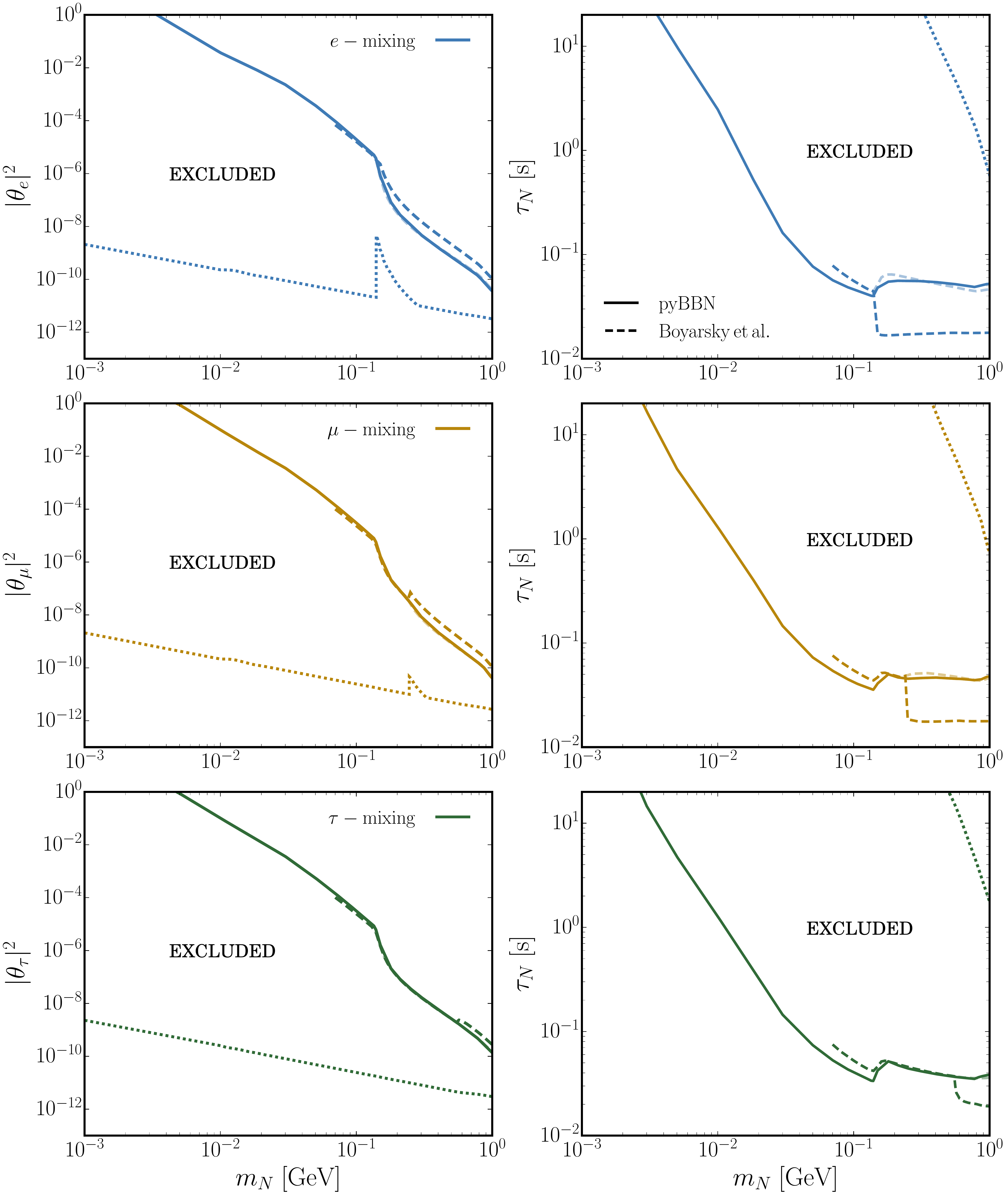}
\end{center}
\vspace{-0.1cm}
\caption{Bounds from BBN on the mixing angles (\emph{left panels}) and lifetime (\emph{right panels}) of HNLs that mix purely with electron (\emph{top}), muon (\emph{middle}) and tau neutrinos (\emph{bottom}). The bounds on the mixing angles are for two degenerate Majorana HNLs. Solid lines are the results of this work (at $2\sigma$) and dashed lines are from~\cite{Boyarsky:2020dzc} (using a correction to the helium abundance of 2.5\% for the sake of comparison, as their main results are obtained with a different error on the helium abundance). The more transparent dashed lines are the bounds from~\cite{MV_paper} without the meson effect included. The dotted lines roughly denote the boundaries beyond which our assumptions are no longer valid (see Section~\ref{subsec:bounds_mixing_lifetime} and~\cite{MV_paper} for more details).}
\label{fig:bounds_angles}
\end{figure}

\paragraph{Current Bounds.}
The bounds obtained in this work are presented in Figure~\ref{fig:bounds_angles} for the mass range 3 MeV $-$ 1 GeV. Note that the bounds on the lifetime in the range $10 - 100$ MeV are the updated ones as compared to what is displayed in Figure~\ref{fig:low_mass_bounds}. The constraints on the mixing angles are shown for two degenerate Majorana HNLs. In the case of Dirac HNLs, the bounds on $|\theta_\alpha|^2$ in this figure should be multiplied by a factor of 2. In this same figure we also show the constraints obtained in~\cite{MV_paper, Boyarsky:2020dzc}. Overall, we find an excellent agreement at high masses and a good agreement for masses below 200 MeV, with a maximum difference of ${\sim}30\%$ in all three mixing cases. See also Section~\ref{subsec:comparison_previous_CMB} for a comment on the approach used in~\cite{MV_paper} and the source of this deviation. Finally, the study~\cite{Gelmini:2020ekg} reports BBN bounds for HNLs with masses 150 $-$ 450 MeV below the seesaw limit, i.e., for very small mixing angles and when they possibly never enter equilibrium during the Universe's evolution.

\newpage

\enlargethispage{0.6cm}

\section{Contraints from the CMB}
\label{app:CMB_constraints}
\begin{figure}[h!]
\begin{center}
\includegraphics[width=\textwidth]{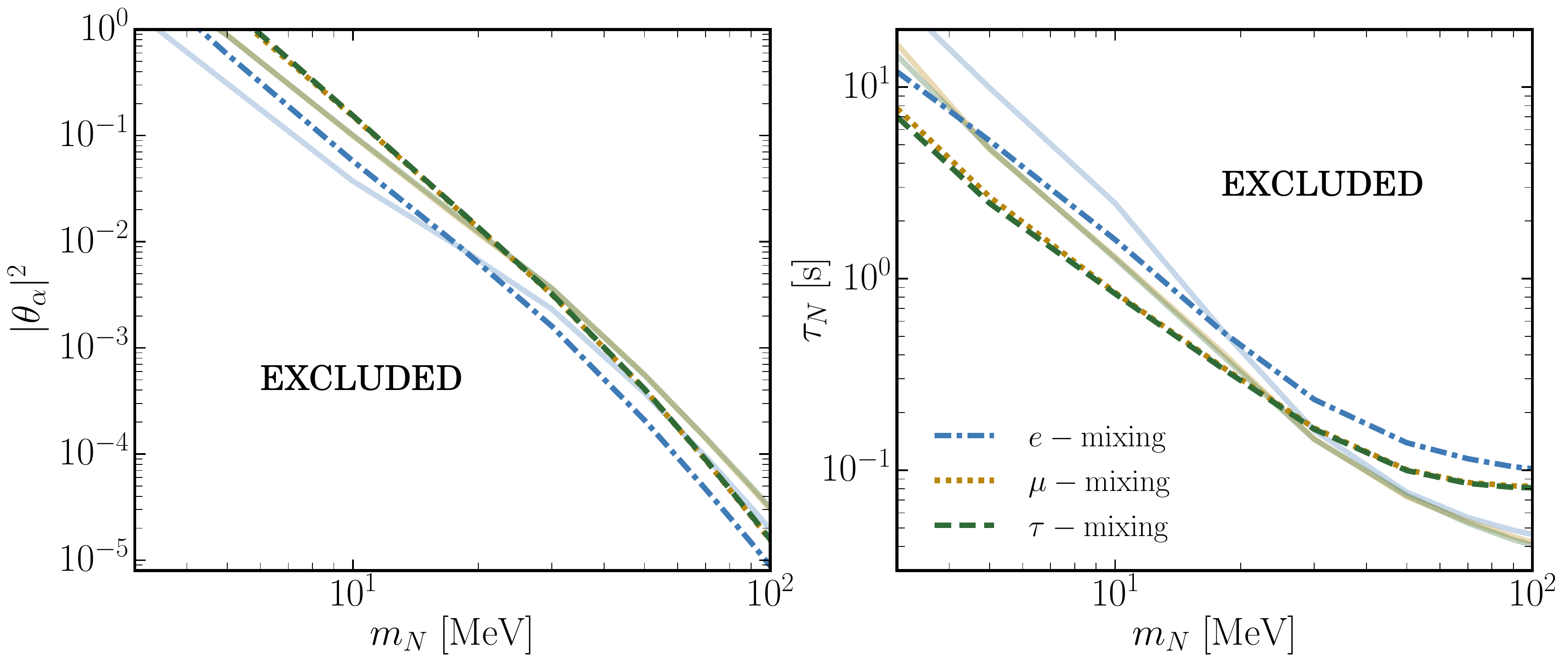}
\end{center}
\caption{Constraints from the CMB on the mixing angles (\emph{left}) and lifetime (\emph{right}) of HNLs up to a mass of 100 MeV. Mixing with electron, muon or tau neutrinos only is assumed. The bounds on the mixing angles are for two degenerate Majorana HNLs, while the bounds on the lifetime are independent of the Dirac/Majorana nature. The BBN bounds (solid lines) are included for comparison.}
\label{fig:CMB_bound}
\end{figure}

Since HNLs can affect $Y_\mathrm{P}$ and $N_\mathrm{eff}$ by their decay into neutrinos and charged leptons (see Section~\ref{subsec:spectral_dilution}), they can also be directly probed by the CMB. To this end we follow the same approach as in~\cite{Sabti:2019mhn} and introduce a Gaussian likelihood of the form:
\begin{align}\label{eq:chi2_CMB}
\chi_\mathrm{CMB}^2  &= \left[\Theta - \Theta_\mathrm{Obs}\right]^T \,  \Sigma_\mathrm{CMB}^{-1} \,\left[\Theta - \Theta_\mathrm{Obs}\right]\\
\Theta &= (\Omega_\mathrm{b} h^2,\,N_\mathrm{eff},\,Y_\mathrm{P})\\
\Sigma_\mathrm{CMB} &= 
 \left(
\begin{array}{ccc}
\sigma_1^2 & \sigma_1 \sigma_2 \rho_{12} & \sigma_1 \sigma_3 \rho_{13}  \\
\sigma_1 \sigma_2 \rho_{12}     & \sigma_2^2 & \sigma_2 \sigma_3 \rho_{23} \\
\sigma_1 \sigma_3 \rho_{13}     & \sigma_2 \sigma_3 \rho_{23}        & \sigma_3^2 \\
\end{array}
\right)\, ,
\end{align}
where the parameters in the covariance matrix $\Sigma_\mathrm{CMB}$ are obtained from~\cite{Aghanim:2018eyx,Aghanim:2019ame} and read:
\begin{align}
\label{eq:CovarianceForm}
(\Omega_\mathrm{b} h^2,\,N_\mathrm{eff},\,Y_\mathrm{P})|_\mathrm{Obs} &= (0.02225,\, 2.89,\, 0.246)\nonumber  \\
(\sigma_{1},\,\sigma_{2},\,\sigma_{3}) &= (0.00022,\, 0.31,\, 0.018)  \\
(\rho_{12},\,\rho_{13},\,\rho_{23}) &= ( 0.40,\, 0.18,\, -0.69) \, . \nonumber
\end{align}
The CMB bound is obtained in a similar way as described in Section~\ref{subsec:bbn_data_analysis} and shown in Figure~\ref{fig:CMB_bound} for HNLs up to 100 MeV. With the current precision in the determination of the primordial abundances and $N_\mathrm{eff}$, the CMB provides stronger bounds than BBN in the lower mass regions. This is mainly because in this mass range $N_\mathrm{eff}$ increases strongly~\cite{HNLs_Neff}. For higher masses, however, $N_\mathrm{eff}$ decreases only marginally and the BBN bounds remain stronger, as the error in measurements of $Y_\mathrm{P}$ by Planck (Eq.~\ref{eq:CovarianceForm}) is currently larger than the one in Eq.~\ref{eq:Yp}. Note that a similar lower/upper boundary exists for the mixing angles/lifetime as in Figure~\ref{fig:bounds_angles}.

\pagebreak
\enlargethispage{0.5cm}
\vspace*{-0.9cm}
\section{Temperature Evolution}
\label{app:TemperatureEvolution}

In this appendix an explicit formula for the temperature evolution equation of the background cosmology is derived. Consider a plasma consisting of four particle species representing the different types of contents in the Universe. As an example of these four species, we consider here photons $\gamma$ (radiation in equilibrium), electrons $e$ (massive particles in equilibrium), active neutrinos $\nu$ (massless non-equilibrium) and HNLs $N$ (massive non-equilibrium). The addition of other species, e.g. muons, will then follow a similar procedure. In what follows, we will use comoving coordinates: $y = pa, \ \widetilde{T} = aT, \ \widetilde{E} = aE$, where $a$ is the scale factor and $p,\ T,\ E$ are the momentum, temperature and energy of a particle respectively. The total energy density $\rho_\mathrm{tot}$ and total pressure $p_\mathrm{tot}$ are given by
\vspace{0.2cm}
\begin{align}
\begin{tabular}{ll}
$\rho_\mathrm{tot} = \rho_\gamma + \rho_e + \rho_\nu + \rho_N$\hspace{2.5 cm} & $p_\mathrm{tot} = p_\gamma + p_e + p_\nu + p_N$\\ \\
$\rho_\gamma = g_\gamma\frac{\pi^2}{30}\frac{1}{a^4}\widetilde{T}^4$ & $p_\gamma = \frac{1}{3}\rho_\gamma$\\ \\
$\rho_e = \frac{g_e}{2\pi^2}\frac{1}{a^4}\int\mathrm{d}y y^2\frac{\widetilde{E}_e}{e^{\widetilde{E}_e / \widetilde{T}}+1}$ & $p_e = \frac{g_e}{6\pi^2}\frac{1}{a^4}\int\mathrm{d}y\frac{y^4}{\widetilde{E}_e}\frac{1}{e^{\widetilde{E}_e/\widetilde{T}}+1}$\\ \\
$\rho_\nu = \frac{g_\nu}{2\pi^2}\frac{1}{a^4}\int\mathrm{d}y y^3 f_\nu$ & $p_\nu = \frac{1}{3}\rho_\nu$\\ \\
$\rho_N = \frac{g_N}{2\pi^2}\frac{1}{a^4}\int\mathrm{d}y y^2\sqrt{y^2+a^2m_N^2}f_N$ & $p_N = \frac{g_N}{6\pi^2}\frac{1}{a^4}\int\mathrm{d}y\frac{y^4}{\widetilde{E}_N}f_N$\ ,
\end{tabular}
\end{align}
\vspace{0.3cm}

where $f_i$ is the distribution function and $g_i$ the number of internal degrees of freedom of a particle. The energy conservation law reads
\begin{align}
\frac{\mathrm{d} \rho_\mathrm{tot}}{\mathrm{d} \ln{a}}\frac{\mathrm{d}\ln{a}}{\mathrm{d}t} + 3H\left(\rho_\mathrm{tot} + p_\mathrm{tot}\right) =0 \longrightarrow \frac{\mathrm{d} \rho_\mathrm{tot}}{\mathrm{d} \ln{a}} + 3\left(\rho_\mathrm{tot} + p_\mathrm{tot}\right) = 0\ ,
\end{align}
where $H\equiv\dot{a}/a$ is the Hubble parameter. The individual contributions are
\begin{align}
\frac{\mathrm{d}\rho_\gamma}{\mathrm{d}\ln{a}} + 3\left(\rho_\gamma + p_\gamma\right) &= 4\frac{\rho_\gamma}{\widetilde{T}}\frac{\mathrm{d}\widetilde{T}}{\mathrm{d}\ln{a}}\\
\frac{\mathrm{d}\rho_e}{\mathrm{d}\ln{a}} + 3\left(\rho_e + p_e\right) &= \frac{1}{a^4}\left[-\frac{a^2m_e^2}{\widetilde{T}}R_1 + \frac{1}{\widetilde{T}^2}\left\{R_2 + a^2m_e^2R_1\right\}\frac{\mathrm{d}\widetilde{T}}{\mathrm{d}\ln{a}}\right]\\
\frac{\mathrm{d}\rho_\nu}{\mathrm{d}\ln{a}} + 3\left(\rho_\nu + p_\nu\right) &= \frac{g_\nu}{2\pi^2}\frac{1}{a^4}\int\mathrm{d}y y^3 \frac{1}{H}I_\nu\\
\frac{\mathrm{d}\rho_N}{\mathrm{d}\ln{a}} + 3\left(\rho_N + p_N\right) &= \frac{g_N}{2\pi^2}\frac{1}{a^4}\int\mathrm{d}y y^2 \widetilde{E}_N \frac{1}{H}I_N\ , 
\end{align}
with
\begin{align}
R_n &= \frac{g_e}{2\pi^2}\int\mathrm{d}y y^{2n} \frac{e^{\frac{\widetilde{E}_e}{\widetilde{T}}}}{\left(e^{\frac{\widetilde{E}_e}{\widetilde{T}}} + 1\right)^2}\ .
\end{align}
Adding all these equations together and solving for the derivative of temperature results in

\begin{align}
\label{eqapp:temperature_evolution}
\frac{\mathrm{d}\widetilde{T}}{\mathrm{d}\ln{a}} = \frac{\frac{a^2m_e^2}{\widetilde{T}}R_1 - \frac{g_\nu}{2\pi^2}\int\mathrm{d}y y^3 \frac{1}{H}I_\nu - \frac{g_N}{2\pi^2}\int\mathrm{d}y y^2 \widetilde{E}_N \frac{1}{H}I_N}{\frac{2\pi^2g_\gamma}{15}\widetilde{T}^3 + \frac{1}{\widetilde{T}^2}R_2 + \frac{a^2m_e^2}{\widetilde{T}^2}R_1}
\end{align}

\newpage

Note that the terms containing a collision integral in the numerator of Eq.~\eqref{eqapp:temperature_evolution} describe the energy injection rate of reactions. For particles involved in the same reaction, these terms cancel out due to conservation of energy during each reaction. Indeed, the decay of a decoupled particle into another decoupled particle should not heat up the system. When particles in  equilibrium are involved, the contributions of particles in initial and final state might not cancel completely (since we do not include the collision integral for equilibrium species) and hence those reactions will change the temperature of the plasma.

\section{Matrix Elements}
\label{app:MatrixElements}

In this appendix all the relevant tree-level matrix elements used in the BBN simulations are summarized. Section \ref{appsec:SMMatrixElements} contains the reactions involving SM particles only, Section \ref{appsec:HNLAboveQCD} the reactions involving HNLs above QCD-scale and Section \ref{appsubsec:HNLBelowQCD} the reactions involving HNLs below QCD-scale. The matrix elements listed here are \emph{not} averaged over any helicities, the symmetry factor $S$ in the expressions below takes care of identical particles in the final state. In this work, only HNL decay channels with a branching ratio larger than $\sim$1\% for some mass below $\sim$1 GeV are considered. If the HNL is a Dirac particle, then there is an additional set of matrix elements involving the charge-conjugated channels. However, these are not relevant for the BBN phenomenology considered here, as both the Dirac HNL and its anti-particle are treated in the same way.

HNLs with high masses decay into mesons that in their turn can decay into other particles. The explicit determination of SM matrix elements involving multiple mesons can be very challenging. Therefore, an approximation has been made by assuming that such particles decay isotropically, i.e., with a uniform probability in the whole phase space.
The matrix element, which is assumed to be constant, can then be obtained from the measured decay width $\Gamma$ (see e.g.~\cite{pdg}):
\begin{align}
\Gamma = \frac{1}{2gM}\int\left(\prod_{i}\frac{\mathrm{d}^3y_i}{(2\pi)^32E_i}\right)|\mathcal{M}|^2(2\pi)^4\delta^4(P-\sum_i P_i)\, .
\end{align}
For two-body decays this method gives the exact matrix element. Throughout this appendix, the following notations are used: $\alpha,\beta \in \{e,\mu,\tau\}$ with $\alpha \neq \beta$, $g_R = \sin^2\theta_\mathrm{W}, \ g_L = 1/2 + \sin^2\theta_\mathrm{W}\ \mathrm{and}\ \widetilde{g_L} = -1/2 + \sin^2\theta_\mathrm{W}$, with $\theta_\mathrm{W}$ the Weinberg angle such that $\sin^2\theta_\mathrm{W} = 0.2312$. 
The values of the meson decay constants used in Section \ref{appsubsec:HNLBelowQCD} are from \cite{Bondarenko:2018ptm} and summarized in Table \ref{tab:meson_decay_constants}.

\begin{table}[h!]
\begin{center}
{\renewcommand{\arraystretch}{1.35}
\begin{tabular}{p{1.25cm}|p{1.25cm}|p{1.25cm}|p{1.25cm}|p{1.25cm}}
\hline\hline
\multicolumn{5}{c}{\textbf{Meson Decay Constants [MeV]}}\\
\hline\hline
\hfil $f_{\pi}$ &\hfil $f_{\eta}$ &\hfil $f_{\rho}$ &\hfil $f_{\omega}$ &\hfil $f_{\eta'}$\\
\hline 
\hfil 130.2 &\hfil 81.7 &\hfil 208.9 &\hfil 195.5 &\hfil -94.7\\
\hline\hline
\end{tabular}
}
\end{center}
\caption{Meson decay constants.}
\label{tab:meson_decay_constants}
\end{table}

The matrix elements are rigorously checked and compared with those in \cite{Dolgov:1997mb, Dolgov:2000pj, Ruchayskiy:2012si, Grohs:2015tfy}. We find that our expressions are consistent with \cite{Grohs:2015tfy}, but differ from those in \cite{Dolgov:1997mb, Dolgov:2000pj, Ruchayskiy:2012si} for the reactions $\nu_{\mu/\tau} + \overbar{\nu_{\mu/\tau}} \rightarrow e^+ + e^-$ and $N + \overbar{\nu_{\mu/\tau}} \rightarrow e^+ + e^-$.

\newpage
\subsection{Matrix Elements in the SM}
\label{appsec:SMMatrixElements}
\begin{table}[H]
\begin{center}
{\renewcommand{\arraystretch}{1.45}
\caption*{\large \bf Four-particle reactions with leptons}
\begin{tabular}{c|c|l}
\hline\hline
\textbf{Reaction} $\boldsymbol{(1 + 2 \rightarrow 3 + 4)}$ & $\boldsymbol{S}$ & \multicolumn{1}{c}{$\boldsymbol{SG_\mathrm{F}^2a^{-4}\left|\mathcal{M}\right|^2}$}\\
\hline \hline
$\nu_\alpha + \nu_\beta \rightarrow \nu_\alpha + \nu_\beta$ & 1 & $32\left(Y_1\cdot Y_2\right)\left(Y_3\cdot Y_4\right)$\\ 
\hline
$\nu_\alpha + \overbar{\nu_\beta} \rightarrow \nu_\alpha + \overbar{\nu_\beta}$ & 1 & $32\left(Y_1\cdot Y_4\right)\left(Y_2\cdot Y_3\right)$\\
\hline
$\nu_\alpha + \nu_\alpha \rightarrow \nu_\alpha + \nu_\alpha$ & $\frac{1}{2}$ & $64\left(Y_1\cdot Y_2\right)\left(Y_3\cdot Y_4\right)$\\
\hline
$\nu_\alpha + \overbar{\nu_\alpha} \rightarrow \nu_\alpha + \overbar{\nu_\alpha}$ & 1 & $128\left(Y_1\cdot Y_4\right)\left(Y_2\cdot Y_3\right)$\\
\hline
$\nu_\alpha + \overbar{\nu_\alpha} \rightarrow \nu_\beta + \overbar{\nu_\beta}$ & 1 & $32\left(Y_1\cdot Y_4\right)\left(Y_3\cdot Y_2\right)$\\
\hline
\multirow{2}{*}{$\nu_e + \overbar{\nu_e} \rightarrow e^+ + e^-$} & \multirow{2}{*}{1} & $128\Big[g_L^2\left(Y_1\cdot Y_3\right)\left(Y_2\cdot Y_4\right)+ g_R^2\left(Y_1\cdot Y_4\right)\left(Y_2\cdot Y_3\right) $ \\ & & $+ g_Lg_Ra^2m_e^2\left(Y_1\cdot Y_2\right) \Big]$\\
\hline
\multirow{2}{*}{$\nu_e + e^- \rightarrow \nu_e + e^-$} & \multirow{2}{*}{1} & $128\Big[g_L^2\left(Y_1\cdot Y_2\right)\left(Y_3\cdot Y_4\right)+ g_R^2\left(Y_1\cdot Y_4\right)\left(Y_3\cdot Y_2\right) $ \\ & & $- g_Lg_Ra^2m_e^2\left(Y_1\cdot Y_3\right) \Big]$\\
\hline
\multirow{2}{*}{$\nu_e + e^+ \rightarrow \nu_e + e^+$} & \multirow{2}{*}{1} & $128\Big[g_L^2\left(Y_1\cdot Y_4\right)\left(Y_3\cdot Y_2\right)+ g_R^2\left(Y_1\cdot Y_2\right)\left(Y_3\cdot Y_4\right)  $ \\ & & $- g_Lg_Ra^2m_e^2\left(Y_1\cdot Y_3\right) \Big]$\\
\hline
\multirow{2}{*}{$\nu_{\mu/\tau} + \overbar{\nu_{\mu/\tau}} \rightarrow e^+ + e^-$} & \multirow{2}{*}{1} & $128\Big[\widetilde{g_L}^2\left(Y_1\cdot Y_3\right)\left(Y_2\cdot Y_4\right)+ g_R^2\left(Y_1\cdot Y_4\right)\left(Y_2\cdot Y_3\right) $ \\ & & $+ \widetilde{g_L}g_Ra^2m_e^2\left(Y_1\cdot Y_2\right) \Big]$\\
\hline
\multirow{2}{*}{$\nu_{\mu/\tau} + e^- \rightarrow \nu_{\mu/\tau} + e^-$} & \multirow{2}{*}{1} & $128\Big[\widetilde{g_L}^2\left(Y_1\cdot Y_2\right)\left(Y_3\cdot Y_4\right)+ g_R^2\left(Y_1\cdot Y_4\right)\left(Y_3\cdot Y_2\right)  $ \\ & & $- \widetilde{g_L}g_Ra^2m_e^2\left(Y_1\cdot Y_3\right) \Big]$\\
\hline
\multirow{2}{*}{$\nu_{\mu/\tau} + e^+ \rightarrow \nu_{\mu/\tau} + e^+$} & \multirow{2}{*}{1} & $128\Big[\widetilde{g_L}^2\left(Y_1\cdot Y_4\right)\left(Y_3\cdot Y_2\right)+ g_R^2\left(Y_1\cdot Y_2\right)\left(Y_3\cdot Y_4\right)  $ \\ & & $- \widetilde{g_L}g_Ra^2m_e^2\left(Y_1\cdot Y_3\right) \Big]$\\
\hline\hline
\end{tabular}
}
\end{center}
\caption{Squared matrix elements for weak processes involving active neutrinos and electrons/positrons.}
\label{table:SM1}
\end{table}

\begin{table}[H]
\begin{center}
{\renewcommand{\arraystretch}{1.45}
\caption*{\large \bf Pion decays}
\begin{tabular}{c|c|l}
\hline\hline
\textbf{Reaction} $\boldsymbol{(1 \rightarrow 2 + 3)}$ & $\boldsymbol{S}$ & \multicolumn{1}{c}{$\boldsymbol{\left|\mathcal{M}\right|^2}$}\\
\hline \hline
$\pi^0 \rightarrow \gamma + \gamma$ & 1 & $\alpha_{\mathrm{em}}^2m_{\pi}^4 \left[2\pi^2f_{\pi}^2\right]^{\hspace*{-0.03 cm}-1}$\\
\hline
$\pi^+ \rightarrow \mu^+ + \nu_\mu$ & 1 & $2G_\mathrm{F}^2\left|V_{ud}\right|^2f_{\pi}^2m_ \mu^4\left[\frac{m_{\pi}^2}{m_\mu^2}-1\right]$\\
\hline\hline
\end{tabular}
}
\end{center}
\caption{Squared matrix elements for pion decays.}
\end{table}

\newpage
\subsection{Matrix Elements for HNLs Above QCD-scale}
\label{appsec:HNLAboveQCD}
\begin{table}[H]
\label{table:tableHNLlep1}
\begin{center}
{\renewcommand{\arraystretch}{1.45}
\caption*{\large \bf Four-particle reactions with leptons only}
\begin{tabular}{c|c|l}
\hline\hline
\textbf{Reaction} $\boldsymbol{(1 + 2 \rightarrow 3 + 4)}$ & $\boldsymbol{S}$ & \multicolumn{1}{c}{$\boldsymbol{SG_\mathrm{F}^{-2}a^{-4}\left|\mathcal{M}\right|^2}$}\\
\hline \hline
$N + \nu_\beta \rightarrow \nu_\alpha + \nu_\beta$ & 1 & $32\left|\theta_\alpha\right|^2\left(Y_1\cdot Y_2\right)\left(Y_3\cdot Y_4\right)$\\ 
\hline
$N + \overbar{\nu_\beta} \rightarrow \nu_\alpha + \overbar{\nu_\beta}$ & 1 & $32\left|\theta_\alpha\right|^2\left(Y_1\cdot Y_4\right)\left(Y_2\cdot Y_3\right)$\\
\hline
$N + \nu_\alpha \rightarrow \nu_\alpha + \nu_\alpha$ & $\frac{1}{2}$ & $64\left|\theta_\alpha\right|^2\left(Y_1\cdot Y_2\right)\left(Y_3\cdot Y_4\right)$\\
\hline
$N + \overbar{\nu_\alpha} \rightarrow \nu_\alpha + \overbar{\nu_\alpha}$ & 1 & $128\left|\theta_\alpha\right|^2\left(Y_1\cdot Y_4\right)\left(Y_2\cdot Y_3\right)$\\
\hline
$N + \overbar{\nu_\alpha} \rightarrow \nu_\beta + \overbar{\nu_\beta}$ & 1 & $32\left|\theta_\alpha\right|^2\left(Y_1\cdot Y_4\right)\left(Y_3\cdot Y_2\right)$\\
\hline
\multirow{2}{*}{$N + \overbar{\nu_e} \rightarrow e^+ + e^-$} & \multirow{2}{*}{1} & $128\left|\theta_e\right|^2\Big[g_L^2\left(Y_1\cdot Y_3\right)\left(Y_2\cdot Y_4\right)+ g_R^2\left(Y_1\cdot Y_4\right)\left(Y_2\cdot Y_3\right) $ \\ & & $+ g_Lg_R a^2m_e^2\left(Y_1\cdot Y_2\right) \Big]$\\
\hline
\multirow{2}{*}{$N + e^- \rightarrow \nu_e + e^-$} & \multirow{2}{*}{1} & $128\left|\theta_e\right|^2\Big[g_L^2\left(Y_1\cdot Y_2\right)\left(Y_3\cdot Y_4\right)+ g_R^2\left(Y_1\cdot Y_4\right)\left(Y_3\cdot Y_2\right) $ \\ & & $- g_Lg_R a^2m_e^2\left(Y_1\cdot Y_3\right) \Big]$\\
\hline
\multirow{2}{*}{$N + e^+ \rightarrow \nu_e + e^+$} & \multirow{2}{*}{1} & $128\left|\theta_e\right|^2\Big[g_L^2\left(Y_1\cdot Y_4\right)\left(Y_3\cdot Y_2\right)+ g_R^2\left(Y_1\cdot Y_2\right)\left(Y_3\cdot Y_4\right) $ \\ & & $- g_Lg_R a^2m_e^2\left(Y_1\cdot Y_3\right) \Big]$\\
\hline
\multirow{2}{*}{$N + \overbar{\nu_{\mu/\tau}} \rightarrow e^+ + e^-$ }& \multirow{2}{*}{1} & $128\left|\theta_{\mu/\tau}\right|^2\Big[\widetilde{g_L}^2\left(Y_1\cdot Y_3\right)\left(Y_2\cdot Y_4\right)+ g_R^2\left(Y_1\cdot Y_4\right)\left(Y_2\cdot Y_3\right) $ \\ & & $+ \widetilde{g_L}g_Ra^2m_e^2\left(Y_1\cdot Y_2\right) \Big]$\\
\hline
\multirow{2}{*}{$N + e^- \rightarrow \nu_{\mu/\tau} + e^-$} & \multirow{2}{*}{1} & $128\left|\theta_{\mu/\tau}\right|^2\Big[\widetilde{g_L}^2\left(Y_1\cdot Y_2\right)\left(Y_3\cdot Y_4\right)+ g_R^2\left(Y_1\cdot Y_4\right)\left(Y_3\cdot Y_2\right)  $ \\ & & $- \widetilde{g_L}g_Ra^2m_e^2\left(Y_1\cdot Y_3\right) \Big]$\\
\hline
\multirow{2}{*}{$N + e^+ \rightarrow \nu_{\mu/\tau} + e^+$} & \multirow{2}{*}{1} & $128\left|\theta_{\mu/\tau}\right|^2\Big[\widetilde{g_L}^2\left(Y_1\cdot Y_4\right)\left(Y_3\cdot Y_2\right)+ g_R^2\left(Y_1\cdot Y_2\right)\left(Y_3\cdot Y_4\right)  $ \\ & & $- \widetilde{g_L}g_Ra^2m_e^2\left(Y_1\cdot Y_3\right) \Big]$\\
\hline
$N + \overbar{\nu_\mu} \rightarrow e^- + \mu^+$ & 1 & $128\left|\theta_e\right|^2\left(Y_1\cdot Y_4\right)\left(Y_2\cdot Y_3\right)$\\
\hline
$N + \overbar{\nu_e} \rightarrow e^+ + \mu^-$ & 1 & $128\left|\theta_\mu\right|^2\left(Y_1\cdot Y_3\right)\left(Y_2\cdot Y_4\right)$\\
\hline
$N + e^- \rightarrow \nu_e + \mu^-$ & 1 & $128\left|\theta_\mu\right|^2\left(Y_1\cdot Y_2\right)\left(Y_3\cdot Y_4\right)$\\
\hline
$N + e^+ \rightarrow \nu_\mu + \mu^+$ & 1 & $128\left|\theta_e\right|^2\left(Y_1\cdot Y_4\right)\left(Y_3\cdot Y_2\right)$\\
\hline
\multirow{2}{*}{$N + \overbar{\nu_\mu} \rightarrow \mu^+ + \mu^-$} & \multirow{2}{*}{1} & $128\left|\theta_\mu\right|^2\Big[g_L^2\left(Y_1\cdot Y_3\right)\left(Y_2\cdot Y_4\right)+ g_R^2\left(Y_1\cdot Y_4\right)\left(Y_2\cdot Y_3\right) $ \\ & & $+ g_Lg_Ra^2m_\mu^2\left(Y_1\cdot Y_2\right) \Big]$\\
\hline
\multirow{2}{*}{$N + \overbar{\nu_{e/\tau}} \rightarrow \mu^+ + \mu^-$} & \multirow{2}{*}{1} & $128\left|\theta_{e/\tau}\right|^2\Big[\widetilde{g_L}^2\left(Y_1\cdot Y_3\right)\left(Y_2\cdot Y_4\right)+ g_R^2\left(Y_1\cdot Y_4\right)\left(Y_2\cdot Y_3\right) $ \\ & & $+ \widetilde{g_L}g_Ra^2m_\mu^2\left(Y_1\cdot Y_2\right) \Big]$\\
\hline\hline
\end{tabular}
}
\end{center}
\caption{Squared matrix elements for weak processes involving HNLs and leptons only.}
\end{table}

\begin{table}[H]
\vspace*{-2 cm}
\begin{center}
{\renewcommand{\arraystretch}{1.45}
\caption*{\large \bf Four-particle decays into leptons}
\vspace{-0.2cm}
\begin{tabular}{c|c|l}
\hline\hline
\textbf{Reaction} $\boldsymbol{(1 \rightarrow 2 + 3 + 4)}$ & $\boldsymbol{S}$ & \multicolumn{1}{c}{$\boldsymbol{SG_\mathrm{F}^{-2}a^{-4}\left|\mathcal{M}\right|^2}$}\\
\hline \hline
$N \rightarrow \nu_\alpha + \nu_\beta + \overbar{\nu_\beta}$ & 1 & $32\left|\theta_\alpha\right|^2\left(Y_1\cdot Y_4\right)\left(Y_2\cdot Y_3\right)$\\ 
\hline
$N \rightarrow \nu_\alpha + \nu_\alpha + \overbar{\nu_\alpha}$ & $\frac{1}{2}$ & $64\left|\theta_\alpha\right|^2\left(Y_1\cdot Y_4\right)\left(Y_2\cdot Y_3\right)$\\
\hline
\multirow{2}{*}{$N \rightarrow \nu_e + e^+ + e^-$} & \multirow{2}{*}{1} & $128\left|\theta_e\right|^2\Big[g_L^2\left(Y_1\cdot Y_3\right)\left(Y_2\cdot Y_4\right)+ g_R^2\left(Y_1\cdot Y_4\right)\left(Y_2\cdot Y_3\right) $ \\ & & $+ g_Lg_Ra^2m_e^2\left(Y_1\cdot Y_2\right) \Big]$\\
\hline
\multirow{2}{*}{$N \rightarrow \nu_{\mu/\tau} + e^+ + e^-$} & \multirow{2}{*}{1} & $128\left|\theta_{\mu/\tau}\right|^2\Big[\widetilde{g_L}^2\left(Y_1\cdot Y_3\right)\left(Y_2\cdot Y_4\right)+ g_R^2\left(Y_1\cdot Y_4\right)\left(Y_2\cdot Y_3\right) $ \\ & & $+ \widetilde{g_L}g_Ra^2m_e^2\left(Y_1\cdot Y_2\right) \Big]$\\
\hline
$N \rightarrow \nu_\mu + e^- + \mu^+$ & 1 & $128\left|\theta_e\right|^2\left(Y_1\cdot Y_4\right)\left(Y_2\cdot Y_3\right)$\\
\hline
$N \rightarrow \nu_e + e^+ + \mu^-$ & 1 & $128\left|\theta_\mu\right|^2\left(Y_1\cdot Y_3\right)\left(Y_2\cdot Y_4\right)$\\
\hline
\multirow{2}{*}{$N \rightarrow \nu_\mu + \mu^+ + \mu^-$} & \multirow{2}{*}{1} & $128\left|\theta_\mu\right|^2\Big[g_L^2\left(Y_1\cdot Y_3\right)\left(Y_2\cdot Y_4\right)+ g_R^2\left(Y_1\cdot Y_4\right)\left(Y_2\cdot Y_3\right) $ \\ & & $+ g_Lg_Ra^2m_\mu^2\left(Y_1\cdot Y_2\right) \Big]$\\
\hline
\multirow{2}{*}{$N \rightarrow \nu_{e/\tau} + \mu^+ + \mu^-$} & \multirow{2}{*}{1} & $128\left|\theta_{e/\tau}\right|^2\Big[\widetilde{g_L}^2\left(Y_1\cdot Y_3\right)\left(Y_2\cdot Y_4\right)+ g_R^2\left(Y_1\cdot Y_4\right)\left(Y_2\cdot Y_3\right) $ \\ & & $+ \widetilde{g_L}g_Ra^2m_\mu^2\left(Y_1\cdot Y_2\right) \Big]$\\
\hline\hline
\end{tabular}
}
\end{center}
\vspace{-0.6 cm}
\caption{Squared matrix elements for HNL decays into leptons. Low temperatures are assumed here. At high temperatures, reactions such as $N + \mu^-\rightarrow e^- + \nu_\mu$ are possible. The corresponding matrix elements can be trivially deduced from the ones given here.}
\vspace{-0.5 cm}
\begin{center}
{\renewcommand{\arraystretch}{1.45}
\caption*{\large \bf Four-particle reactions with leptons and quarks}
\vspace{-0.2cm}
\begin{tabular}{c|c|l}
\hline\hline
\textbf{Reaction} $\boldsymbol{(1 + 2 \rightarrow 3 + 4)}$ & $\boldsymbol{S}$ & \multicolumn{1}{c}{$\boldsymbol{S\left|\theta_\alpha\right|^{-2}G_\mathrm{F}^{-2}a^{-4}\left|\mathcal{M}\right|^2}$}\\
\hline \hline
$N + \ell_\alpha^+ \rightarrow U + \overbar{D}$ & 1 & $128\left|V_{ud}\right|^2\left(Y_1\cdot Y_4\right)\left(Y_2\cdot Y_3\right)$\\
\hline
$N + D\rightarrow \ell_\alpha^- + U$ & 1 & $128\left|V_{ud}\right|^2\left(Y_1\cdot Y_2\right)\left(Y_3\cdot Y_4\right)$\\
\hline
$N + \overbar{U}\rightarrow \ell_\alpha^- + \overbar{D}$ & 1 & $128\left|V_{ud}\right|^2\left(Y_1\cdot Y_4\right)\left(Y_3\cdot Y_2\right)$\\
\hline
\multirow{2}{*}{$N + \overbar{\nu_\alpha} \rightarrow \overbar{U} + U$} & \multirow{2}{*}{1} & $\frac{32}{9}\Big[16g_R^2\left(Y_1\cdot Y_4\right)\left(Y_2\cdot Y_3\right) + \left(3-4g_R\right)^2\left(Y_1\cdot Y_3\right)\left(Y_2\cdot Y_4\right) $ \\ & & $+  4g_R\left(4g_R-3\right)a^2m_U^2\left(Y_1\cdot Y_2\right) \Big]$\\
\hline
\multirow{2}{*}{$N + U \rightarrow \nu_\alpha + U$} & \multirow{2}{*}{1} & $\frac{32}{9}\Big[16g_R^2\left(Y_1\cdot Y_4\right)\left(Y_2\cdot Y_3\right) + \left(3-4g_R\right)^2\left(Y_1\cdot Y_2\right)\left(Y_3\cdot Y_4\right) $ \\ & & $-  4g_R\left(4g_R-3\right)a^2m_U^2\left(Y_1\cdot Y_3\right) \Big]$\\
\hline
\multirow{2}{*}{$N + \overbar{U} \rightarrow \nu_\alpha + \overbar{U}$} & \multirow{2}{*}{1} & $\frac{32}{9}\Big[16g_R^2\left(Y_1\cdot Y_2\right)\left(Y_3\cdot Y_4\right) + \left(3-4g_R\right)^2\left(Y_1\cdot Y_4\right)\left(Y_3\cdot Y_2\right) $ \\ & & $-  4g_R\left(4g_R-3\right)a^2m_U^2\left(Y_1\cdot Y_3\right) \Big]$\\
\hline
\multirow{2}{*}{$N + \overbar{\nu_\alpha} \rightarrow \overbar{D} + D$} & \multirow{2}{*}{1} & $\frac{32}{9}\Big[4g_R^2\left(Y_1\cdot Y_4\right)\left(Y_2\cdot Y_3\right) + \left(3-2g_R\right)^2\left(Y_1\cdot Y_3\right)\left(Y_2\cdot Y_4\right) $ \\ & & $+  2g_R\left(2g_R-3\right)a^2m_D^2\left(Y_1\cdot Y_2\right) \Big]$\\

\hline
\multirow{2}{*}{$\hspace{1cm}N + D \rightarrow \nu_\alpha + D\hspace{0.9cm}$} & \multirow{2}{*}{1} & $\frac{32}{9}\Big[4g_R^2\left(Y_1\cdot Y_4\right)\left(Y_2\cdot Y_3\right) + \left(3-2g_R\right)^2\left(Y_1\cdot Y_2\right)\left(Y_3\cdot Y_4\right) $ \\ & & $-  2g_R\left(2g_R-3\right)a^2m_D^2\left(Y_1\cdot Y_3\right) \Big]$\\
\hline
\multirow{2}{*}{$N + \overbar{D} \rightarrow \nu_\alpha + \overbar{D}$} & \multirow{2}{*}{1} & $\frac{32}{9}\Big[4g_R^2\left(Y_1\cdot Y_2\right)\left(Y_3\cdot Y_4\right) + \left(3-2g_R\right)^2\left(Y_1\cdot Y_4\right)\left(Y_3\cdot Y_2\right) $ \\ & & $-  2g_R\left(2g_R-3\right)a^2m_D^2\left(Y_1\cdot Y_3\right) \Big]$\\
\hline\hline
\end{tabular}
}
\vspace{-0.2 cm}
\caption{Squared matrix elements for weak scattering processes involving HNLs, leptons and quarks. $U$ are up-type quarks, $D$ down-type quarks.}
\end{center}
\end{table}

\begin{table}[H]
\begin{center}
{\renewcommand{\arraystretch}{1.45}
\caption*{\large \bf Four-particle decays into leptons and quarks}
\begin{tabular}{c|c|l}
\hline\hline
\textbf{Reaction} $\boldsymbol{(1 \rightarrow 2 + 3 + 4)}$ & $\boldsymbol{S}$ & \multicolumn{1}{c}{$\boldsymbol{S\left|\theta_\alpha\right|^{-2}G_\mathrm{F}^{-2}a^{-4}\left|\mathcal{M}\right|^2}$}\\
\hline \hline
$N \rightarrow \ell_\alpha^- + U + \overbar{D}$ & 1 & $128\left|V_{ud}\right|^2\left(Y_1\cdot Y_4\right)\left(Y_2\cdot Y_3\right)$\\
\hline
\multirow{2}{*}{$N \rightarrow \nu_\alpha + \overbar{U} + U$} & \multirow{2}{*}{1} & $\frac{32}{9}\Big[16g_R^2\left(Y_1\cdot Y_4\right)\left(Y_2\cdot Y_3\right) + \left(3-4g_R\right)^2\left(Y_1\cdot Y_3\right)\left(Y_2\cdot Y_4\right) $ \\ & & $+  4g_R\left(4g_R-3\right)a^2m_U^2\left(Y_1\cdot Y_2\right) \Big]$\\
\hline
\multirow{2}{*}{$N \rightarrow \nu_\alpha + \overbar{D} + D$} & \multirow{2}{*}{1} & $\frac{32}{9}\Big[4g_R^2\left(Y_1\cdot Y_4\right)\left(Y_2\cdot Y_3\right) + \left(3-2g_R\right)^2\left(Y_1\cdot Y_3\right)\left(Y_2\cdot Y_4\right) $ \\ & & $+  2g_R\left(2g_R-3\right)a^2m_D^2\left(Y_1\cdot Y_2\right) \Big]$\\
\hline\hline
\end{tabular}
}
\end{center}
\caption{Squared matrix elements for weak decay processes involving HNLs, leptons and quarks. $U$ are up-type quarks, $D$ down-type quarks.}
\end{table}

\subsection{Matrix Elements for HNLs Below QCD-scale}
\label{appsubsec:HNLBelowQCD}
In addition to interactions with leptons, HNLs will also decay into mesons.
\begin{table}[H]
\begin{center}
{\renewcommand{\arraystretch}{1.75}
\begin{tabular}{c|c|l}
\hline\hline
\textbf{Reaction} $\boldsymbol{(1 \rightarrow 2 + 3)}$ & $\boldsymbol{S}$ & \multicolumn{1}{c}{$\boldsymbol{SG_\mathrm{F}^{-2}m_N^{-4}\left|\mathcal{M}\right|^2}$}\\
\hline \hline
$N \rightarrow \nu_\alpha + \pi^0$ & 1 & $\left|\theta_\alpha\right|^2f_\pi^2\left[1 - \frac{m_\pi^2}{m_N^2}\right]$\\
\hline
$N \rightarrow \ell_\alpha^\mp + \pi^\pm$ & 1 & $2\left|\theta_\alpha\right|^2\left|V_{ud}\right|^2f_\pi^2\left[\left(1 - \frac{m_{\ell_\alpha}^2}{m_N^2}\right)^2-\frac{m_\pi^2}{m_N^2}\left(1+\frac{m_{\ell_\alpha}^2}{m_N^2}\right)\right]$\\
\hline
$N \rightarrow \nu_\alpha + \eta$ & 1 & $\left|\theta_\alpha\right|^2f_\eta^2\left[1 - \frac{m_\eta^2}{m_N^2}\right]$\\
\hline
$N \rightarrow \nu_\alpha + \rho^0$ & 1 & $\left|\theta_\alpha\right|^2\left(1-2\sin^2\theta_\mathrm{W}\right)^2f_\rho^2\left[1+2\frac{m_\rho^2}{m_N^2}\right]\left[1 - \frac{m_\rho^2}{m_N^2}\right]$\\
\hline
$N \rightarrow \ell_\alpha^\mp + \rho^\pm$ & 1 & $2\left|\theta_\alpha\right|^2\left|V_{ud}\right|^2f_\rho^2\left[\left(1 - \frac{m_{\ell_\alpha}^2}{m_N^2}\right)^2 + \frac{m_{\rho}^2}{m_N^2}\left(1 + \frac{m_{\ell_\alpha}^2}{m_N^2}\right)-2\frac{m_\rho^4}{m_N^4}\right]$\\
\hline
$N \rightarrow \nu_\alpha + \omega$ & 1 & $\left|\theta_\alpha\right|^2\left(\frac{4}{3}\sin^2\theta_\mathrm{W}\right)^2f_\omega^2\left[1+2\frac{m_\omega^2}{m_N^2}\right]\left[1 - \frac{m_\omega^2}{m_N^2}\right]$\\
\hline
$N \rightarrow \nu_\alpha + \eta^\prime$ & 1 & $\left|\theta_\alpha\right|^2f_{\eta^\prime}^2\left[1 - \frac{m_{\eta^\prime}^2}{m_N^2}\right]$\\
\hline\hline
\end{tabular}
}
\end{center}
\caption{Squared matrix elements for HNL decays into mesons.}
\end{table}


%
%

\section{Collision Integrals}
\label{app:CollisionIntegral}
The collision integral in Eq.~\eqref{eq:CollisionIntegral} is six- and nine-dimensional in the case of three- and four-particle interactions respectively. Here we show a procedure based on \cite{Dolgov:1997mb, Blaschke:2016xxt} to reduce it to a lower-dimensional integral, such that it can be easily integrated in a numerical code. 

\pagebreak
The full collision integral in comoving momenta ($y = pa$) reads:
\begin{align}
I_\mathrm{coll} = \frac{a^{7-2\gamma}N_W}{2g_1\widetilde{E}_1}\sum_{\mathrm{in, out}}\int\left(\prod_{i=2}^{\gamma}\frac{\mathrm{d}^3y_i}{(2\pi)^32\widetilde{E}_i}\right)S|\mathcal{M}|^2F[f](2\pi)^4\delta^4(Y_\mathrm{in}-Y_\mathrm{out})\, ,
\end{align}
where $\widetilde{E}_i = aE_i$ is the comoving energy, $Y_i = aP_i$ the comoving four-momentum and $\gamma$ is the number of particles participating in the reaction. The delta function can be rewritten as
\begin{align}
\delta^4(Y_\mathrm{in}-Y_\mathrm{out}) = \delta^4(s_1Y_1  + s_2Y_2 + ... + s_\gamma Y_\gamma)\ ,
\end{align}
with $s_i = \{-1, 1\}$ if particle $i$ is on the \{left, right\}-hand side of the reaction.

\subsection{Three-Particle Collision Integral}

\begin{align}
I_\mathrm{coll} = \frac{a}{2g_1\widetilde{E}_1}\int\frac{\mathrm{d}^3y_2\mathrm{d}^3y_3}{(2\pi)^62\widetilde{E}_22\widetilde{E}_3}S|\mathcal{M}|^2F[f](2\pi)^4\delta^4(s_1Y_1 + s_2Y_2 + s_3Y_3)
\end{align}
\subsubsection{Case $y_1 \neq 0$}
Since a homogeneous and isotropic universe is assumed, only absolute values of momenta are relevant. Moreover, the matrix element in three-particle interactions is independent of the four-momenta. The collision integral then becomes:

\begin{align}
I_\mathrm{coll} = \frac{S|\mathcal{M}|^2a}{8(2\pi)^2g_1\widetilde{E}_1}\int\frac{\mathrm{d}y_2\mathrm{d}y_3 \mathrm{d}\Omega_2\mathrm{d}\Omega_3 y_2^2y_3^2}{\widetilde{E}_2\widetilde{E}_3}F[f]\delta^4(s_1Y_1 + s_2Y_2 + s_3Y_3)\, .
\end{align} 
Using the identity
\begin{align}
\delta^3(s_1\mathbf{y_1} + s_2\mathbf{y_2} + s_3\mathbf{y_3}) = \frac{1}{(2\pi)^3}\int\mathrm{d}\lambda\mathrm{d}\Omega_\lambda\lambda^2e^{i(s_1\mathbf{y_1} + s_2\mathbf{y_2} + s_3\mathbf{y_3})\cdot \boldsymbol{\lambda}}\ ,
\end{align}
where $\mathbf{y_i}$ is the 3-momentum vector of particle $i$, gives
\begin{align}
\label{eq:ThreeParticleCollisionIntegralSimplified}
I_\mathrm{coll} = &\ \frac{S|\mathcal{M}|^2a}{8(2\pi)^5g\widetilde{E_1}}\int\frac{\mathrm{d}y_2\mathrm{d}y_3 y_2^2y_3^2}{\widetilde{E}_2\widetilde{E}_3}F[f]\delta(s_1\widetilde{E}_1 + s_2\widetilde{E}_2 + s_3\widetilde{E}_3)\ \cdot \nonumber \\
& \cdot \int\mathrm{d}\lambda\lambda^2\int\mathrm{d}\Omega_\lambda e^{is_1y_1\lambda\cos\theta_\lambda}\int\mathrm{d}\Omega_2 e^{is_1y_2\lambda\cos\theta_2}\int\mathrm{d}\Omega_3 e^{is_1y_3\lambda\cos\theta_3}\nonumber \\ 
= &\ \frac{S|\mathcal{M}|^2a}{8(2\pi)^5g\widetilde{E_1}}\int\frac{\mathrm{d}y_2\mathrm{d}y_3  y_2^2y_3^2}{\widetilde{E}_2\widetilde{E}_3}F[f]\delta(s_1\widetilde{E}_1 + s_2\widetilde{E}_2 + s_3\widetilde{E}_3)\ \nonumber \cdot\\
& \cdot \int\mathrm{d}\lambda\lambda^2\left(4\pi\frac{\sin(y_1\lambda)}{y_1\lambda}\right)\left(4\pi\frac{\sin(y_2\lambda)}{y_2\lambda}\right)\left(4\pi\frac{\sin(y_3\lambda)}{y_3\lambda}\right) \nonumber \\
= &\ \frac{S|\mathcal{M}|^2a}{(2\pi)^2g\widetilde{E_1}y_1}\int\frac{\mathrm{d}y_2\mathrm{d}y_3 y_2y_3}{\widetilde{E}_2\widetilde{E}_3}F[f]\delta(s_1\widetilde{E}_1 + s_2\widetilde{E}_2 + s_3\widetilde{E}_3)\ \nonumber \cdot\\
& \cdot \int\frac{\mathrm{d}\lambda}{\lambda}\sin(y_1\lambda)\sin(y_2\lambda)\sin(y_3\lambda)\, .
\end{align}

The delta function of energies can be rewritten as:
\begin{align}
\label{eq:DeltaFunctionRewrite}
\int\frac{\mathrm{d}y_3y_3}{\widetilde{E}_3}\delta(s_1\widetilde{E}_1 + s_2\widetilde{E}_2 + s_3\widetilde{E}_3) = &\int\mathrm{d}y_3\frac{y_3}{\widetilde{E}_3}\frac{\delta\left(y_3 - y_3^*\right)}{y_3^*/\widetilde{E}_3^*}\theta\left(\left(s_1\widetilde{E}_1+s_2\widetilde{E}_2\right)^2-a^2m_3^2\right)\nonumber \\
= &\int\mathrm{d}y_3\frac{y_3}{\widetilde{E}_3}\frac{\widetilde{E}_3^*}{y_3^*}\delta(y_3-y^*_3)\ \theta\left(\left(\widetilde{E}_3^*\right)^2-a^2m_3^2\right)\ ,
\end{align}
where $\left(\widetilde{E}_3^*\right)^2 = \left(y_3^*\right)^2 + a^2m_3^2 = \left(s_1\widetilde{E}_1+s_2\widetilde{E}_2\right)^2$ and $y_3^* = \sqrt{(s_1\widetilde{E}_1+s_2\widetilde{E}_2)^2-a^2m_3^2}$.\\ \\
Plugging Eq.~\eqref{eq:DeltaFunctionRewrite} in Eq.~\eqref{eq:ThreeParticleCollisionIntegralSimplified} gives:
\begin{align}
I_\mathrm{coll} = \frac{S|\mathcal{M}|^2a}{(2\pi)^2g_1\widetilde{E}_1y_1}\int\frac{\mathrm{d}y_2 y_2}{\widetilde{E}_2}F[f] \int\frac{\mathrm{d}\lambda}{\lambda}\sin(y_1\lambda)\sin(y_2\lambda)\sin(y_3^*\lambda)\theta\left(\left(\widetilde{E}_3^*\right)^2-a^2m_3^2\right)
\end{align}
 
\noindent Now, the integral over $\lambda$ is equal to:
\begin{align}
\label{eq:ThreeParticleTriangle}
\mathcal{X} = \frac{\pi}{8}\left(-\mathrm{Sgn}[y_1-y_2-y_3^*]+\mathrm{Sgn}[y_1+y_2-y_3^*]+\mathrm{Sgn}[y_1-y_2+y_3^*]-1\right)\ ,
\end{align}
with Sgn the signum function and where $y_1\geq y_2\geq y_3$ is assumed. The final form is then:
\begin{align}
I_\mathrm{coll} = &\,\frac{S|\mathcal{M}|^2a}{(2\pi)^2g_1\widetilde{E}_1y_1}\int\frac{\mathrm{d}y_2 y_2}{\widetilde{E}_2}\mathcal{X} \theta\left(\left(s_1\widetilde{E}_1+s_2\widetilde{E}_2\right)^2-a^2m_3^2\right)\left(F[f]\right)\bigg|_{y_3 = y_3^*}
\end{align}

\subsubsection{Case $y_1 = 0$} 
\begin{align}
I_\mathrm{coll} = &\ \frac{S|\mathcal{M}|^2a}{8(2\pi)^2g_1am_1}\int\frac{\mathrm{d}^3y_2\mathrm{d}^3y_3}{\widetilde{E}_2\widetilde{E}_3}F[f]\delta\left(s_1am_1 + s_2\widetilde{E}_2 + s_3\widetilde{E}_3\right)\delta^3\left(s_2\mathbf{y_2} + s_3\mathbf{y_3}\right)\nonumber \\
= &\ \frac{S|\mathcal{M}|^2}{8(2\pi)^2gm_1}\int\mathrm{d}^3y_2F[f]\delta\left(s_1am_1 + s_2\sqrt{y_2^2 + \left(am_2\right)^2} + s_3\sqrt{y_2^2+\left(am_3\right)^2}\right)\ \nonumber \cdot \\
&\ \cdot \left(\sqrt{y_2^2 + \left(am_2\right)^2}\sqrt{y_2^2+\left(am_3\right)^2}\right)^{-1}\nonumber \\
= &\ \frac{S|\mathcal{M}|^2}{8\pi gm_1}\int\mathrm{d}y_2 y_2^2F[f]\delta\left(y_2 - y_2^*\right)\left|\frac{s_2y_2^*}{\sqrt{(y_2^*)^2 + a^2m_2^2}} + \frac{s_3y_2^*}{\sqrt{(y_2^*)^2 + a^2m_3^2}}\right|^{-1}\ \cdot \nonumber \\
&\ \cdot \left(\sqrt{y_2^2 + \left(am_2\right)^2}\sqrt{y_2^2+\left(am_3\right)^2}\right)^{-1}\theta\left(\left(s_1am_1+s_2\widetilde{E}_2\right)^2-a^2m_3^2\right) \nonumber \\
= &\ \frac{S|\mathcal{M}|^2}{8\pi gm_1}y_2^*\left|s_2\sqrt{\left(y_2^*\right)^2 + \left(am_3\right)^2} + s_3\sqrt{\left(y_2^*\right)^2+\left(am_2\right)^2}\right|^{-1}\theta\left(\left(\widetilde{E}_3^*\right)^2-a^2m_3^2\right)\ \cdot \nonumber \\
& \cdot \ \left(F[f]\right)\bigg|_{y_1 = 0,\ y_2 = y_2^*,\ y_3 = -y_2^*}\nonumber \\
= &\ \frac{S|\mathcal{M}|^2}{8\pi gm_1}y_2^*\left|s_1s_2s_3am_1\right|^{-1}\theta\left(\left(\widetilde{E}_3^*\right)^2-a^2m_3^2\right)\left(F[f]\right)\bigg|_{y_1 = 0,\ y_2 = y_2^*,\ y_3 = -y_2^*}\nonumber \\
= &\ \frac{S|\mathcal{M}|^2}{8\pi gm_1^2}\frac{y_2^*}{a}\theta\left(\left(\widetilde{E}_3^*\right)^2-a^2m_3^2\right)\left(F[f]\right)\bigg|_{y_1 = 0,\ y_2 = y_2^*,\ y_3 = -y_2^*}\ \ \  ,
\end{align} 
with $\left(\widetilde{E}_3^*\right)^2 = \left(s_1am_1+s_2\widetilde{E}_2\right)^2$ and $y_2^* = a\sqrt{\frac{\left(m_1^2-m_2^2-m_3^2\right)^2-4m_2^2m_3^2}{4m_1^2}}$\ .

\subsection{Four-Particle Collision Integral}
\begin{align}
\hspace{-0.2 cm}I_\mathrm{coll} = \frac{N_W}{2g_1a\widetilde{E}_1}\int\frac{\mathrm{d}^3y_2\mathrm{d}^3y_3\mathrm{d}y_4^3}{(2\pi)^98\widetilde{E}_2\widetilde{E}_3\widetilde{E}_4}S|\mathcal{M}|^2F[f](2\pi)^4\delta^4(s_1Y_1 + s_2Y_2 + s_3Y_3 + s_4Y_4)
\end{align}
As can be seen in Appendix \ref{app:MatrixElements}, $|\mathcal{M}|^2$ has the following form:
\begin{align}
|\mathcal{M}|^2 = \frac{1}{a^4}\sum_{i\neq j\neq k\neq l}\left[K_1(Y_i\cdot Y_j)(Y_k\cdot Y_l)+K_2a^2m_im_j(Y_k\cdot Y_l)\right]\, ,
\end{align}
with $K_1$ and $K_2$ reaction-specific constants. A similar procedure as with the three-particle case is followed here. 
\subsubsection{Case $y_1 \neq 0$}
\begin{align}
I_\mathrm{coll} = &\ \frac{SN_W}{16(2\pi)^5g_1\widetilde{E}_1a}\int\frac{\mathrm{d}y_2\mathrm{d}y_3\mathrm{d}y_4 y_2^2y_3^2y_4^2}{\widetilde{E}_2\widetilde{E}_3\widetilde{E}_4}F[f]\delta\left(s_1\widetilde{E}_1 + s_2\widetilde{E}_2 + s_3\widetilde{E}_3+s_4\widetilde{E}_4\right)\ \cdot \nonumber \\
& \cdot \int\mathrm{d}\Omega_2\mathrm{d}\Omega_3\mathrm{d}\Omega_4\left|\mathcal{M}\right|^2|\delta^3\left(s_1\mathbf{y_1}+s_2\mathbf{y_2}+s_3\mathbf{y_3}+s_4\mathbf{y_4}\right)\nonumber \\
= &\ \frac{SN_W}{64\pi^3g_1\widetilde{E}_1y_1a^5}\int\frac{\mathrm{d}y_2\mathrm{d}y_3\mathrm{d}y_4 y_2y_3y_4}{\widetilde{E}_2\widetilde{E}_3\widetilde{E}_4}F[f]\delta\left(s_1\widetilde{E}_1 + s_2\widetilde{E}_2 + s_3\widetilde{E}_3s_4\widetilde{E}_4\right)\ \nonumber \cdot\\
& \cdot \mathcal{D}(Y_1, Y_2, Y_3, Y_4)\ ,
\end{align}
with 
\begin{align}
\mathcal{D}(Y_1, Y_2, Y_3, Y_4) = &\ \frac{y_1y_2y_3y_4}{64\pi^5}\int\mathrm{d}\Omega_2\mathrm{d}\Omega_3\mathrm{d}\Omega_4\left|\mathcal{M}\right|^2|\delta^3\left(s_1\mathbf{y_1}+s_2\mathbf{y_2}+s_3\mathbf{y_3}+s_4\mathbf{y_4}\right)\nonumber \\
= &\ \frac{y_1y_2y_3y_4}{64\pi^5}\int\mathrm{d}\lambda\lambda^2\int\mathrm{d}\Omega_\lambda e^{is_1\mathbf{y_1}\cdot\boldsymbol{\lambda}}\int\mathrm{d}\Omega_2 e^{is_2\mathbf{y_2}\cdot\boldsymbol{\lambda}}\int\mathrm{d}\Omega_3 e^{is_3\mathbf{y_3}\cdot\boldsymbol{\lambda}}\ \nonumber \cdot \\
&\ \cdot\ \int\mathrm{d}\Omega_4 e^{is_4\mathbf{y_4}\cdot\boldsymbol{\lambda}}\sum_{i\neq j\neq k\neq l}\left[K_1(Y_i\cdot Y_j)(Y_k\cdot Y_l)+K_2a^2m_im_j(Y_k\cdot Y_l)\right]\nonumber \\
= &\ \frac{y_1y_2y_3y_4}{64\pi^5}\sum_{i\neq j\neq k\neq l}\int\mathrm{d}\lambda\lambda^2\int\mathrm{d}\Omega_\lambda e^{is_iy_i\lambda\cos\theta_i}\int\mathrm{d}\Omega_j e^{is_jy_j\lambda\cos\theta_j}\ \cdot \nonumber \\
&\ \cdot\ \int\mathrm{d}\Omega_k e^{is_ky_k\lambda\cos\theta_k}\int\mathrm{d}\Omega_l e^{is_ly_l\lambda\cos\theta_l}\left[K_1(Y_i\cdot Y_j)(Y_k\cdot Y_l)\right.+\ \nonumber \\
&\ + \left.K_2a^2m_im_j(Y_k\cdot Y_l)\right]\ .
\end{align}
The inner products can be worked out:
\begin{align}
Y_i\cdot Y_j &= \widetilde{E}_i\widetilde{E}_j - \mathbf{y_i}\cdot\mathbf{y_j} = \widetilde{E}_i\widetilde{E}_j - y_iy_j\cos\theta_{ij} \nonumber \\
&= \widetilde{E}_i\widetilde{E}_j - y_iy_j\left(\cos\theta_i\cos\theta_j + \cos(\phi_i-\phi_j)\sin\theta_i\sin\theta_j\right)\ ,
\end{align}

\noindent where $\theta_{ij}$ is the angle between vectors $\mathbf{y_i}$ and $\mathbf{y_j}$. Next, using the identity
\begin{align}
\int_0^\pi\int_0^{2\pi}\mathrm{d}\theta_i\mathrm{d}\phi_i e^{is_iy_i\lambda\cos\theta_i}\sin^2\theta_i\cos(\phi_i-\phi_j) = 0
\end{align}
gives
\begin{align}
\mathcal{D}(Y_1, Y_2, Y_3, Y_4) = &\ \frac{y_1y_2y_3y_4}{64\pi^5}\sum_{i\neq j\neq k\neq l}\int\mathrm{d}\lambda\lambda^2\int\mathrm{d}\theta_i\mathrm{d}\phi_i\sin\theta_i e^{is_iy_i\lambda\cos\theta_i}\ \cdot \\
& \cdot\ 
\int\mathrm{d}\theta_j\mathrm{d}\phi_j\sin\theta_j e^{is_jy_j\lambda\cos\theta_j}\int\mathrm{d}\theta_k\mathrm{d}\phi_k\sin\theta_k e^{is_ky_k\lambda\cos\theta_k}\ \cdot \nonumber \\
&\cdot \ \int\mathrm{d}\theta_l\mathrm{d}\phi_l\sin\theta_l e^{is_ly_l\lambda\cos\theta_l} \left[K_1\left(\widetilde{E}_i\widetilde{E}_j - y_iy_j\cos\theta_i\cos\theta_j\right)\right. \ \cdot \nonumber \\ & \cdot \ \left.\left(\widetilde{E}_k\widetilde{E}_l - y_ky_l\cos\theta_k\cos\theta_l\right)+K_2a^2m_im_j\left(\widetilde{E}_k\widetilde{E}_l - y_ky_l\cos\theta_k\cos\theta_l\right)\right]\, .\nonumber
\end{align}
The integrals over the angles can be evaluated:
\begin{align}
\int_0^\pi\int_0^{2\pi}\mathrm{d}\theta\mathrm{d}\phi \sin\theta e^{isy\lambda\cos\theta} &= 4\pi\frac{\sin(y\lambda)}{y\lambda}\\
\int_0^\pi\int_0^{2\pi}\mathrm{d}\theta\mathrm{d}\phi \sin\theta\cos\theta e^{isy\lambda\cos\theta} &= \frac{4\pi}{isy\lambda}\left[\cos(y\lambda)-\frac{\sin(y\lambda)}{y\lambda}\right]\, .
\end{align}
Working out all the brackets gives:
\begin{align}
\mathcal{D}(Y_1, Y_2, Y_3, Y_4) = &\sum_{i\neq j\neq k\neq l}\left[K_1\left\{\widetilde{E}_1\widetilde{E}_2\widetilde{E}_3\widetilde{E}_4D_1\left(y_1, y_2, y_3, y_4\right)+\widetilde{E}_i\widetilde{E}_jD_2\left(y_i, y_j, y_k, y_l\right)\right.\right. + \nonumber \\ 
& \hspace{-0.14 cm}+ \left.\left.\widetilde{E}_k\widetilde{E}_lD_2\left(y_k, y_l, y_i, y_j\right) + D_3\left(y_1, y_2, y_3, y_4\right)\right\}\right.+ \nonumber \\ 
& \hspace{-0.14 cm}+ \left. K_2a^2m_im_j\left\{\widetilde{E}_k\widetilde{E}_lD_1\left(y_1, y_2, y_3, y_4\right) + D_2\left(y_i, y_j, y_k, y_l\right)\right\}\right],
\end{align}
with
\begin{align}
D_1\left(y_i, y_j, y_k, y_l\right) = &\ \frac{4}{\pi}\int\frac{\mathrm{d}\lambda}{\lambda^2}\sin(y_i\lambda)\sin(y_j\lambda)\sin(y_k\lambda)\sin(y_l\lambda)\, ,\\
D_2\left(y_i, y_j, y_k, y_l\right) = &\ s_ks_l\frac{4y_ky_l}{\pi}\int\frac{\mathrm{d}\lambda}{\lambda^2}\sin(y_i\lambda)\sin(y_j\lambda)\left[\cos(y_k\lambda)-\frac{\sin(y_k\lambda)}{y_k\lambda}\right]\ \cdot \nonumber \\
& \cdot\ \left[\cos(y_l\lambda)-\frac{\sin(y_l\lambda)}{y_l\lambda}\right]\, ,\\
D_3\left(y_i, y_j, y_k, y_l\right) = &\ s_is_js_ks_l\frac{4y_iy_jy_ky_l}{\pi}\int\frac{\mathrm{d}\lambda}{\lambda^2}\left[\cos(y_i\lambda)-\frac{\sin(y_i\lambda)}{y_i\lambda}\right]\left[\cos(y_j\lambda)-\frac{\sin(y_j\lambda)}{y_j\lambda}\right]\ \cdot \nonumber \\
& \cdot\ \left[\cos(y_k\lambda)-\frac{\sin(y_k\lambda)}{y_k\lambda}\right]\left[\cos(y_l\lambda)-\frac{\sin(y_l\lambda)}{y_l\lambda}\right]\, .
\end{align}

\noindent All these three functions are symmetric under the exchange $y_i \leftrightarrow y_j$ and $y_k \leftrightarrow y_l$, which then allows us to take $y_i > y_j$ and $y_k > y_l$. Integrating out $\lambda$ gives the functions in terms of polynomials for all possible values of momenta (factors $s_ks_l$ and $s_is_js_ks_l$ are omitted for convenience):
\begin{itemize}
\item $y_i > y_j + y_k + y_l$ or $y_k > y_i + y_j + y_l$:
\begin{align}
D_1 = D_2 = D_3 = 0
\end{align}
\item $y_i + y_j > y_k + y_l$ and $y_i + y_l < y_j + y_k$:
\begin{align}
D_1 = &\ y_l\hspace{6.1 cm}\nonumber\\
D_2 = &\ \frac{1}{3}y_l^3\hspace{6.1 cm}\\
D_3 = &\ \frac{1}{30}y_l^3\left[5\left(y_i^2+y_j^2+y_k^2\right)-y_l^2\right]\hspace{6.1 cm}\nonumber
\end{align}
\item $y_i + y_j > y_k + y_l$ and $y_i + y_l > y_j + y_k$:
\begin{align}
D_1 = &\ \frac{1}{2}(y_j + y_k + y_l - y_i)\nonumber\\
D_2 = &\ \frac{1}{12}\left[(y_i-y_j)\left\{(y_i-y_j)^2-3\left(y_k^2+y_l^2\right)\right\}+2\left(y_k^3+y_l^3\right)\right]\\
D_3 = &\ \frac{1}{60}\left[y_i^5-y_j^5-y_k^5-y_l^5+5\left(-y_i^3y_j^2+y_i^2y_j^3-y_i^3y_k^2+y_i^2y_k^3\right.\right.\nonumber\\ 
&\ \ \ \ \ \ -\left.\left.y_i^3y_l^2+y_i^2y_l^3+y_j^3y_k^2+y_j^2y_k^3+y_j^3y_l^2+
y_j^2y_l^3+y_k^3y_l^2+y_k^2y_l^3\right)\right]\nonumber
\end{align}
\item $y_i + y_j < y_k + y_l$ and $y_i + y_l > y_j + y_k$:
\begin{align}
D_1 &= \ y_j\hspace{5.13 cm}\nonumber\\
D_2 &= \ \frac{1}{6}y_j\left[3\left(y_k^2+y_l^2-y_i^2\right)-y_j^2\right]\hspace{5.13 cm}\\
D_3 &= \ \frac{1}{30}y_j^3\left[5\left(y_i^2+y_k^2+y_l^2\right)-y_j^2\right]\hspace{5.13 cm}\nonumber
\end{align}
\item $y_i + y_j < y_k + y_l$ and $y_i + y_l < y_j + y_k$:
\begin{align}
D_1 = &\ \frac{1}{2}(y_i + y_j + y_l - y_k)\nonumber\\
D_2 = &\ -\frac{1}{12}\left[(y_i+y_j)\left\{(y_i+y_j)^2-3\left(y_k^2+y_l^2\right)\right\}+2\left(y_k^3-y_l^3\right)\right]\\
D_3 = &\ \frac{1}{60}\left[y_k^5-y_i^5-y_j^5-y_l^5+5\left(-y_k^3y_l^2+y_k^2y_l^3-y_k^3y_i^2+y_k^2y_i^3\right.\right.\nonumber\\ 
&\ \ \ \ \ \ -\left.\left.y_k^3y_j^2+y_k^2y_j^3+y_l^3y_i^2+y_l^2y_i^3+
y_l^3y_j^2+ y_l^2y_j^3+y_i^3y_j^2+y_i^2y_j^3\right)\right]\nonumber
\end{align}
\end{itemize}
\noindent Going back to the collision integral, the same trick as before can be applied to the delta function of energies, which eventually gives:
\begin{align}
I_\mathrm{coll} = &\ \frac{SN_W}{64\pi^3g_1\widetilde{E}_1y_1a^5}\int\mathrm{d}y_2\mathrm{d}y_3\frac{y_2y_3}{\widetilde{E}_2\widetilde{E}_3}\mathcal{D}(Y_1, Y_2, Y_3, Y_4)\ \cdot\nonumber \\
& \cdot \theta\left(\left((s_1\widetilde{E}_1+s_2\widetilde{E}_2+s_3\widetilde{E}_3\right)^2-a^2m_4^2\right)\left(F[f]\right)\bigg|_{y_4 = y_4^*}\ \ \ ,
\end{align}
with $y_4^* = \sqrt{\left(s_1\widetilde{E}_1 + s_2\widetilde{E}_2 + s_3\widetilde{E}_3\right)^2 - a^2m_4^2}\ .$

\pagebreak
\subsubsection{Case $y_1 = 0$}
\begin{align}
I_\mathrm{coll} = &\ \frac{SN_W}{64\pi^3g_1am_1}\frac{1}{a^5}\int\frac{\mathrm{d}y_2\mathrm{d}y_3\mathrm{d}y_4 y_2y_3y_4}{\widetilde{E}_2\widetilde{E}_3\widetilde{E}_4}F[f]\delta\left(s_1am_1 + s_2\widetilde{E}_2 + s_3\widetilde{E}_3 + s_4\widetilde{E}_4\right)\ \cdot\nonumber \\
& \cdot \mathcal{B}(Y_1, Y_2, Y_3, Y_4)\ ,
\end{align}
with
\begin{align}
\mathcal{B}(Y_1, Y_2, Y_3, Y_4) =  &\ \frac{y_2y_3y_4}{64\pi^5}\int\mathrm{d}\Omega_2\mathrm{d}\Omega_3\mathrm{d}\Omega_4\left|\mathcal{M}\right|^2|\delta^3\left(s_2\mathbf{y_2}+s_3\mathbf{y_3}+s_4\mathbf{y_4}\right)\nonumber \\
= &\ \frac{y_2y_3y_4}{64\pi^5}\int\mathrm{d}\lambda\lambda^2\mathrm{d}\Omega_\lambda
\int\mathrm{d}\theta_2\mathrm{d}\phi_2\sin\theta_2 e^{is_2y_2\lambda\cos\theta_2}\ \cdot\nonumber \\
& \cdot\ \int\mathrm{d}\theta_3\mathrm{d}\phi_3\sin\theta_3 e^{is_3y_3\lambda\cos\theta_3}\int\mathrm{d}\theta_4\mathrm{d}\phi_4\sin\theta_4 e^{is_4y_4\lambda\cos\theta_4}\ \cdot\nonumber \\
& \cdot\  \sum_{i\neq j\neq k\neq l}\left[K_1\left(\widetilde{E}_i\widetilde{E}_j - y_iy_j\cos\theta_i\cos\theta_j\right)\cdot \left(\widetilde{E}_k\widetilde{E}_l - y_ky_l\cos\theta_k\cos\theta_l\right)\right. + \nonumber \\ &+ K_2\left.a^2m_im_j\left(\widetilde{E}_k\widetilde{E}_l - y_ky_l\cos\theta_k\cos\theta_l\right)\right]
\end{align}
Consider the case $i=1$ in one of the terms of $|\mathcal{M}|^2$. Then the $\mathcal{B}$-function can be written as:
\begin{align}
\mathcal{B}_{i=1}(Y_1, Y_2, Y_3, Y_4) = &\ \frac{y_2y_3y_4}{64\pi^5}4\pi\int\mathrm{d}\lambda\lambda^2\sum_{j\neq k\neq l}
\int\mathrm{d}\theta_j\mathrm{d}\phi_j\sin\theta_j e^{is_jy_j\lambda\cos\theta_j}\ \cdot\nonumber \\
& \cdot\ \int\mathrm{d}\theta_k\mathrm{d}\phi_k\sin\theta_k e^{is_ky_k\lambda\cos\theta_k}\int\mathrm{d}\theta_l\mathrm{d}\phi_l\sin\theta_l e^{is_ly_l\lambda\cos\theta_l}\ \cdot\nonumber \\
& \cdot\ \left[K_1am_1\widetilde{E}_j\cdot \left(\widetilde{E}_k\widetilde{E}_l - y_ky_l\cos\theta_k\cos\theta_l\right)\right. + \nonumber \\ &+ K_2\left.a^2m_1m_j\left(\widetilde{E}_k\widetilde{E}_l - y_ky_l\cos\theta_k\cos\theta_l\right)\right]\nonumber\\
=&\ K_1am_1\sum_{j\neq k\neq l}\left[\widetilde{E_j}\widetilde{E_k}\widetilde{E_l}B_1\left(y_j, y_k, y_l\right) + \widetilde{E_j}B_2\left(y_j, y_k, y_l\right)\right] + \nonumber \\
& + K_2am_1\sum_{j\neq k\neq l}am_j\left[\widetilde{E_k}\widetilde{E_l}B_1\left(y_j, y_k, y_l\right) + B_2\left(y_j, y_k, y_l\right)\right],
\end{align} 
with $B_1\left(y_j, y_k, y_l\right)$ given by Eq. (\ref{eq:ThreeParticleTriangle}) and
\begin{align} 
\hspace{-0.21cm}B_2\left(y_j, y_k, y_l\right) = &\ s_ks_l\frac{4y_ky_l}{\pi}\int\frac{\mathrm{d}\lambda}{\lambda}\sin(y_j\lambda)\left[\cos(y_k\lambda)-\frac{\sin(y_k\lambda)}{y_k\lambda}\right]\left[\cos(y_l\lambda)-\frac{\sin(y_l\lambda)}{y_l\lambda}\right]\nonumber\\
= &\begin{cases}
\frac{1}{2}\left[y_k^2 + y_l^2 - y_j^2\right], \ \ \ y_j + y_k \geq y_l\ \& \ y_j + y_l \geq y_k\ \& \ y_k + y_l \geq y_j \\
0, \ \ \ \mathrm{otherwise}
\end{cases}.
\end{align}
This procedure is repeated for all the other terms in $\left|\mathcal{M}\right|^2$. If $j=1$, the result is obtained by $i\leftrightarrow j$. Note that if $k=1$ or $l=1$, there is no $B_2$-term in the part with $K_2$. Finally:
\begin{align}
I_\mathrm{coll} = &\frac{SN_W}{64\pi^3g_1m_1a^6}\int\mathrm{d}y_2\mathrm{d}y_3\frac{ y_2y_3}{\widetilde{E}_2\widetilde{E}_3}\mathcal{B}(Y_1, Y_2, Y_3, Y_4)\ \cdot\nonumber \\
& \cdot \theta\left(\left((s_1am_1+s_2\widetilde{E}_2+s_3\widetilde{E}_3\right)^2-a^2m_4^2\right)\left(F[f]\right)\bigg|_{y_4 = y_4^*}\ \ \ ,
\end{align}
with $y_4^* = \sqrt{\left(s_1am_1 + s_2\widetilde{E}_2 + s_3\widetilde{E}_3\right)^2 - a^2m_4^2}\ $.

\section{Neutron-Proton Conversion Rates}
\label{app:weak_rates}
Here we list the rates of the neutron-proton conversion reactions as used in \texttt{pyBBN}. Moreover, we include a number of relevant corrections to these rates due to finite nucleon size and Coulomb interactions. The relevant conversion reactions are
\begin{align}
    \label{eq:nnu_to_pe}
    n + \nu_e &\leftrightarrow p + e^-\\
    \label{eq:ne_to_pnu}
    n + e^+ &\leftrightarrow p + \overline{\nu_e}\\
    \label{eq:n_to_penu}
    n &\leftrightarrow p + e^- + \overline{\nu_e}\ ,
\end{align}
for which the corresponding averaged squared matrix element is given by:
\begin{align}
    |\mathcal{M}|^2 =&\ 16G_\mathrm{F}^2|V_\mathrm{ud}|^2\Big[(1+g_\mathrm{A})^2(P_e\cdot P_p)(P_\nu\cdot P_n)+(1-g_\mathrm{A})^2(P_e\cdot P_n)(P_\nu\cdot P_p)\nonumber\\ 
    & \hspace{2 cm}+(g^2_\mathrm{A}-1)m_nm_p(P_e\cdot P_\nu)\Big]\ ,
\end{align}
with $V_\mathrm{ud}$ the CKM-matrix element, $g_A$ the axial vector coupling constant and $P_i$ the momentum 4-vector of particle $i$. Note that for the backward reaction in~\eqref{eq:nnu_to_pe} and the forward reaction in~\eqref{eq:ne_to_pnu} there is an additional multiplicative factor of $\frac{1}{2}$ which, however, cancels with the spin degrees of freedom in the rates. The differential cross-section for a reaction of the form $A + B \rightarrow C + D$ can be obtained by:
\begin{align}
    \label{eq:diff_cross_section}
     \frac{\mathrm{d}\sigma}{\mathrm{d}E_D} = \frac{\mathrm{d}\sigma}{\mathrm{d}t}\frac{\mathrm{d}t}{\mathrm{d}E_D} = \frac{|\mathcal{M}|^2}{64\pi s \left(p_B^\mathrm{CM}\right)^2}\frac{\mathrm{d}t}{\mathrm{d}E_D}\ ,
\end{align}
where $s$ and $t$ are the Lorentz-invariant Mandelstam variables and $p_B^\mathrm{CM}$ is the momentum of particle $B$ in the center-of-mass frame. This equation only holds when particle $A$ is at rest (this would correspond to the neutron or proton). The lower and upper integration bounds are:
\begin{align}
    \hspace{-0.5 cm}E_D|_\mathrm{min}^\mathrm{max} = \frac{1}{2m_A}\left[\left(\frac{m_A^2 - m_B^2 - m_C^2 + m_D^2}{2\sqrt{s}}\right)^2 - \left(p_B^\mathrm{CM}\mp p_D^\mathrm{CM}\right)^2 + m_A^2 - m_C^2 + 2m_AE_B\right]\ .
\end{align}

The bare rates for the forward and backward reactions \eqref{eq:nnu_to_pe} -- \eqref{eq:n_to_penu} used in \texttt{pyBBN} are summarized below. In what follows, $f_\nu$ and $f_e$ are the neutrino and electron/positron distribution functions respectively.

\begin{itemize}
    \item $\boldsymbol{n\rightarrow pe^-\overline{\nu_e}}$
            \begin{align}
               \label{eq:n_decay_rate}
                \Gamma &= \frac{1}{(4\pi)^3m_n}\int|\mathcal{M}|^2(1-f_\nu)(1-f_e)\mathrm{d}E_e\mathrm{d}E_\nu\\
                E_e^\mathrm{min}&=m_e\\
                 E_e^\mathrm{max}&=\frac{m_n^2-m_p^2+m_e^2}{2m_n}\\
                E_\nu^\mathrm{min} &= \frac{\frac{1}{2}(m_n^2-m_p^2+m_e^2)-m_nE_e}{m_n-E_e+\sqrt{E_e^2-m_e^2}}\\
                E_\nu^\mathrm{max} &= \frac{\frac{1}{2}(m_n^2-m_p^2+m_e^2)-m_nE_e}{m_n-E_e-\sqrt{E_e^2-m_e^2}}
                \end{align}
                
    \item $\boldsymbol{pe^-\overline{\nu_e}\rightarrow n}$\\
    This rate is highly suppressed compared to the rates of the other reactions~\cite{Esposito:1998rc}. Therefore, for convenience, we make use of the infinite-mass approximation to simplify the computation.
            \begin{align}
                \Gamma &= \frac{G_\mathrm{F}^2|V_\mathrm{ud}|^2(1+3g_\mathrm{A}^2)}{(2\pi)^3\pi}\int E_e\sqrt{E_e^2-m_e^2}f_\nu f_e\mathrm{d}^3p_\nu\\
                E_e &= m_n-m_p-p_\nu\\
                p_\nu^\mathrm{min}&=0\\
                p_\nu^\mathrm{max}&=m_n-m_p-m_e
            \end{align}
    \item $\boldsymbol{n\nu_e\rightarrow pe^-}$
            \begin{align}
                \Gamma &= \frac{1}{(2\pi)^3}\int \frac{\mathrm{d}\sigma}{\mathrm{d}E_e} f_\nu(1-f_e)\mathrm{d}E_e\mathrm{d}^3p_\nu\\
                p_\nu^\mathrm{min} &= 0\\
                p_\nu^\mathrm{max} &= \infty
            \end{align}
    \item $\boldsymbol{pe^-\rightarrow n\nu_e}$   
            \begin{align}
                \Gamma &= \frac{1}{(2\pi)^3}\int \frac{\mathrm{d}\sigma}{\mathrm{d}E_\nu} \sqrt{1-\left(\frac{m_e}{E_e}\right)^2}f_e(1-f_\nu)\mathrm{d}E_\nu\mathrm{d}^3p_e\\
                p_e^\mathrm{min} &= \sqrt{\left(\frac{m_n^2-m_p^2-m_e^2}{2m_p}\right)^2-m_e^2}\\
                p_e^\mathrm{max} &= \infty
            \end{align}
    \item $\boldsymbol{ne^+\rightarrow p\overline{\nu_e}}$
               \begin{align} 
              \Gamma &=\frac{1}{(2\pi)^3}\int \frac{\mathrm{d}\sigma}{\mathrm{d}E_\nu} \sqrt{1-\left(\frac{m_e}{E_e}\right)^2}f_e(1-f_\nu)\mathrm{d}E_\nu\mathrm{d}^3p_e\\
              p_e^\mathrm{min} &= 0\\
              p_e^\mathrm{max} &= \infty
            \end{align}
    \item $\boldsymbol{p\overline{\nu_e}\rightarrow ne^+}$
            \begin{align}
                \Gamma =& \frac{1}{(2\pi)^3}\int \frac{\mathrm{d}\sigma}{\mathrm{d}E_e} f_\nu(1-f_e)\mathrm{d}E_e\mathrm{d}^3p_\nu\\
                p_\nu^\mathrm{min} =& \frac{(m_n+m_e)^2-m_p^2}{2m_p}\\
                p_\nu^\mathrm{max} =& \infty
            \end{align}
\end{itemize}
\vfill

\subsection{Corrections to Conversion Rates}
\label{app:corrections_to_rates}
The bare rates already account for the finite mass of the nucleons. We include finite-size nucleon corrections and Coulomb corrections in our analysis as we find that these could change the primordial helium abundance significantly. Other corrections, such as zero- and finite-temperature radiative corrections, are usually derived in the literature under the basic assumptions of point-like and infinite-mass nucleons, which makes their validity questionable in this case. Even regardless of this, they provide only minor, sub-percent corrections to the primordial abundances~\cite{Dicus:1982bz, Lopez:1998vk, Esposito:1998rc, Fukugita:2004cq, Pitrou:2018cgg}. Since our analysis is nearly insensitive to such corrections, we approximate their contribution by multiplying all rates with $\tau_n\Gamma\approx 1.05$, where $\tau_n = 880.2$ s and $\Gamma$ is given in~\eqref{eq:n_decay_rate}.

\subsubsection{Nucleon Electromagnetic Form Factors}
The approximation of treating neutrons and protons as point-like particles becomes no longer valid once they scatter with particles of energies close to their mass. Indeed, HNLs with a high mass decay into neutrinos with energies $E_\nu\sim m_N/2\sim \mathcal{O}(m_p)$, which means that such neutrinos can probe the internal structure of the nucleons. The charge distribution is encoded in the nucleon electromagnetic form factors, which alter the cross-sections for the reactions
$n + \nu_e \rightarrow p + e^-$ and $p + \overline{\nu_e} \rightarrow n + e^+$ (see e.g. \cite{Kuzmin:2007kr}). We follow a similar procedure as in~\cite{Kuzmin:2007kr, leitner2005neutrino, LlewellynSmith:1971uhs, Singh:1992dc}, where the nucleon current is given by:
\begin{align}
    J_{np}^\mu =&\ V_\mathrm{ud}\overline{u}_p(k')\Big[\gamma^\mu\left(F_1^V(Q^2)-\gamma^5F^A(Q^2)\right)+\frac{i}{2m_p}\sigma^{\mu\nu}q_\nu\left(F_2^V(Q^2)-\gamma^5F^T(Q^2)\right)\nonumber\\ &+\frac{q^\mu}{m_p}\left(F^S(Q^2)-\gamma^5F^P(Q^2)\right)\Big]u_n(k)\ ,
\end{align}
with $\sigma^{\mu\nu} = -i\left[\gamma^\mu,\gamma^\nu\right]/4$ and $Q^2 = -q^2 = -(P_n-P_p)^2 = -(P_{\nu_e}-P_e)^2$. The form factors in the dipole approximation read:
\begin{align}
    F^S &= F^T = 0\\
    F^A(Q^2) &= \frac{g_A}{\left(1+\frac{Q^2}{m_A^2}\right)^2}\\
    F^P(Q^2) &= \frac{2m_p^2}{q^2+m_\pi^2}F^A(Q^2)\\
    F_1^V(Q^2) &= \frac{1+\mu_n\frac{a\tau}{1+b\tau}+\tau\left(\mu_p-\mu_n\right)}{1+\tau}\left(1+\frac{Q^2}{m_V^2}\right)^{-2}\\
    F_2^V(Q^2) &= \frac{\mu_p-\mu_n-1-\mu_n\frac{a\tau}{1+b\tau}}{1+\tau}\left(1+\frac{Q^2}{m_V^2}\right)^{-2}\ ,
\end{align}
with $\tau = \frac{Q^2}{4m_p^2}$, $g_A = 1.27$, $\mu_n = -1.913$ and $\mu_p = 2.793$ the neutron and proton magnetic moments respectively, $m_A = 1.026$ GeV, $m_V = 0.843$ GeV, $m_\pi = 0.13957$ GeV, $a = 0.942$ and $b = 4.61$. Note that the proton and neutron masses are not distinguished here, which means that this approximation becomes invalid at low energies, when $E_\nu\sim m_n-m_p$~\cite{Strumia:2003zx}. The differential cross-sections are then obtained by~\cite{Kuzmin:2007kr, leitner2005neutrino, LlewellynSmith:1971uhs, Singh:1992dc}:
\begin{align}
    \label{eq:diff_sigma_highE}
    \frac{\mathrm{d}\sigma}{\mathrm{d}Q^2} =&\ \frac{m_p^2G_\mathrm{F}^2|V_\mathrm{ud}|^2}{8\pi E_\nu^2}\left[X(Q^2)\pm\frac{s-u}{m_p^2}Y(q^2)+\frac{(s-u)^2}{m_p^4}Z(Q^2)\right]\\
    X(Q^2) =&\ \frac{m_e^2+Q^2}{m_p^2}\Bigg[(1+\tau)F_A^2-(1-\tau)(F_1^V)^2+\tau(1-\tau)(F_2^V)^2+4\tau F_1^VF_2^V\nonumber\\ &-\frac{m_e^2}{4m_p^2}\left\{(F_1^V+F_2^V)^2+(F^A+2F^P)^2-\left(\frac{Q^2}{m_p^2}+4\right)(F^P)^2\right\}\Bigg]\\
    Y(Q^2) =&\ \frac{Q^2}{m_p^2}F^A(F_1^V+F_2^V)\\
    Z(Q^2) =&\ \frac{1}{4}\left((F^A)^2+(F_1^V)^2+\tau(F_2^V)^2\right)\ ,
\end{align}
where in \eqref{eq:diff_sigma_highE} the `+' sign holds for $n + \nu_e \rightarrow p + e^-$ and the `-' sign for $p + \overline{\nu_e} \rightarrow n + e^+$. The total cross-section reads:
\begin{align}
    \sigma =& \underset{{Q_\mathrm{min}^2}}{\overset{Q_\mathrm{max}^2}{\int}}\mathrm{d}Q^2\frac{\mathrm{d}\sigma}{\mathrm{d}Q^2}(1-f_e)\ ,
\end{align}
with integration bounds
\begin{align}
    Q_\mathrm{min}^2 =& \frac{2E_\nu^2m_p-m_e^2m_p-E_\nu m_e^2-E_\nu\left((s-m_e^2)^2-2(s+m_e^2)m_p^2+m_p^4\right)^{1/2}}{2E_\nu+m_p}\\
    Q_\mathrm{min}^2 =& \frac{2E_\nu^2m_p-m_e^2m_p-E_\nu m_e^2+E_\nu\left((s-m_e^2)^2-2(s+m_e^2)m_p^2+m_p^4\right)^{1/2}}{2E_\nu+m_p}\ .
\end{align}
In summary, for the reactions $n + \nu_e \rightarrow p + e^-$ and $p + \overline{\nu_e} \rightarrow n + e^+$ we use the cross-sections as described at the beginning of this appendix in the low-energy limit (see Eq.~\eqref{eq:diff_cross_section}) and replace them by the ones above in the high-energy limit. The threshold between the two regimes is chosen as $p_\nu = 20$ MeV, as we find that this is when the cross-sections in the two regimes overlap with each other.

\subsubsection{Coulomb Corrections}
The Coulomb correction is applied by multiplying the integrands of the rates for the reactions $n+\nu_e \leftrightarrow p+e^-$ and $n\leftrightarrow p+e^-+\overline{\nu}_e$ by the non-relativistic Fermi factor~\cite{Dicus:1982bz}:
\begin{align}
    C = \frac{2\pi\alpha/\beta}{1-e^{-2\pi\alpha/\beta}}\ ,
\end{align}
where $\alpha$ is the fine-structure constant and $\beta = \sqrt{1-m_e^2/E_e^2}$ the electron velocity. We note that we resort to this form of the Coulomb correction as opposed to its relativistic counterpart, as our analysis is insensitive to the difference between the two approaches~\cite{Smith:2009bt, Pitrou:2018cgg}.



\pagebreak
\enlargethispage{0.8cm}
\section{Numerical Methods}
\label{app:numerical-methods}
In this appendix we give a short overview of the Boltzmann code and the numerical schemes utilized to simulate BBN in the presence of HNLs. A more comprehensive user guide will be available on the \texttt{GitHub} page\footnote{\href{https://github.com/ckald/pyBBN}{https://github.com/ckald/pyBBN}}.

\subsection{pyBBN: Boltzmann Code for BBN with HNLs}
The main framework of \texttt{pyBBN} is written in \texttt{Python}. Since the computation of collision integrals is a very time consuming process, this part is done in \texttt{C++}. The Boltzmann code is able to simulate BBN in the presence of HNLs at the level of $10^{-3}-10^{-2}$, starting from temperatures of several GeV down to temperatures below keV. While the current version only accommodates for HNLs, it is generally possible to extend it to include other BSM particles. As explained in Section~\ref{sec:methodology}, since the baryon-to-photon ratio is very small, this allows us to separate the system in a part involving nuclear physics only and a background cosmology. The nuclear reaction network computations are done in the modified \texttt{KAWANO} code~\cite{Kawano:1992ua, Ruchayskiy:2012si}. The Boltzmann equation in real space has a term that is proportional to $\partial f/\partial p$ (see Eq.~\eqref{eq:BoltzmannEquation}) and is rather difficult to deal with. Therefore, \texttt{pyBBN} is written in comoving coordinates, which makes this term vanish and allows for an easier way to solve the Boltzmann equation. In comoving coordinates, the following rules apply: $p\rightarrow y = pa$, $m\rightarrow ma$ and $E\rightarrow aE$ for the momentum, mass and energy respectively.

\subsubsection{General Structure}
Each simulation is divided in two steps: 
\begin{enumerate}
    \item The background cosmology and the rates of the reactions
    \begin{align}
    n + e^{+} \leftrightarrow p + \overline{\nu_{e}}, \quad n + \nu_e \leftrightarrow p + e^-,\quad n \leftrightarrow p + e^- + \overline{\nu_{e}}\, , 
    \end{align}
    are computed in \texttt{pyBBN}. This involves solving the system of equations for the evolution of temperature, scale factor and distribution functions of decoupled species like active neutrinos, HNLs and unstable HNL decay products (see Section~\ref{sec:system_of_equations}). 
    \item The relevant cosmological quantities together with the aforementioned rates are tabulated and passed on to the modified \texttt{KAWANO} code, that takes care of the nuclear physics part of the simulation and outputs the light element abundances. We note that the modified \texttt{KAWANO} code used in this work is slightly updated with respect to the one used in~\cite{Ruchayskiy:2012si}, in that the code now automatically accounts for non-standard thermal histories by adjusting the initial baryon-to-photon ratio at the start of simulation, such that the final baryon-to-photon ratio is equal to $\eta\approx 6.09\times10^{-10}$~\cite{Aghanim:2018eyx}.
\end{enumerate}

\subsubsection{The 5-step Computational Scheme}
Each simulation step in \texttt{pyBBN} consists of five parts:
\begin{enumerate}
\item The current regime of each particle is determined (in-equilibrium or decoupled) and its parameter set is updated with the latest cosmological and thermodynamical variables, such as temperature, number density, energy density and pressure. 
\item The interactions for decoupled species are initialized and the relevant collision integrals are determined.
\item The collision integrals are computed. Neutrino oscillations -- if enabled -- are taken into account by mixing the collision integrals (see Eq. \eqref{eq:BoltzmannEquationActiveNeutrinos}). 
\item The Boltzmann equations are integrated and distribution functions are updated.
\item The temperature evolution equation is evaluated and integrated to update the latest cosmological quantities of the Universe at the end of the time step. 
\end{enumerate}

\subsubsection{Approximation Schemes}
Here we mention some of the approximations implemented in the code that simplify the treatment of certain particles and improve the agility of the code. More details can be found in the user guide.
\paragraph{Decoupling Temperatures.}
To properly account for particle decoupling, we define the decoupling temperature in the code as the temperature at which the particle is guaranteed to be in equilibrium and start to compute the kinetic equations from that point onward. Naturally, this decoupling temperature is chosen as close to the real decoupling temperature as possible. Photons and electrons are always treated as equilibrium particles.
\begin{itemize}
\item \textbf{HNLs.} The decoupling temperature in the ultra-relativistic limit $T_\mathrm{rel}$ can be obtained from the relation $H = \Gamma_N = |\theta|^2 G_\mathrm{F}^2T^5$. If this temperature is higher than the HNL mass $m_N$ by a certain threshold, then it is taken as the HNL decoupling temperature. A similar reasoning is used in the case the ultra-relativistic decoupling temperature is smaller than the HNL mass. Between these two regimes, we also implement an intermediate regime. We find that a prefactor of 1.5 works well enough for defining the threshold between the regimes. In summary, we use the following scheme:
\begin{align}
T_\mathrm{dec} = 
\begin{cases}
1.5 \cdot T_\mathrm{rel} \qquad &\mathrm{if\ } 1.5\cdot m_N <  1.5\cdot T_\mathrm{rel} \\
1.5 \cdot m_N \qquad &\mathrm{if\ } m_N \leq  1.5\cdot T_\mathrm{rel} \leq 1.5\cdot m_N\\
m_N \qquad &\mathrm{if\ }  1.5\cdot T_\mathrm{rel} < m_N\\
\end{cases}\ .
\end{align}
In principle, there is no harm in overestimating the decoupling temperature, as the Boltzmann equation keeps the particle in equilibrium when it should be. This temperature is also taken as the initial temperature of the simulation. 
\item \textbf{Active neutrinos.}
The decoupling temperature of active neutrinos in SBBN is theoretically estimated to be around $2\,\mathrm{MeV}$ \cite{Dolgov:2002wy}. In the code the default decoupling temperature is set equal to 5 MeV, in order to properly account for spectral distortions induced by HNL decays that can influence the decoupling process. We have checked that setting it equal to 10 MeV does not change the results.
\item \textbf{Unstable HNL decay products.}
The unstable HNL decay products considered in this work are muons, pions, $\eta$-, $\rho$- and $\omega$-mesons. Muons and pions are treated as equilibrium particles down to temperatures around the neutrino decoupling temperature. After that, 
their distribution functions is set equal to zero.
The distributions of the other mesons are set to zero right after QCD transition and their Boltzmann equation is solved close to neutrino decoupling. Thus, the assumption is made that around neutrino decoupling any muons and mesons present in the system originate from HNL decays.
\end{itemize}

\paragraph{Momentum-space discretization.}
The code uses evenly-spaced grids in comoving momentum space that have only 2 adjustable parameters: maximum momentum and number of samples. We have found that a default grid with 400 samples up to $100\,\mathrm{MeV}$ is enough for convergence of the primordial abundances in SBBN.
In simulations involving HNLs, the maximum momentum of the grids of HNL decay products is defined by the HNL mass and the maximum scale factor at which energy injections are expected. The latter we arbitrarily set to $a_\mathrm{max} = 6$, corresponding roughly to $T \approx 0.5\,\mathrm{MeV}$ (HNLs still present at these temperatures are definitely excluded). Hence, the maximum comoving momentum of the grids is estimated as $y_\textrm{max} = a_\textrm{max} m_N$.
The grid of HNLs themselves is cut off when their equilibrium distribution function drops below $10^{-100}$ and has a resolution of 0.67 MeV/sample.

The resolution of the grid of an HNL decay product depends on the type of particle and interactions it is involved in. In general, four-particle reactions require a resolution around 0.25 MeV/sample. On the other hand, the collision integral of a two-body decay is sharply peaked and requires a momentum grid of much higher resolution in order to be resolved. A full list of grid parameters is provided in the user guide.

\paragraph{Grid cut-offs.} We optimize the numerical integration by adjusting the integration region, e.g., by excluding kinematically forbidden combinations of particle momenta.

\paragraph{Interpolation of collision integrals.}
A particle that participates in three-particle reactions is required to have a grid with high resolution. If the same particle also participates in four-particle reactions, then using the same grid points would be computationally expensive, since four-particle reactions require a lower grid resolution. To this end, an interpolation mechanism is implemented that takes a subset of the grid with a resolution of ${\sim}0.25\,\mathrm{MeV/sample}$ and computes the four-particle collision integral only for this subset. Then it uses linear interpolation to obtain the collision integral for all other points of the original grid.

\paragraph{Unstable HNL decay products.}
\label{subsec:dynamicalequilibrium}
Sterile neutrinos of high masses decay into short-lived particles -- e.g., the muon lifetime is ${\sim}10^{-6}\,\mathrm{s}$. Such timescales are orders of magnitude smaller than any practical computational time step for a simulation spanning ${\sim}10^6$ s. It means that for such particles a different approach must be utilized. HNLs with masses $m_N<m_\mu$ will decay into stable particles. HNLs with higher masses will have decay products that are unstable. Some of these unstable decay products will interact with the plasma before they decay. The analysis here will be done for muons, but can be applied to charged pions as well. There are three important things to consider:
\begin{enumerate}
\item $\mu$ is created from an HNL decay.\\
The distribution function of these muons is a non-thermal distribution $f_\mathrm{noneq}$.
\item $\mu$ thermalizes.\\
The muon-photon scattering rate is higher than the muon decay rate: $\Gamma_{\gamma\mu} \sim \frac{\alpha^2}{m_\mu E_\gamma}T_\gamma^3 \sim 10^{-9}\,\mathrm{MeV}$ vs. $\Gamma_{\mu,\mathrm{decay}} \sim 10^{-16}\,\mathrm{MeV}$. Therefore, they will first thermalize and transfer excess energy to the electromagnetic sector of the plasma. After thermalization, the muons will share the same temperature as the equilibrium plasma and will have a thermal distribution 
$f_\mathrm{thermal} = e^{-\frac{m_\mu - \xi}{T}}e^{-\frac{p^2}{2m_\mu T}}$, where $\xi$ is a chemical potential and is determined by the condition that the number density before and after thermalization must be equal. The collision term corresponding to this process is then estimated as:
\begin{align}
I_\mathrm{thermalization} \approx \frac{f_\mathrm{thermal} - f_\mathrm{noneq}}{\Delta t}\ ,
\end{align}
with $\Delta t$ the time step of the simulation. Note that this procedure does not apply for charged $\rho$-mesons as they have a lifetime that is too short for this process to occur.
\item $\mu$ decays.\\
The main decay channel of muons is $\mu^- \rightarrow e^-+\overbar{\nu_e}+\nu_\mu$. The muon has a lifetime that is much smaller than the timestep of the simulation. This poses a problem right away: when the evolution of the distribution function for the muon and active neutrinos is computed roughly as $\Delta f = \mp I_\mathrm{coll}\Delta t$, the behavior of the collision integral $I_\mathrm{coll}$ is not resolved. It is assumed to be constant during the whole timestep $\Delta t$, which is not true: the created muons have already decayed well within this timestep. What therefore happens is that the number of muons that have decayed and the number of neutrinos that are created, are overestimated. 

This issue can be solved by using dynamical equilibrium, where the condition that the same amount of muons is created and destroyed during each time step is imposed. Consider the following chain:\\
\begin{minipage}{\linewidth}
\vspace{0.4 cm}
\begin{center}
\begin{tikzpicture}
  \node[draw,circle,inner sep=5pt] (A) at (0,0) {HNL};
  \node[draw,circle,inner sep=10pt] (B) at (2.75, 0) {$\mu$};
  \node[draw,circle,inner sep=10.8pt] (C) at (5.5,0) {$\nu$};
  \node (D) at (1.3,0.4) {$+\Delta N$};
  \node (E) at (4.05,0.4) {$-\Delta N$};
        
\draw [->,>=stealth] (A) edge (B) (B) edge (C);     
\end{tikzpicture}
\end{center}
\vspace{0.1 cm}
\end{minipage}

The time step $\Delta t$ is much smaller than the lifetime of the HNL, which means that there is approximately a constant inflow of muons \emph{during} each time step. Since the number of muons created, $\Delta N$, decays almost instantaneously, the same number of active neutrinos is created: for each muon that decays, one electron neutrino and one muon neutrino is created. Now, a scaling factor $\alpha$ can be introduced in $\Delta f$ = $\alpha I_\mathrm{coll}\Delta t$ such that $\int\mathrm{d}^3p\Delta f/(2\pi)^3 = \Delta N$. Basically, this is a simple rescaling of the timestep $\Delta t$.
\end{enumerate}
At the end of each time step, the distribution functions of unstable HNL decay products are set equal to zero. In summary, the creation of unstable particles from HNL decays is treated in the regular way, while in their subsequent decays dynamical equilibrium is used to ensure that the same amount of particles have decayed and are created. In addition, muons and charged pions thermalize with the plasma before their decay.

\subsection{Code Testing}
\label{app:pyBBN_tests}
The Boltzmann code \texttt{pyBBN} has been tested in multiple situations modelling both SBBN, as well as nucleosynthesis in the presence of HNLs. Below we summarize the most representative selection of them. Throughout this appendix comoving coordinates are used, such that $y = pa$, $\widetilde{E} = aE$ and $aT$ are the comoving momentum, energy and temperature respectively.

\subsubsection{Electron-Positron Annihilation}
\label{app:ee_annihilation}
\begin{figure}[h!]
\vspace{-0.3cm}
\begin{center}
\includegraphics[width=0.9\textwidth]{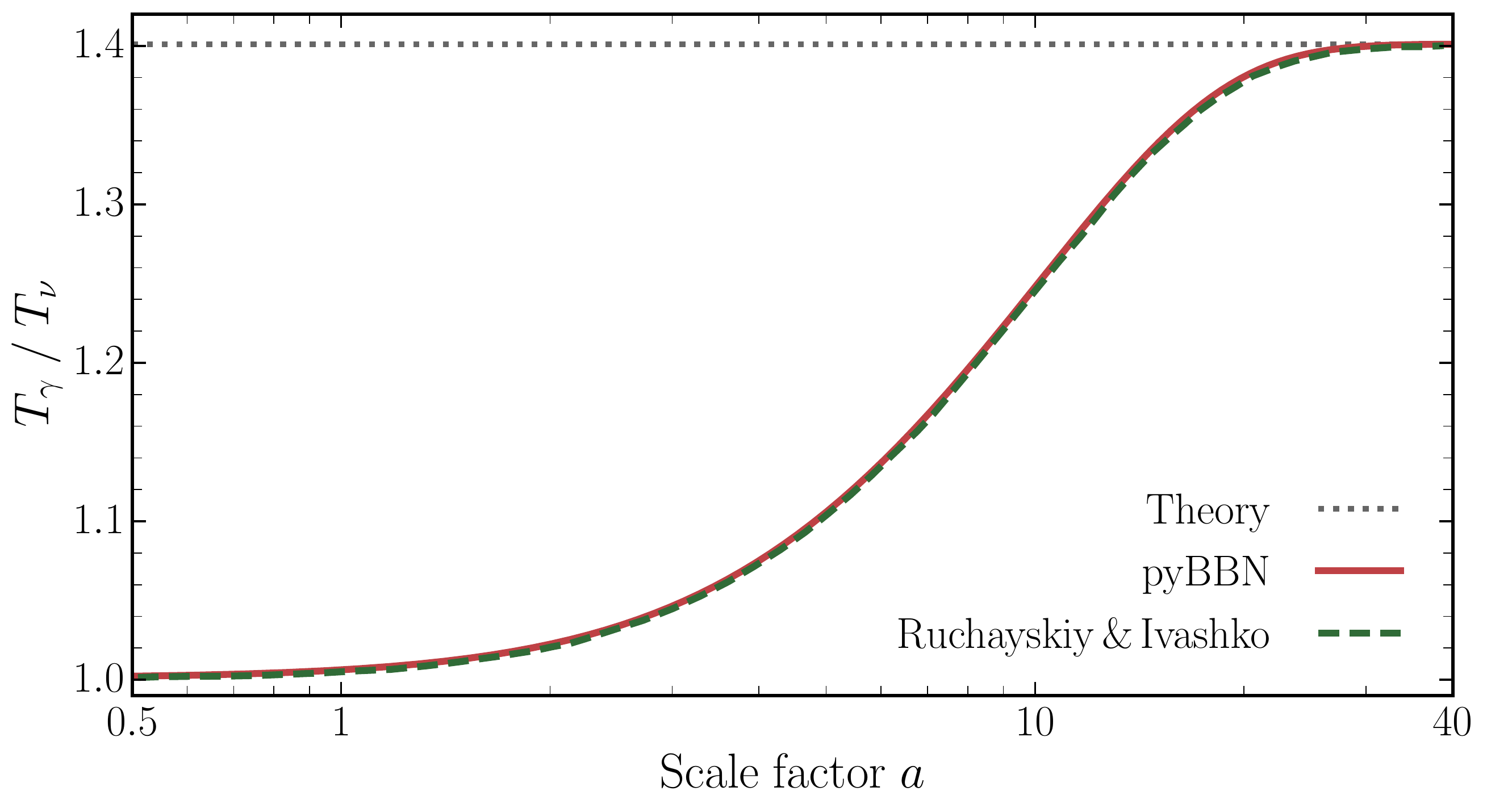}
\end{center}
\vspace{-0.2cm}
\caption{Evolution of the photon-to-neutrino temperature ratio due to electron-positron annihilation. The green dashed line is from \cite{Ruchayskiy:2012si}, the black dotted line is based on Eq.~\eqref{eq:ee_annihilation} and the red solid line is the result of this work.}
\label{fig:ee_annihilation}
\end{figure}

Once the temperature drops below the electron mass, the electrons and positrons in the plasma will annihilate into photons. Since most neutrinos are decoupled at this time, they will not experience this heat-up. A simple approximation based on entropy conservation gives the ratio between the photon and effective neutrino temperature at the end of this process:
\begin{align}
\label{eq:ee_annihilation}
\frac{T_\gamma}{T_\nu} = \left(\frac{g_*(T_\mathrm{before})}{g_*(T_\mathrm{after})}\right)^{\frac{1}{3}} = \left(\frac{\frac{7}{8}\cdot 4 + 2}{2}\right)^{\frac{1}{3}} \approx 1.401
\end{align}
In Figure~\ref{fig:ee_annihilation} we show the evolution of the photon-to-neutrino temperature ratio in SBBN. 

\vspace{-0.15cm}
\subsubsection{Number Conservation During Elastic Scatterings}
\label{subsec:particlenumberconservation}
The Boltzmann equation in comoving coordinates,
\begin{align}
\frac{\partial f}{\partial t} = H\frac{\partial f}{\partial\ln a} = I_\mathrm{coll}\,,
\end{align}
can be integrated over momentum to give:
\begin{align}
\label{eq:elasticscat}
\frac{\partial n_\mathrm{c}}{\partial \ln{a}} = \frac{g}{2\pi^2 H}\int\mathrm{d}y y^2I_\mathrm{coll} \ ,
\end{align}
where $n_\mathrm{c} \equiv na^3$ is the comoving number density. In elastic scatterings the number of particles of each species involved does not change. Therefore, the expectation is that the right-hand side of Eq.~\eqref{eq:elasticscat} vanishes. We test the conservation of particle number in elastic scatterings for active neutrinos in the temperature range from $T = 10\,\mathrm{MeV}$ to $T = 0.1\,\mathrm{MeV}$. The test is performed with various resolutions of the momentum grid (e.g., for the default grid up to $y = 100\,\mathrm{MeV}$ with 400 points and a finer one with 1000 points) and neutrino decoupling temperatures ($5\,\mathrm{MeV}$ and $10\,\mathrm{MeV})$.
We find for all neutrino species a maximum relative change of $\frac{\partial n_\mathrm{c}}{n_\mathrm{c} \partial\ln a} < 10^{-4}$.

\subsubsection{Active Neutrino Decoupled Spectra in SBBN}
\begin{figure}[t!]
\begin{center}
\includegraphics[width=0.9\textwidth]{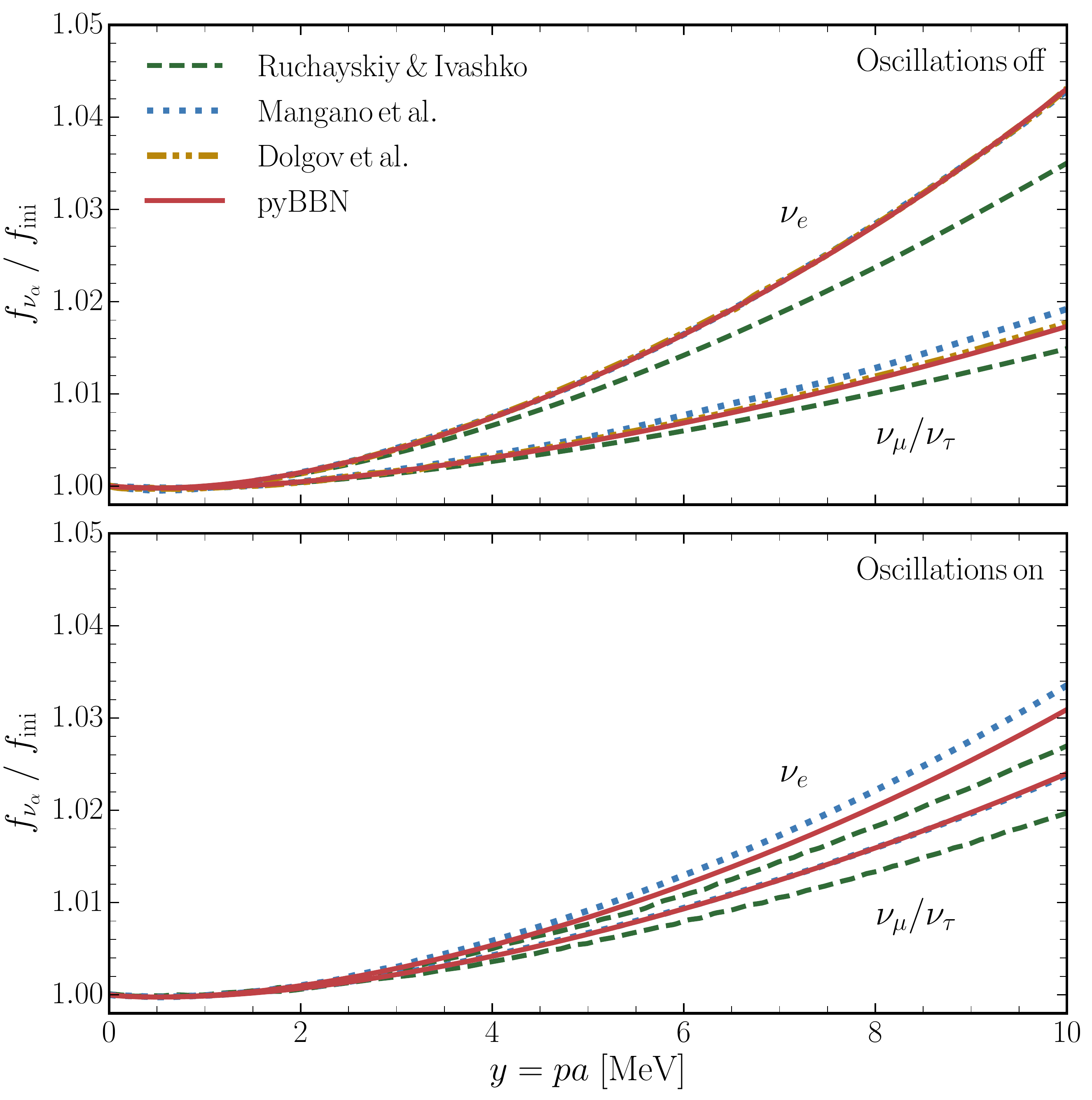}
\end{center}
\caption{Ratio of active neutrino decoupled spectra $f_{\nu_\alpha}$ to their equilibrium distribution before the onset of BBN $f_\mathrm{ini}$ versus comoving momentum $y$. In both panels, the upper curves show the distortion of the electron neutrino spectrum and the lower of muon and tau neutrinos. Dashed curves are from \cite{Ruchayskiy:2012si}, dotted from \cite{Mangano:2005cc}, dash-dotted from~\cite{Dolgov:1997mb} and the solid curves are the result of this work.
\emph{Top}: Neutrino flavour oscillations are not taken into account. \emph{Bottom}: Neutrino oscillations are included with parameters $\sin^2\theta_{12} = 0.3$, $\theta_{13} = 0$ and $\sin^2\theta_{23}=0.5$.}
\label{fig:DecoupledSpectra}
\end{figure}

\begin{figure}[H]
\begin{center}
\includegraphics[width=\textwidth]{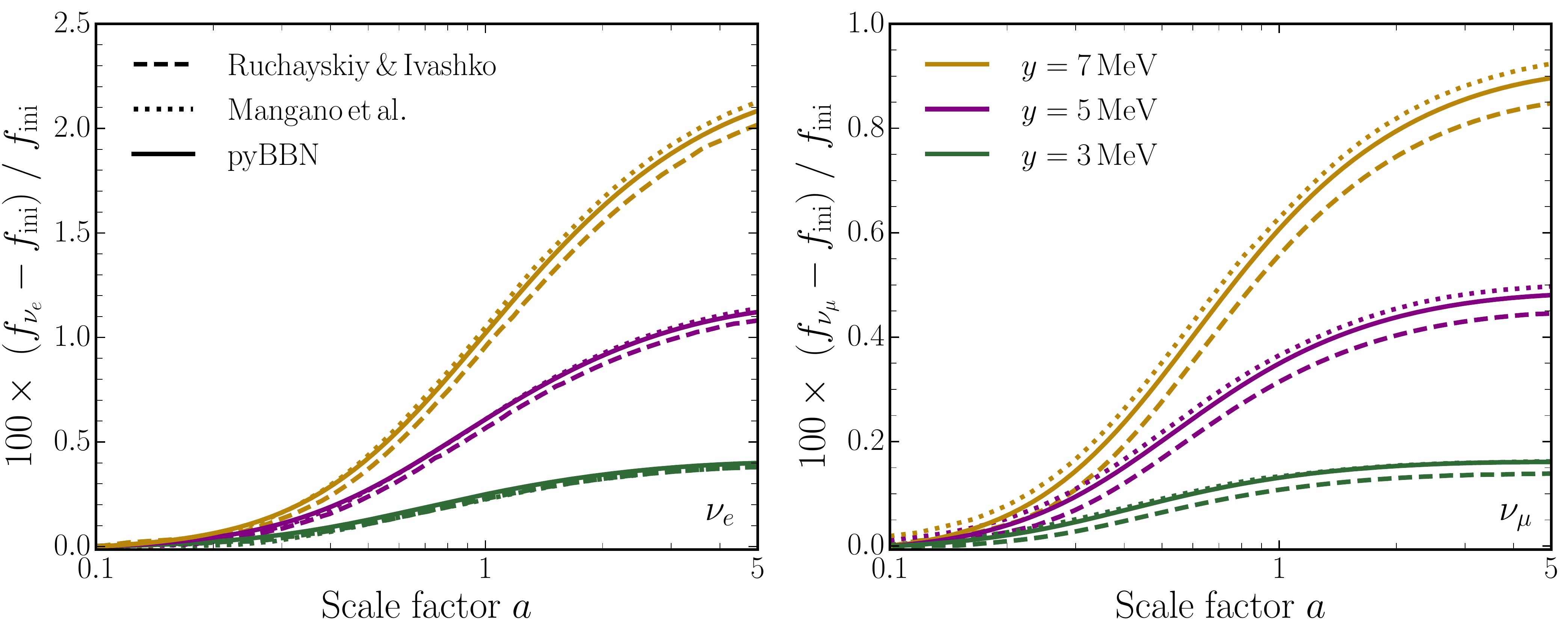}
\end{center}
\caption{Distortion of active neutrino spectra relative to their equilibrium distribution before the onset of BBN $f_\mathrm{ini}$ versus scale factor. The green, purple and gold curves represent comoving momenta y = 3, 5 and 7 MeV respectively. Neutrino flavour oscillations are not taken into account here. Solid curves are from \texttt{pyBBN}, dashed from \cite{Ruchayskiy:2012si} and dotted from \cite{Dolgov:1997mb}. Left plot is for electron neutrino, right plot for muon neutrino.}
\label{fig:DecoupledSpectra2}
\end{figure}

In Fermi theory the cross section increases with momentum as $\sigma \propto p^2$, which means that neutrinos with higher momenta stay longer in equilibrium. Since these neutrinos decouple later, they will briefly experience the heat-up of the plasma due to electron-positron annihilation, shown in Figure~\ref{fig:ee_annihilation}. This heat-up can be characterized by the increase in the quantity $aT$. This means that in comoving coordinates the distribution function of particles that are still in equilibrium, as given by
\begin{align}
f_\nu = \frac{1}{e^\frac{p}{T}+1} = \frac{1}{e^\frac{y}{aT}+1}\ ,
\end{align}
increases then accordingly. At temperatures of $\mathcal{O}(1)\,\mathrm{MeV}$ electron neutrinos interact through both charged and neutral currents, while muon and tau neutrinos only interact through neutral currents. This is because the temperature is too low for muons and tau leptons to be present in the plasma or to be created from muon and tau neutrinos through charged current interactions. The cross section of electron neutrinos is therefore larger and, consequently, they stay longer in equilibrium. The deviation of the distribution functions of active neutrinos from equilibrium is shown in Figures~\ref{fig:DecoupledSpectra} and \ref{fig:DecoupledSpectra2} (see also~\cite{Akita:2020szl} for a calculation of relic neutrino decoupling with next-to-leading order effects included).

\vfill
\newpage

\subsubsection{Neutron-to-Proton Ratio in SBBN}
\label{app:npratio}
\begin{figure}[h!]
\begin{center}
\includegraphics[width=0.9\textwidth]{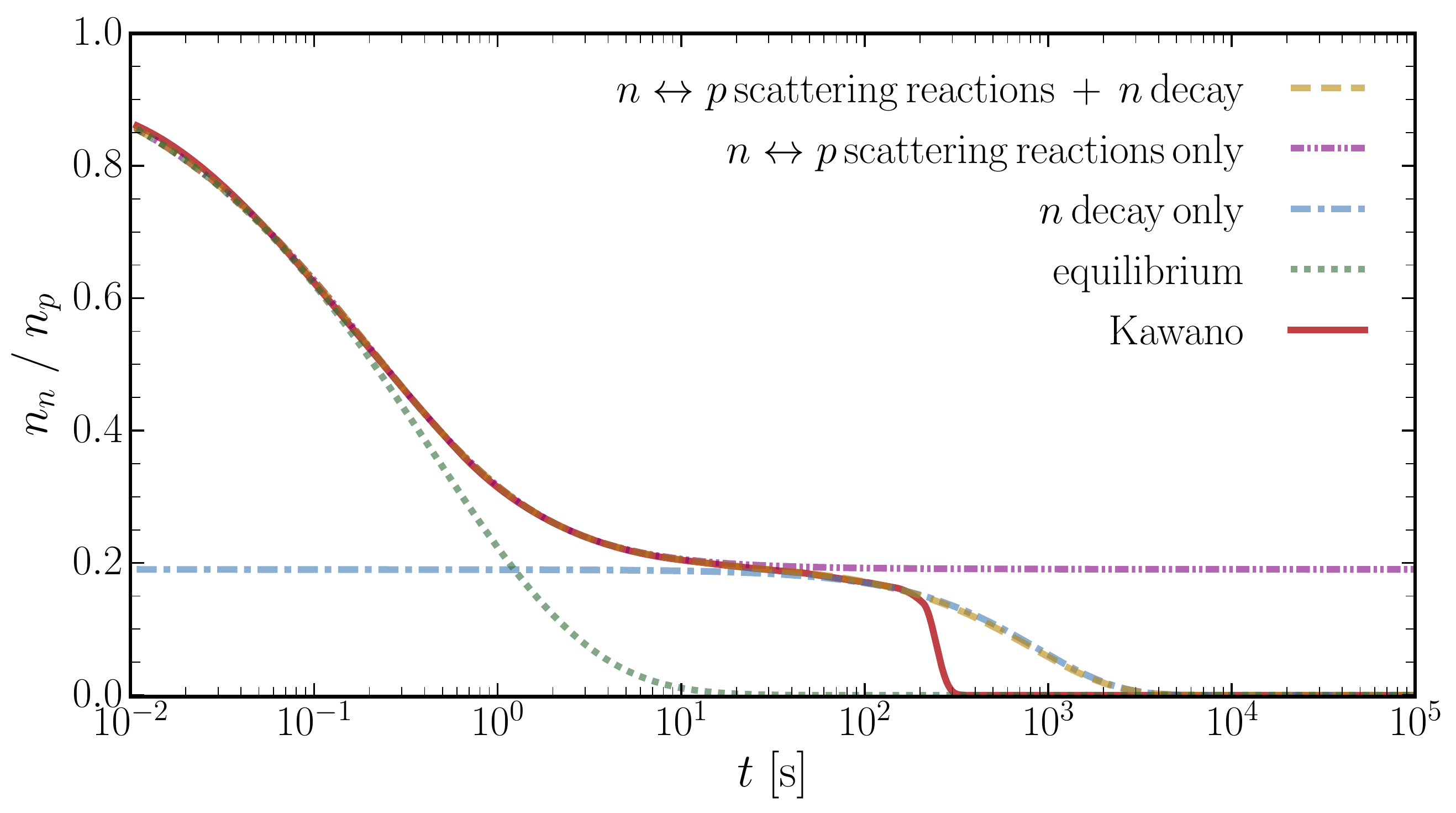}
\end{center}
\caption{Evolution of the neutron-to-proton ratio in SBBN. The solid red line is obtained from the modified \texttt{KAWANO} code, while the yellow dashed curve is calculated by using the semi-analytical approach described in the text (see also Eqs.~\eqref{eq:XNsemianalytical} and \eqref{eq:npfromXn}). The blue and purple dash-dotted lines are a decomposition of the yellow dashed line and account for neutron decay only and neutron-proton scattering reactions only respectively. The green dotted line shows the case if all particles were in equilibrium.}
\label{fig:np_ratio}
\end{figure}

In our numerical scheme, the rates of the neutron-proton conversion reactions Eq.~\eqref{eq:np_reactions} are computed in \texttt{pyBBN} and subsequently passed on to the modified \texttt{KAWANO} code. Here we compare the evolution of the neutron-to-proton ratio and the primordial helium abundance in SBBN as output by \texttt{KAWANO} and by using the semi-analytical approach described in \cite{Kolb:1990vq, Mukhanov:2005sc}. In the latter, the evolution equation for the relative concentration of neutrons, $X_n \equiv \frac{n_n}{n_n + n_p}$, is given by:
\begin{align}
\label{eq:XNsemianalytical}
\frac{\mathrm{d}X_n}{\mathrm{d}t} =& -\left[\Gamma_{n\nu_e\rightarrow pe^-} + \Gamma_{ne^+\rightarrow p\overbar{\nu_e}} + \Gamma_{n\rightarrow pe^-\overbar{\nu_e}}\right]X_n\nonumber\\ 
& + \left[\Gamma_{pe^-\rightarrow n\nu_e} + \Gamma_{p\overbar{\nu_e}\rightarrow ne^+} + \Gamma_{pe^-\overbar{\nu_e}\rightarrow n}\right](1-X_n)\ ,
\end{align}
where the $\Gamma$'s are the rates of the corresponding reactions. The neutron-to-proton ratio is then obtained by:
\begin{align}
\label{eq:npfromXn}
\frac{n_n}{n_p} = \frac{X_n}{1 - X_n}\ .
\end{align}
The results are shown in Figure~\ref{fig:np_ratio}. The sudden dip of the solid red line around $t \sim 200\,\mathrm{s}$ is due to the onset of primordial nucleosynthesis. The neutron-to-proton ratio at this point, together with the simplified formula for the primordial helium abundance, $Y_\mathrm{P} = \frac{2n_n/n_p}{n_n/n_p+1}$, gives a value that is consistent with the output of \texttt{KAWANO}.

\subsubsection{Helium-4 and Deuterium Abundances in SBBN}
\label{app:Abundances_comparison}
In this test we compare the outputs of the modified \texttt{KAWANO} and \texttt{PArthENoPE2.\hspace{-0.5mm}0} \cite{Consiglio:2017pot} codes. The \texttt{KAWANO} code is somewhat dated and does not account for the latest nuclear reaction rates. However, we show in Table~\ref{tab:HeDHComp} that this induces only minor deviations in the primordial helium and deuterium abundances.
The values presented are obtained for neutron lifetime $\tau_\mathrm{n} = 880.2\,\mathrm{s}$ \cite{pdg} and baryon-to-photon ratio $\eta = 6.09\times10^{-10}$ \cite{Aghanim:2018eyx}. Moreover, using the theoretical and observational errors as described in Section~\ref{subsec:bbn_data_analysis}, we find that our predicted helium and deuterium abundances are well within $1\sigma$ from the measured abundances in Eqs.~\eqref{eq:Yp} and \eqref{eq:DH}.

\begin{table}[H]
\begin{center}
{\def\arraystretch{1.35}
\begin{tabular}{l|c|c}
\hline\hline
\multicolumn{1}{c|}{\textbf{Code}} & $\boldsymbol{Y_\mathrm{P}}$ & $\boldsymbol{10^5\times D/H|_\mathrm{P}}$\\
\hline\hline
\texttt{PArthENoPE2.\hspace{-0.5mm}0} & 0.24691 &  2.6156\\
\hline
\texttt{pyBBN} & 0.24657 & 2.6082\\
\hline
Rel. Diff. & 0.14\% & 0.3\% \\
\hline\hline
\end{tabular}
}
\end{center}
\caption{The primordial helium and deuterium abundances in SBBN obtained from \texttt{pyBBN} with the modified \texttt{KAWANO} code and from \texttt{PArthENoPE2.\hspace{-0.5mm}0}~\cite{Consiglio:2017pot}.}
\label{tab:HeDHComp}
\end{table}


\subsubsection{HNL Decay Width}
\label{subsec:appHNLdecaywidth}
Here we compare the HNL decay width in vacuum with the expected theoretical value in the case of three-body and two-body decays.
In what follows, we consider an HNL of mass $m_N$ that mixes only with electron neutrinos with a mixing angle $\theta_e$.
\paragraph{Three-body decay width.} For HNL masses lower than the muon mass, there are four decay channels:
\begin{align}
\label{eq:HNLnuemixingdecays4p}
\begin{matrix}
N \rightarrow \nu_e + \nu_e + \overbar{\nu_e} & \hspace{1.3 cm} 
N \rightarrow \nu_e + \nu_\mu + \overbar{\nu_\mu}\\
N \rightarrow \nu_e + \nu_\tau + \overbar{\nu_\tau} & \hspace{1.6cm}
N \rightarrow \nu_e + e^+ + e^-\ ,
\end{matrix}
\end{align}

from which the decay width in vacuum can be computed:
\begin{align}
\label{eq:HNLdecaywidth4p}
\Gamma_N = \frac{G_F^2\left|\theta_e\right|^2m_N^5}{192\pi^3} \left[\frac{1}{4}\left(1+4\sin^2\theta_W+8\sin^4\theta_W\right)+1\right]\ .
\end{align}
The decay rate as computed in \texttt{pyBBN} is given by:
\begin{align}
\Gamma_\mathrm{code} = -I_\mathrm{coll} \frac{E_N H}{f_N m_N}\ ,
\end{align}
with $E_N$ the energy of the HNL, $f_N$ its distribution function and $H$ the Hubble parameter. For an HNL of mass $m_N = 30$ MeV and mixing angle $|\theta_e|^2 = 10^{-4}$, we find the maximum discrepancy between the theoretical and numerical decay rates to be of the order $10^{-3}$ in the temperature range from $T = 30\,\mathrm{MeV}$ down to $T= 0.1\,\mathrm{MeV}$.

\paragraph{Two-body decay width.}
A similar procedure can be followed for two-body decays. In this test, the decay $N \rightarrow \nu_e + \pi^0$ is considered. The corresponding decay width is given by:
\begin{align}
\label{eq:HNLdecaywidth3p}
\Gamma_N = \frac{G_\mathrm{F}^2f_\pi^2m_N^3|\theta_e|^2}{32\pi}\left[1-\frac{m_\pi^2}{m_N^2}\right]^2\ .
\end{align}
For an HNL of mass $m_N = 150\,\mathrm{MeV}$ and mixing angle $|\theta_e|^2 = 10^{-6}$ we find a maximum relative discrepancy between the theoretical and numerical values at the level of $10^{-7}$.

\subsubsection{Number of Decayed and Created Particles}
In this test we check whether the number of electron neutrinos and muon neutrinos created from HNL decays matches the number of HNLs that have decayed. Consider the reactions
\begin{align}
\label{eq:HNL_4pdecayeqnum}
&N \rightarrow \nu_e + \nu_\mu + \overline{\nu_\mu}\\
\label{eq:HNL_3pdecayeqnum}
&N \rightarrow \nu_e + \pi^0\\
\label{eq:HNL_4pscateqnum}
&\nu_e + \nu_e \leftrightarrow \nu_e + \nu_e\\
\label{eq:HNL_4pscateqnum2}
&\nu_e + \overline{\nu_e} \leftrightarrow \nu_e + \overline{\nu_e}\ .
\end{align}
For each HNL that decays in Eq.~\eqref{eq:HNL_4pdecayeqnum}, one electron neutrino and two muon neutrinos (one from this decay, other from charge conjugated channel) are created. Same holds for the reaction in Eq.~\eqref{eq:HNL_3pdecayeqnum}: For each HNL that decays, one electron neutrino is created. The addition of the reactions in Eqs.~\eqref{eq:HNL_4pscateqnum} and \eqref{eq:HNL_4pscateqnum2} conserve the number of created particles. The number of electron- and muon-neutrinos created is:
\begin{align}
    N_{\nu_e} &= \mathrm{BR}_{N \rightarrow \nu_e + \nu_\mu + \overline{\nu_\mu}}\Delta N + \mathrm{BR}_{N \rightarrow \nu_e + \pi^0}\Delta N = \Delta N\\ 
    N_{\nu_\mu} &= 2\mathrm{BR}_{N \rightarrow \nu_e + \nu_\mu + \overline{\nu_\mu}}\Delta N\ ,
\end{align}
where $\Delta N$ is the number of HNLs that have decayed. In this test, we choose $m_N = 143$ MeV and $\tau_N = 0.1$ s, such that the branching ratios are $\mathrm{BR}_{N \rightarrow \nu_e + \nu_\mu + \overline{\nu_\mu}} = 0.3$ and $\mathrm{BR}_{N \rightarrow \nu_e + \pi^0} = 0.7$.

\begin{figure}[t!]
\begin{center}
\includegraphics[width=0.9\textwidth]{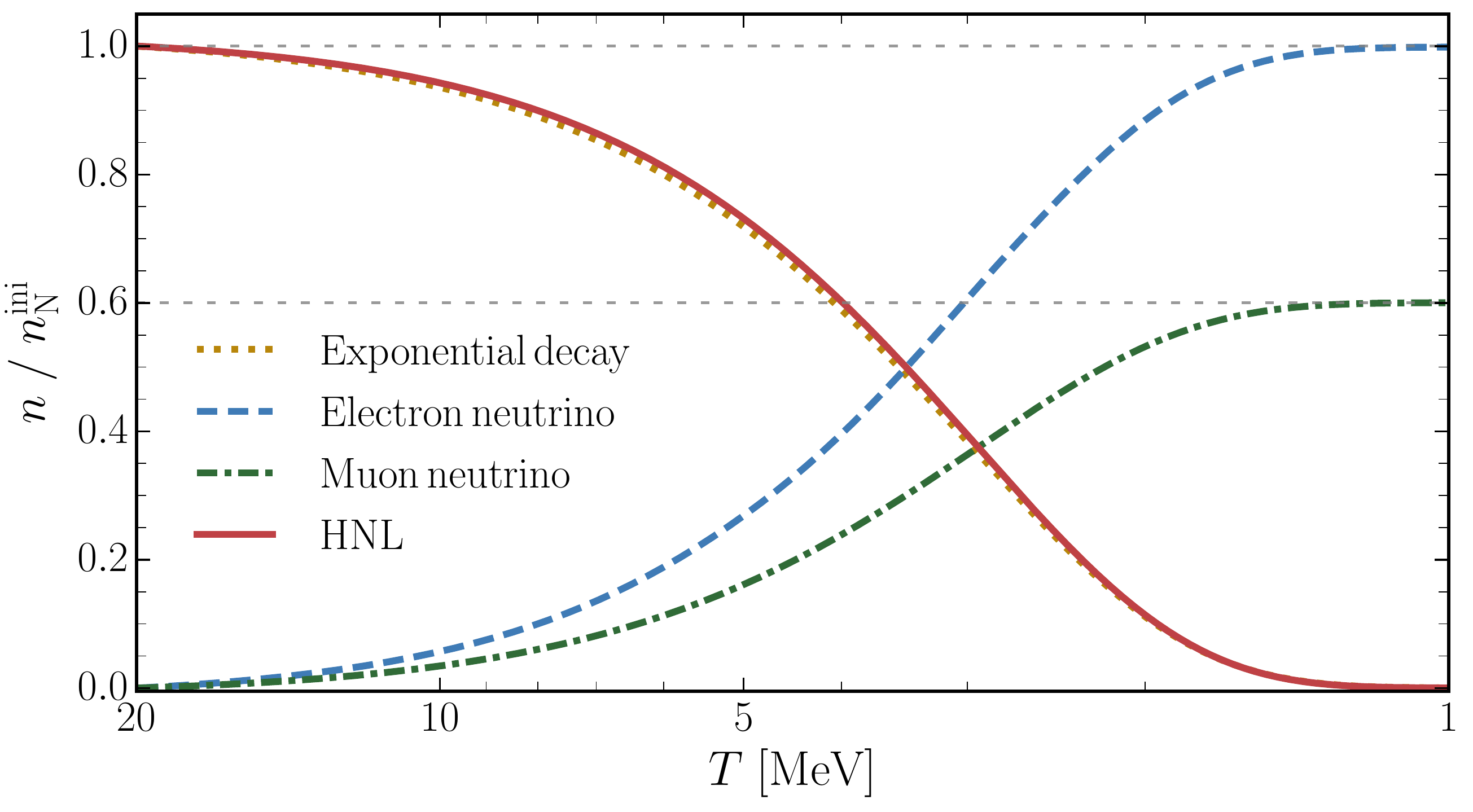}
\end{center}
\caption{Comoving number density $n$ of HNLs, electron neutrinos and muon neutrinos, normalized by the initial HNL comoving number density $n_N^\mathrm{ini}$, where the reactions~\eqref{eq:HNL_4pdecayeqnum} -- \eqref{eq:HNL_4pscateqnum2} are considered. The mass of the HNL here is $m_N = 143\,\mathrm{MeV}$ and the lifetime is $\tau_N = 0.1\,\mathrm{s}$. The initial densities of the electron and muon neutrinos are taken 0 for convenience. The dotted line is the exponential decay according to Eq.~\eqref{eq:HNLexponentialdecay}.}
\label{fig:decay_and_creation}
\end{figure}

Assuming HNLs are stationary and decay in vacuum, the number density falls off exponentially:
\begin{align}
\label{eq:HNLexponentialdecay}
n_N(t) = n_{N}^\mathrm{ini}e^{-{t}/{\tau_N}}\ ,
\end{align}
with $n_{N}^\mathrm{ini}$ the initial HNL number density and $\tau_N$ the lifetime of the HNL. Note that the first assumption slightly overestimates the decay width, as a fraction of the HNLs have nonzero momentum. Figure~\ref{fig:decay_and_creation} shows the decrease of the HNL abundance and the corresponding increase of the electron neutrino and muon neutrino abundances in terms of comoving number densities.

\subsubsection{Reheating due to Secondary Interactions}
Some of the HNL decay products are unstable and will decay in their turn. Such secondary decays can inject entropy into the electromagnetic sector of the plasma and heat it up. Consider the decays
\begin{align}
\label{eq:HNLneutralpiondecay}
N &\rightarrow \pi^0 + \nu_e\\
\label{eq:neutralpiondecay}
\pi^0 &\rightarrow \gamma\gamma\ .
\end{align}
For each HNL that decays, one neutral pion is created. If the HNL has energy very close to that of the neutral pion, then the neutral pion created will be highly non-relativistic. The energy $\Delta\rho$ injected in the plasma due to neutral pion decay during each time step is therefore approximately:
\begin{align}
\Delta\rho \approx 2m_{\pi^0}\Delta n\ ,
\end{align} 
with $\Delta n$ the number density of HNLs that have decayed during the time step.
The factor of 2 comes from the fact that the charge conjugated channel also creates a neutral pion. Note that the heat-up in \texttt{pyBBN} is quantified by $\widetilde{T} = aT$ (see Appendix \ref{app:TemperatureEvolution}). The comoving photon energy density then becomes:
\begin{align}
\rho_{\gamma,\mathrm{c}} = g_\gamma\frac{\pi^2}{30}(aT)^4_\mathrm{new} = \rho_{\gamma,\mathrm{c}}^{\mathrm{old}} + a^4\Delta\rho\ ,
\end{align}
and hence:
\begin{align}
\label{eq:NeutralPionDecayaT}
(aT)_\mathrm{new} = \left[\frac{30}{g_\gamma\pi^2}\left(\rho_{\gamma,\mathrm{c}}^{\mathrm{old}} + 2m_{\pi^0}a^4\Delta n\right)\right]^{1/4}\ .
\end{align}
For an HNL of mass $m_N = 135$ MeV and mixing angle $|\theta_e|^2 = 10^{-4}$, we find a maximum relative discrepancy between the theoretical and numerical values at the level of $10^{-4}$.

\begin{figure}[t!]
\begin{center}
\includegraphics[width=0.9\textwidth]{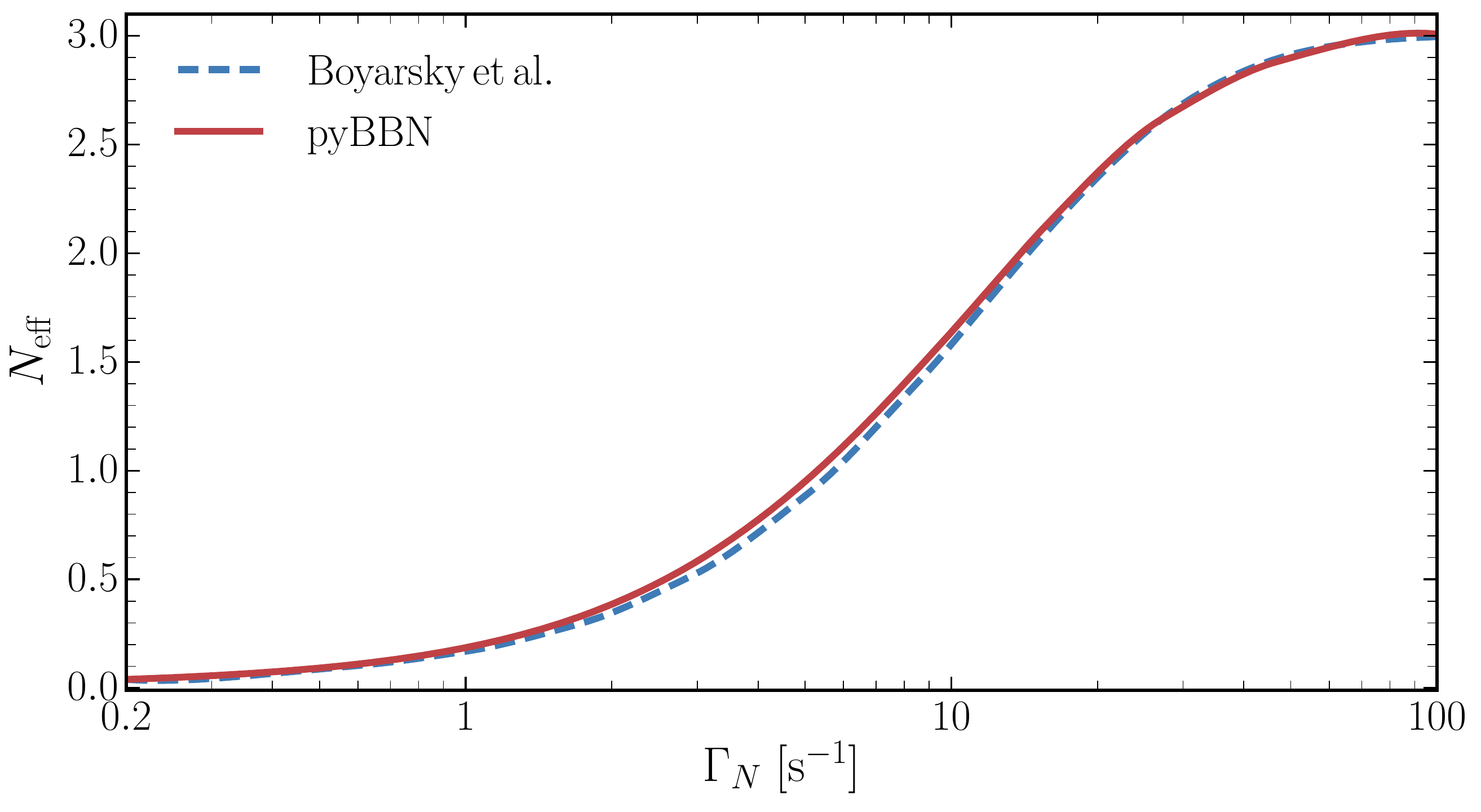}
\end{center}
\caption{Dependence of the effective number of relativistic species $N_\mathrm{eff}$ (Eq.~\eqref{eq:Neffneutrinos}) on the HNL decay width, where only the reheating due to the electrons/positrons created in the decay~\eqref{eq:Nneutrinoelecposdecay} is considered. The solid curve is the result of this work and is obtained by considering an HNL of mass $m_N = 100$ MeV with a decoupling temperature set equal to $T = 300\,\mathrm{MeV}$. The dashed curve is from \cite{MV_paper}.}
\label{fig:HNL_Neff}
\end{figure}

\subsubsection{Effect of Late Reheating on N{\footnotesize eff}}
\label{app:reheating_Neff}
Decays of HNLs into the electromagnetic sector heat up the plasma. This will dilute the abundance of neutrinos, if such decays happen during or after neutrino decoupling. This effect can be quantified by the decrease of the effective number of extra relativistic species $N_\mathrm{eff}$. In this test we consider a plasma that consists of photons, neutrinos, electrons/positrons and HNLs. The HNLs considered here have a mass of 100 MeV and are instantly decoupled at $T = 300$ MeV (see~\cite{Hasegawa:2020ctq} for impact of keV-scale HNLs on $N_\mathrm{eff}$). We include only one decay channel:
\begin{align}
\label{eq:Nneutrinoelecposdecay}
N \rightarrow \nu_e + e^+ + e^-\ .
\end{align}
The created neutrinos are artificially removed and it is assumed that all the energy is injected into the electromagnetic sector. Besides this interaction, we also include all SM interactions (see Appendix~\ref{appsec:SMMatrixElements}).
The effective number of extra relativistic species, defined as
\begin{align}
\label{eq:Neffneutrinos}
N_\mathrm{eff} =  \frac{8}{7}\left(\frac{11}{4}\right)^{4/3}\left(\frac{\rho_\mathrm{tot} - \rho_\gamma}{\rho_\gamma}\right)\ ,
\end{align}
depends on the HNL decay width, since the temperature at which the HNLs decay determines whether neutrinos are interacting strongly enough to stay in equilibrium. The results are shown in Figure~\ref{fig:HNL_Neff}, where we also compare with~\cite{MV_paper}. We observe that for high decay rate of HNLs, the number of relativistic species $N_\mathrm{eff}$ approaches the SM value ${\sim}3$, since the decays happen while neutrinos still strongly interact and are therefore in equilibrium with the plasma. For smaller decay rates, injection happens around or after neutrino decoupling, which severely dilutes the SM neutrino background. Finally, we note that the influence of late reheating on $N_\mathrm{eff}$ has been also studied in, e.g., \cite{Kawasaki:2000en, Hannestad:2004px, Ruchayskiy:2012si, Hasegawa:2019jsa}, but we miss a few key details on their simulations to make a reliable comparison possible. For what it is worth, the agreement between \texttt{pyBBN} (Figure~\ref{fig:HNL_Neff}) and these references is reasonable.